\documentclass[11pt,a4paper]{article}
\pdfoutput=1
\usepackage{jheppub}  
\usepackage{amsfonts,amsbsy,amsmath,amssymb}
\usepackage{color}
\usepackage{graphicx}
\usepackage[colorinlistoftodos]{todonotes}
\usepackage{booktabs}
\usepackage{subcaption} 
\usepackage{multirow} 
\usepackage{multicol}
\usepackage{here}
\usepackage[section]{placeins}
\usepackage[normalem]{ulem}
\usepackage{todonotes}
\usepackage{longtable}

\usepackage{bm}             %lettere greche in grassetto \bm{"lettera"}
\usepackage{physics}
\usepackage{dsfont}         %insiemi numerici fighi \mathbb{robadascrivere}
\usepackage{amsopn}         %dichiarare nuovi operatori
\usepackage{mathrsfs}       %caratteri matematici fighi \mathscr{roba}
\usepackage{xfrac}          %frazioni nel testo \sfrac{num}{den}
\usepackage{bbold}          %numeri tipo la matrice identità \mathbb{1}
\usepackage{slashed}        %notazione di feynmann es: \slashed{\partial}
\usepackage{mathtools}      %per fare le frecce con sopra testo \xrightarrow[testosopra]{testosotto}

\usepackage{pgf}
\usepackage{pgfplots}
\usepgfplotslibrary{dateplot}

 \newcommand{\be}{\begin{equation}}
\newcommand{\ee}{\end{equation}}
\newcommand{\ba}{\begin{align}}
\newcommand{\ea}{\end{align}}
\newcommand{\baa}{\begin{array}}
\newcommand{\eaa}{\end{array}}
\newcommand{\bea}{\begin{eqnarray}}
\newcommand{\eea}{\end{eqnarray}}
\newcommand{\half}{\frac{1}{2}}
\newcommand{\hmu}{\hat \mu}
\newcommand{\hnu}{\hat \nu}
\newcommand{\ad}{\mathrm{adj}}
\newcommand{\tek}{\mathrm{TEK}}
\newcommand{\MS}{{\overline{\rm MS}}}
\newcommand{\GF}{{\rm gf}}
\newcommand{\tl}{\tilde{l}}

\newcommand{\refst}{Gonzalez-Arroyo:2012euf}

\providecommand*{\eu}{\ensuremath{\mathrm{e}}}
\providecommand*{\iu}{\ensuremath{\mathrm{i}}}

\title{Scale setting for large-$N$ SUSY Yang-Mills on the lattice}

\author[a,b]{Pietro Butti,}
\author[a]{Margarita Garc\'{i}a P\'erez,}
\author[a,b]{Antonio Gonz\'alez-Arroyo,}
\author[c,d]{\\ Ken-Ichi Ishikawa,}
\author[d]{Masanori Okawa}

\affiliation[a]{Instituto de F\'{i}sica Te\'orica UAM-CSIC, Nicol\'as
  Cabrera 13-15, Universidad Aut\'onoma de Madrid, Cantoblanco, E-28049 Madrid, Spain}
\affiliation[b]{Departamento de F\'{i}sica Te\'orica, 
M\'odulo 15,  Universidad Aut\'onoma de Madrid, Cantoblanco, E-28049 Madrid, Spain}

\affiliation[c]{Core of Research  for the Energetic Universe, Graduate School of Advanced Science and Engineering, Hiroshima University, Higashi-Hiroshima, Hiroshima 739-8526, Japan}
\affiliation[d]{Graduate School of Advanced Science and Engineering, Hiroshima University,  Higashi-Hiroshima, Hiroshima 739-8526, Japan}

\emailAdd{pietro.butti@uam.es}
\emailAdd{margarita.garcia@uam.es}
\emailAdd{antonio.gonzalez-arroyo@uam.es}
\emailAdd{ishikawa@theo.phys.sci.hiroshima-u.ac.jp}
\emailAdd{okawa@hiroshima-u.ac.jp}

\abstract{ 
In this paper we study the large $N$ limit of four-dimensional Supersymmetric Yang-Mills 
on the lattice using twisted reduced models. We have generated  configurations with
dynamical massive gluinos and various lattice 't Hooft couplings, and verified that the Pfaffian remains positive. We have determined the lattice spacing in terms of various observables obtaining compatible results. Extrapolating to the massless gluino limit we obtain  the lattice spacing dependence on the bare couplings for the supersymmetric theory. The observed dependence goes along the expected behaviour predicted by perturbation theory.
}

\preprint{%
{%\flushright 
IFT-UAM/CSIC-22-39, HUPD-2206
}}

\begin{document}

    \maketitle
    
    \section{Introduction}
    \label{s:introduction}
        This paper deals with $\mathcal{N}=1$ Supersymmetric Yang-Mills theory in the large-$N$ limit. Hence, it  combines two of the ingredients which generally add
a simplification to the pure Yang-Mills theory and  appear in
less traditional  approaches to the study of quantum field
theories~\cite{Witten:1997ep}.
Our work, however, is based on  the more standard Lattice Gauge Theory
approach,  which has proven well suited to address the non-perturbative 
dynamics of QCD and other gauge theories.  In this context, however, 
both Supersymmetry and the large $N$ limit pose additional problems to be 
faced within this approach\footnote{We skip here the extensive
literature in this field and focus upon $\mathcal{N}=1$ SUSY Yang-Mills.}. 
Supersymmetry is broken by the lattice formulation and has to be recovered 
in the continuum limit (For an introduction to Supersymmetry on the lattice 
the reader can consult ref.~\cite{Kaplan:2007zz}). The advantage in our case is that the
$\mathcal{N}=1$ SUSY Yang-Mills theory is just pure Yang-Mills with an
additional massless Majorana fermion in the adjoint representation:
the gluino. The argument made by Kaplan~\cite{Kaplan:1983sk} and Curci
and Veneziano~\cite{Curci:1986sm} is that in the limit in which the mass
of the gluino vanishes  one  recovers
SUSY invariance. With this idea in mind several authors  have studied the
problem in the past, usually for small rank groups as SU(2) and SU(3). 
Given the important role played by chiral invariance in this context a
good deal of work has focused upon overlap and domain-wall fermion
lattice formulations \cite{Nishimura:1997vg, Neuberger:1997bg,
Kaplan:1999jn, Fleming:2000fa, Endres:2008tz, Kim:2011fw,Piemonte:2020wkm}. Another well-explored possibility is
to use Wilson fermions~\cite{Curci:1986sm,Montvay:1995ea, Donini:1996nr}.
One disadvantage of this last approach is that the positive-definite
character of the Pfaffian is not guaranteed and has to be checked or
corrected for.   There
is an extensive study done for SU(2) and SU(3) by a
collaboration involving Desy, M\"unster, Regensburg and Jena Universities~\cite{Ali:2019agk,Ali:2018fbq,Ali:2018dnd,Munster:2014cja,Bergner:2015adz}. 
We  will essentially follow the same discretization and basic 
methodology. 

The other  ingredient in our work is the study of the large $N$ limit of
the theory which, as mentioned at the beginning of the first paragraph, is a further
element in matching with results accessible to other methodologies.
Although both perturbation theory and strong coupling
results~\cite{Gabrielli:1999zt} simplify in the large $N$ limit,  
dealing with  the increasing number of degrees of freedom in Monte
Carlo methods in a theory like this with dynamical fermions represents a 
formidable challenge. Our approach to the large $N$ limit is based on volume
reduction~\cite{Eguchi:1982nm}, which makes a more efficient use of
both spatial and internal degrees of freedom. For the pure gauge part we employ the
so-called twisted Eguchi-Kawai
model~\cite{Gonzalez-Arroyo:1982hwr,Gonzalez-Arroyo:1982hyq,Gonzalez-Arroyo:2010omx}
which has
been used successfully to study pure Yang-Mills
theory in the large $N$ limit~\cite{Gonzalez-Arroyo:2012euf,
GarciaPerez:2014azn,Perez:2020vbn}. 
As explained in ref.~\cite{Gonzalez-Arroyo:1982hyq} the perturbative proof of the  reduction mechanism 
also holds for  theories with fermions in the adjoint representation.
This was applied previously to non-supersymmetric theories with
various flavours of adjoint Dirac fermions~\cite{Gonzalez-Arroyo:2013bta}. 
It was used to study the string tension and other
features~\cite{Gonzalez-Arroyo:2012ztz, Gonzalez-Arroyo:2013gpa, Gonzalez-Arroyo:2013dva,Gonzalez-Arroyo:2012vhw}
and most importantly  to analyze the possible infrared conformal behaviour of
these theories and compute its mass anomalous
dimension~\cite{GarciaPerez:2015rda}.
 Here we will apply the same methodology  to the case of one Majorana fermion.
There are some inherent limitations of the volume reduction method which prevent us, for the
time being, of accessing interesting quantities as the meson and
glueball spectrum. However, there are other physical quantities that
can be computed and in particular  the study of the
scale-setting of the theory, a necessary step in connecting lattice
quantities to continuum ones. Some partial results were already
presented at the 2021 Lattice conference~\cite{Butti:2021qre,
Butti:2021rbf}.

In the next sections we will spell out the details of our lattice
approach and computational techniques. Section~\ref{s:reducedmodels} presents the model used in our study, involving a volume-reduced version of Wilson action with twisted boundary conditions and a fermionic part of the action corresponding to Wilson fermions in the adjoint representation. The Majorana character shows up in the appearance of the Pfaffian, which poses potential non-positivity problems. We also explain how for finite $N$, our gluons and gluinos can be thought of as propagating in a four-dimensional box of size $a\sqrt{N}$. This is an important piece of information in  interpreting the range of parameters used in our study. The present work is a rather extensive study involving four values of the lattice coupling, three values of $N$ and several values of the gluino mass (hopping parameter). The methodology used in generating the configurations is explained in Section~\ref{s:generationofconfs}. We also present  our results regarding the sign of the Pfaffian. Then we proceed to locate the value of the hopping parameters corresponding to massless gluinos in Section~\ref{s:chirallimit}. These are precisely the points at which our model will represent the $\mathcal{N}=1$ Supersymmetric Yang-Mills theory.  The following two sections are dedicated to the main problem under consideration: determining the value of the lattice spacing in some units for each one of our simulation parameters. The first part (Section~\ref{s:scalesetting}) explains the underlying philosophy and presents several possible continuum observables that will  be used to set the scale. The following one (Section~\ref{s:results}) elaborates on the lattice counterparts, the technical problems involved and the results obtained. Redundancy is a proof of the robustness of our results. 
The closing section is then devoted to extracting the corresponding scales for the ${\mathcal N}=1$ SUSY Yang-Mills theory, by extrapolating the 
previous results to the massless gluino limit. A comparison with expectations from perturbation theory in the continuum seems to work remarkably well for the range of values explored in the study. We close with a brief summary of the results.

    \section{\texorpdfstring{$\mathcal{N}=1$}{N=1} SUSY Yang-Mills at large-\texorpdfstring{$N$}{N} on the lattice}
    \label{s:reducedmodels}
        % \section{Reduced model for $\mathcal{N}=1$ SUSY Yang-Mills at large-$N$}
%    \label{s:reducedmodels}

In this section we specify  the model which is used to generate our results. The starting point is  the basic Wilson  
discretization of the $\mathcal{N}=1$ SUSY Yang-Mills action for arbitrary $N$. 
For the pure gauge part this is given by  Wilson plaquette action:
\be
S_g= b N \sum_n \sum_{\mu \ne \nu} \Tr \left [ \mathbf{I} - V_\mu (n) V_\nu(n+\hmu) V_\mu^\dagger (n + \hnu) V_\nu^\dagger (n) \right] ,
\ee
where the link variables $V_\mu(n)$ are SU($N$) matrices (in the fundamental representation) and $b$ is the inverse of the lattice 't Hooft coupling.
For the fermionic part it is given by the Wilson-Dirac action for one adjoint Majorana fermion:
\be
S_f= \half \sum_{n,m} \Psi^t(n) C D_W(n,m) \Psi (m) ,
\ee
where $D_W(n,m)$ stands for the Wilson-Dirac operator in the adjoint representation:
\begin{align}\label{eqn:WilsonDiracop}
&D_W(n,m) = \delta(n,m) \, \mathbf{I} \\
&- \kappa_a \sum_\mu \left [ (1 -\gamma_\mu) V_\mu^\ad (n) \delta (n+\hmu, m)
+
(1 + \gamma_\mu) \left(V_\mu^\ad (n-\hmu) \right)^\dagger  \delta (n-\hmu, m) \right].
\nonumber
\end{align}
where $\kappa_a$ is the adjoint Majorana-fermion hopping parameter and $V_\mu^\text{adj}$ is the link variable in the adjoint representation.
The charge conjugation matrix $C$  satisfies the following properties 
\begin{align}
&\gamma_\mu^t C = -C \gamma_\mu \label{eq:cgamma},\\
& C^t= -C.
 \label{eq:ct}
  \end{align}
ensuring the antisymmetric character of the $C D_W$ matrix.

In the large $N$ limit the twisted reduction prescription~\cite{Gonzalez-Arroyo:1982hyq} implies that the model is equivalent to a matrix model 
obtained as follows (see also refs.~\cite{Eguchi:1982ta,Aldazabal:1983ec}). One introduces 4 SU($N$) constant matrices $\Gamma_\mu$ satisfying
\be
\Gamma_\mu \Gamma_\nu = z_{\mu \nu} \Gamma_\nu \Gamma_\mu,
\ee
with
\be
z_{\mu \nu} =   e^{2 \pi i n_{\mu \nu} /N}, 
\ee
given in terms of the twist tensor $n_{\mu \nu}$, which is an antisymmetric tensor of integers modulo $N$. 
The displacement operator by one lattice point in the direction $\mu$ is then replaced by the adjoint action by the matrix $\Gamma_\mu$. For the link matrices 
this is equivalent to the substitution 
\be
V_\nu(n+\hmu)= \Gamma_\mu V_\nu(n) \Gamma_\mu^\dagger.
\ee
After doing these substitutions, dropping the $n$ label and changing variables to $U_\mu=V_\mu \Gamma_\mu$, the pure gauge action converts 
into the one of the twisted Eguchi-Kawai (TEK) model~\cite{Gonzalez-Arroyo:1982hwr,Gonzalez-Arroyo:1982hyq}:
\be
  S_\tek(U) =  b N \sum_{\mu\ne\nu} \mathrm{Tr}\left[ I - z_{\mu\nu}^* U_{\mu}U_{\nu}U_{\mu}^{\dag}U_{\nu}^{\dag}\right].
               \label{eq:TEKaction}
\ee
The same prescription applies to other fields in the adjoint representation such as the gluino fields. There are two ways to represent the 
adjoint fermion fields. If we represent them as traceless $N\times N$ matrices the prescription  is similar to that of the links
\be
\Psi (n+\hmu) = \Gamma_\mu \Psi(n) \Gamma_\mu^\dagger.
\ee
If we represent these fields as $N^2-1$ dimensional vectors the prescription is the equivalent one:
\be
\Psi (n+\hmu) = \Gamma_\mu^\ad \Psi (n).
\ee
In this way one reaches the reduced form of the Wilson-Dirac operator:
\be
D_W = \mathbf{I} - \kappa_a \sum_\mu \left [ (1 -\gamma_\mu) U_\mu^\ad + (1 + \gamma_\mu) \left(U_\mu^\ad\right)^\dagger \right].
\label{eq:AdjWDmatrix}
\ee
This procedure was used already in ref.~\cite{Gonzalez-Arroyo:2013bta} in constructing the SU($N$) gauge theory  coupled to  several flavours of adjoint Dirac fermions. 

After integration over the fermion  fields,  we finally arrive to the model used in our simulations,  whose partition function is given by 
\be
  {\cal Z}  =\int \prod_{\mu=0}^4 dU_{\mu} \, \mathrm{Pf}\left(CD_W \right) e^{-S_\tek(U)},
               \label{eq:partitionfunction}
\ee
where the integration over the group variables $U_{\mu}$ is given by the Haar measure, and $\mathrm{Pf}(M)$ 
denotes the Pfaffian of the antisymmetric matrix $M$. We still need to specify the choice of the twist tensor $n_{\mu \nu}$. In our case, 
we have chosen the so-called symmetric twist given by:
\be
n_{\mu \nu} = \sqrt{N} k \left ( \theta(\nu-\mu) - \theta(\mu-\nu)\right),
\label{eq:symtw}
\ee
with $\sqrt{N}$ and $k$ coprime integers and with $\theta$ representing the Heaviside theta-function.

The presence of the Pfaffian in eq.~\eqref{eq:partitionfunction} poses the question of whether this defines a positive definite probability distribution. 
Indeed, the general properties of the Dirac operator in the continuum ensures its positivity, but some of these properties do not hold for the lattice Wilson-Dirac operator (they do hold for the overlap~\cite{Neuberger:1997bg}). This brings in the menace of the so-called sign problem. Since the lack of positivity could only be driven by lattice artifacts we expect it not to be too severe. We take advantage of the studies done by the Desy-Jena-Regensburg-Munster collaboration which deals with the same situation~\cite{Ali:2019agk,Ali:2018fbq,Ali:2018dnd,Munster:2014cja}. We can resort to the well-known reweighting method to deal with the sign-problem. This amounts to defining a probability density given by substitution of the Pfaffian by its absolute value:
\begin{equation}
 P(U) = |\mathrm{Pf}\left(CD_W \right)| e^{-S_\tek(U)}.
\end{equation}
The expectation value of an observable $\expval{O}$ in  the original
model is then evaluated as follows
\begin{align}
\expval{O}=
\dfrac{\int \prod_{\mu=1}^{4} \dd{U_{\mu}} \, O(U) \,
\mathrm{sign}\qty(\mathrm{Pf}\qty(CD_W)) P(U)}
      {\int \prod_{\mu=1}^{4} \dd{U_{\mu}}
      \mathrm{sign}\qty(\mathrm{Pf}\qty(CD_W)) P(U)}
      =\dfrac{\expval{O\,\mathrm{sign}\qty(\mathrm{Pf}\qty(CD_W))}_{P(U)}}
             {\expval{
	     \mathrm{sign}\qty(\mathrm{Pf}\qty(CD_W))}_{P(U)}},
\end{align}
The right-hand side involves the ratio of two expectation values
with respect to the probability density $P(U)$. 
In the next section we describe our methodology to generate
configurations according to this distribution.  
	                                  
There is one important consideration that has to be made in interpreting the results of our work. It concerns finite $N$ effects. The reduction mechanism applies at infinite $N$, while the numerical results necessarily are performed at finite although very large $N$. Thus, it is very important to understand the nature and size of finite $N$ effects. The perturbative proof of reduction shows that planar diagrams in the reduced model are identical to those of the ordinary model on a finite lattice of size $(\sqrt{N})^4$. This is not the only difference between the finite $N$ versions of the ordinary and reduced model, but very often is the dominant effect. Hence, it is useful to regard $\sqrt{N}$ as the effective lattice size and finite $N$ errors acquire the form of finite volume effects. Our simulations which use $N=169$, $289$ and $361$ correspond to an effective lattice size of $13^4$, $17^4$ and $19^4$. From that viewpoint extended observables are more affected by this type of errors as compared to more local ones. For example, for Wilson loops the effects grow with the size of the loop and for smeared or flowed quantities with the corresponding smearing radius. A detailed study of these finite $N$ errors on Wilson loops has been done both perturbatively~\cite{Perez:2017jyq} and non-perturbatively~\cite{Gonzalez-Arroyo:2014dua}. 
The finite-$N$/finite-effective volume identification is crucial in interpreting our choice of parameters and our results. For example, a  large value of $b$ implies a small value of the lattice spacing $a$ in physical units. This translates into an effective lattice box of side $a \sqrt{N}$ and this should be kept larger than the characteristic correlation lengths of the theory. This limits the maximum values of $b$ in much the same way as the lattice size does for the ordinary lattice simulations. In conclusion, it is clear that going to larger values of $N$ is of course desirable but this is limited by the computational effort involved. 
    
    \section{Generation of configurations}
    \label{s:generationofconfs}
        In this section we will explain the basic characteristics of our data sample as  well as the methodology used to generate it. We employ a rather extensive set of  configurations generated with the probability distribution explained in the previous section and corresponding to a total of 46 different simulation parameters. This involves three different values of the rank of the group $N=169$, $289$ and $361$, four different values of the inverse lattice coupling $b=0.34$, $0.345$, $0.35$ and $0.36$, and four to six values of the gluino mass for each case. This facilitates the extrapolation to vanishing gluino mass, the analysis of the scaling behaviour of the theory and the appropriate control of all sources of errors. The final list of values of the hopping parameter and the number of configurations generated in each case are listed in table~\ref{tab:ParameterAndStatistics} in appendix~\ref{app:simulation}.

To generate the configurations we use the strategy explained in the  previous section based upon making use of the reweighting method to  deal with the problems associated with the sign of the Pfaffian. Hence, we split the problem into two parts. First, we generate configurations according to the positive definite distribution $P(U)$. This is done by making use of  the Rational Hybrid Monte Carlo (RHMC) algorithm~\cite{Kennedy:1998cu,Clark:2003na,Clark:2005sq,Clark:2006wq,Clark:2006wp}, whose specific features for the case at hand will be explained in the next subsection. Then, we will address the problem of the sign of the Pfaffian by analyzing its distribution over our data.  This is explained in the following subsection. In fact our study is based on the analysis of the low-lying eigenvalues of the adjoint Wilson-Dirac operator, which is also  interesting per se, and serves other purposes as  the determination of the SUSY limit, addressed in the following section.   

The interested reader can consult additional interesting technical details of our  simulation which are described in appendix~\ref{app:simulation}. 

\subsection{Rational Hybrid Monte Carlo algorithm}

To generate the ensemble of $U_\mu$ according to the partition function eq.~\eqref{eq:partitionfunction} 
via Markov chain Monte Carlo methods, we have to transform the integrand to be suitable for Monte Carlo simulations taking into account that the 
Pfaffian could be negative. 
By using the identity
\begin{align}
\qty|\mathrm{Pf}\qty(CD_W)| = \qty|\det\qty[Q_W^2]|^{1/4},
\end{align}
where $Q_W \equiv D_W \gamma_5$, we transform the partition function as follows:
\begin{align}
  \mathcal{Z} &= \int \prod_{\mu=1}^{4} dU_{\mu} \, \mathrm{sign}\qty(\mathrm{Pf}\qty(CD_W))
\qty|\det\qty[Q_W^2]|^{1/4} e^{-S_\tek(U)},
\label{eq:RHMC0}
\end{align}
and employ the RHMC algorithm~\cite{Kennedy:1998cu,Clark:2003na,Clark:2005sq,Clark:2006wq,Clark:2006wp} 
for the probability weight:
\begin{align}
P(U)=\qty|\det\qty[Q_W^2]|^{1/4} e^{-S_\tek(U)}.
\end{align}
Using the pseudo-fermionic integral, the factor $\qty|\det\qty(Q_W^2)|^{1/4}$ is represented by:
\begin{align}
  \qty|\det\qty(Q_W^2)|^{1/4} &= \int \dd{\phi} \dd{\phi}^{\dag} e^{-S_Q(U,\phi)},\quad
S_Q(U,\phi) = \Tr\qty[\phi^{\dag} R_{N_R}^{(-1/4)}\qty(Q_W^2) \phi],
\label{eq:pfaction}
\end{align}
where $\phi$ is the pseudo-fermion field in bi-fundamental form for the adjoint representation 
and $\Tr$ is the trace over the colour index.
The matrix $R_{N_R}^{(p)}\qty(Q_W^2)$ ($p=-1/4$) in eq.~\eqref{eq:pfaction} is an approximation 
for $\qty(Q^2_W)^{p}$
defined through the $N_R$-th order rational polynomial approximation to $x^p$ for a real number $x \in [a,b]$:
\begin{align}
x^p \underset{x\in[a,b]}{\simeq} R_{N_R}^{(p)}\qty(x) \equiv \alpha_0^{(p)} + \sum_{j=1}^{N_R}\dfrac{\alpha_j^{(p)}}{x-\beta_j^{(p)}}.
\end{align}
Using the Remez algorithm, sets of the coefficients $\qty{\alpha_{j=0,\dots,N_R}^{(p)}, \beta_{j=1,\dots,N_R}^{(p)}}$
have been prepared and tabulated for various cases: the power $p=1/8,-1/4$, the dynamic range $b/a$ and the order of the approximation $N_R$.

To generate the pseudo-fermion field $\phi$ from the probability $\exp\qty(-S_Q)$ in eq.~\eqref{eq:pfaction},
we set:
\begin{align}
  \phi &= R^{(1/8)}_{N_R}\qty(Q_W^2)\eta,
\end{align}
where $\eta$ is
drawn from a Gaussian distribution $\exp\qty(-\Tr\qty(\eta^{\dag}\eta))$
at the beginning of the molecular dynamics (MD) evolution in the RHMC algorithm. 
The traceless condition is imposed as $\Tr\qty(\eta)=0$ and $\Tr\qty(\phi)=0$ to remove the unwanted U(1) contribution. 

The sequence of $\qty(Q_W^2 - \beta^{(p)}_j I)^{-1}\phi$, required for the rational polynomial approximation,
is evaluated using the multi-shift conjugate gradient algorithm.
To apply $D_W$ to the pseudo-fermion field $\phi$ in bi-fundamental form, we use
\begin{align}
  D_W \phi =
 \phi
- \kappa_{a} \sum_{\mu}\qty[ \qty(1-\gamma_\mu) U_\mu \phi U_\mu^{\dag}
                        +\qty(1+\gamma_\mu) U_\mu^{\dag} \phi U_\mu ].
\end{align}

The coefficients $\qty{\alpha_{j=0,\dots,N_R}^{(p)}, \beta_{j=1,\dots,N_R}^{(p)}}$ in the RHMC algorithm are determined so as to minimize the metric:
\begin{align}
\Delta_R = \underset{x \in [a,b]}{\min{\max{}}}\qty|x^p - R_{N_R}^{(p)}(x)|,
\end{align}
where the interval $[a,b] $ is set by the lowest and highest eigenvalues of $Q_W^2$.
At the beginning and the end of each MD trajectory,
a highly accurate approximation for the pseudo-fermionic field $\eta$ and action $S_Q$
is needed in the HMC Metropolis test to have the correct weight, c.f. eq.~\eqref{eq:pfaction}. For this purpose a set of coefficients satisfying the condition: $\Delta_R < \mathrm{tol}=10^{-14}$, with the smallest $N_R$, is chosen from the table of set-parameters.
The smallest and largest eigenvalues of $Q_W^2$ are
calculated using a thick restart type Lanczos algorithm that is optimized to the matrix $Q_W^2$,
having the property $J\qty(Q_W^2)J^{-1}=\qty(Q_W^2)^{t}$ with $J=C\gamma_5$~\cite{Ishikawa:2020xac}.
For the coefficient set used during the MD evolution, which should be invariant during the period of generation of configurations,
we fix a lower accuracy $\Delta_R < 10^{-5}$--$10^{-8}$
and select a parameter set for ($N_R$, $a$, $b$), with $a$ and $b$ tuned to cover the eigenvalue range determined during the thermalization period.

In addition, we employ the generalized multiple-time step MD integrator scheme~\cite{HAAR2017113,KAMLEH20121993}
with the fourth order Omelyan-Mryglod-Folk scheme~\cite{Takaishi:2005tz,OMELYAN2003272}.
Three different time step sizes are assigned to the gauge action, and to the UV and IR parts of the pseudo-fermion action.
The pseudo-fermion action in eq.~\eqref{eq:pfaction} is separated at an order $N_{\mathrm{split}} \in (1,N_R)$ as
\begin{align}
  S_Q(U, \phi)
  &= \Tr\qty[\phi^{\dag}\qty(\alpha_0^{(-1/4)}
                           + \sum_{j=1}^{N_{\mathrm{split}}-1}\alpha_j^{(-1/4)}\qty(Q_W^2 - \beta_j^{(-1/4)})^{-1})\phi]
    \notag\\
  &\quad
  + \Tr\qty[\phi^{\dag}\qty( \sum_{j=N_{\mathrm{split}}}^{N_R}\alpha_j^{(-1/4)}\qty(Q_W^2 - \beta_j^{(-1/4)})^{-1})\phi],
\end{align}
where $\beta_j^{(-1/4)}$ are ordered as: $\beta_{N_R}^{(-1/4)} <  \dots < \beta_{2}^{(-1/4)} < \beta_{1}^{(-1/4)}< 0$.
This action defines the Molecular Dynamics force.
Since the coefficients $\alpha_{j}^{(-1/4)}$ align as $0< \alpha_{0}^{(-1/4)} <  \alpha_{1}^{(-1/4)} < \dots  < \alpha_{N_R}^{(-1/4)}$,
we assign the first portion to the IR part and the last one to the UV part,
resulting in a hierarchy in the magnitude of the MD forces~\cite{Clark:2005sq}.
The finest MD time step size is assigned to the gauge action, 
the next finer one to the UV part of the pseudo-fermion,
and the coarsest step size to the IR part, respectively.

\newcommand{\figscale}{0.045}
\begin{figure}[t]
    \centering
    \includegraphics[clip,scale=\figscale]{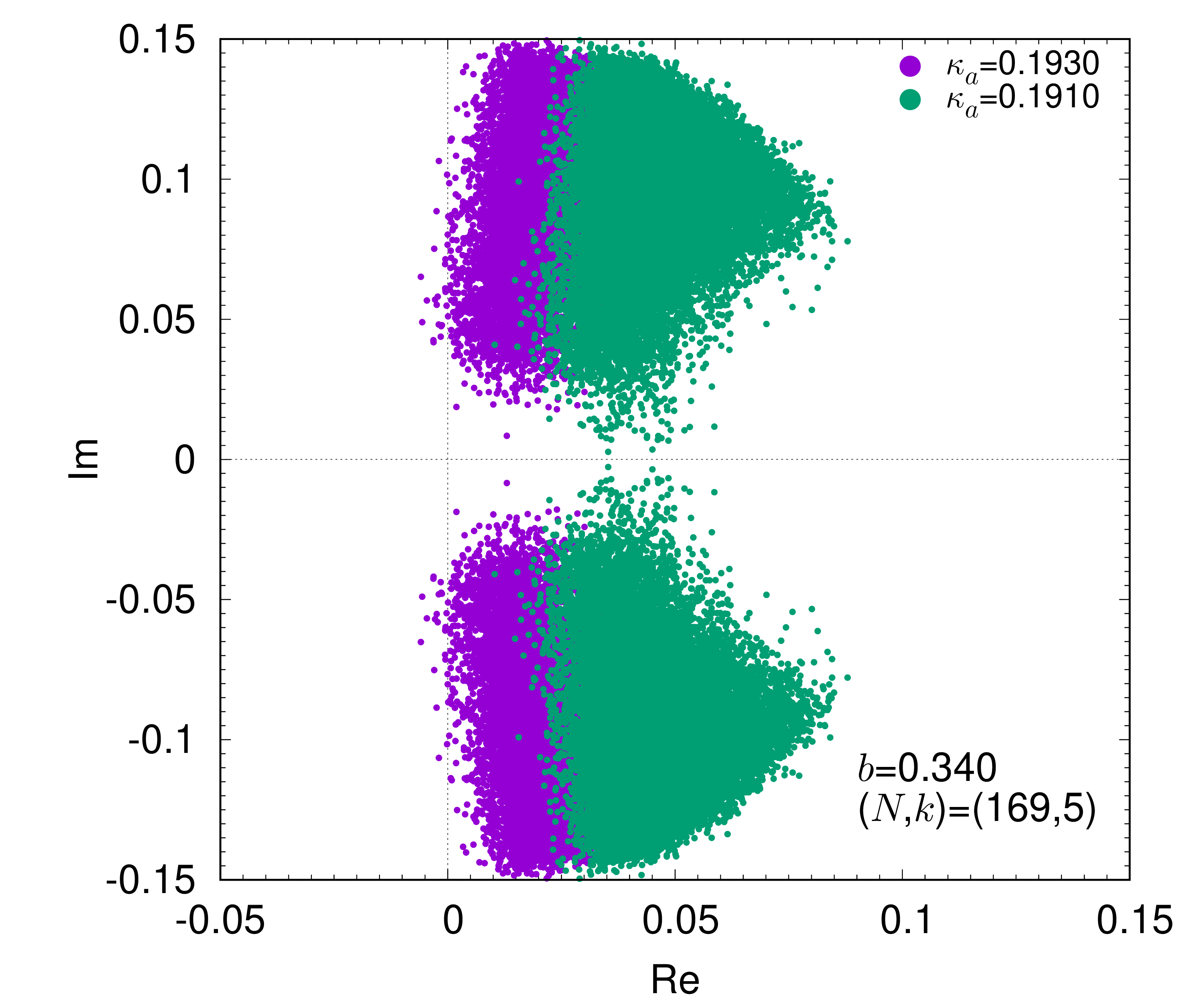}
    \includegraphics[clip,scale=\figscale]{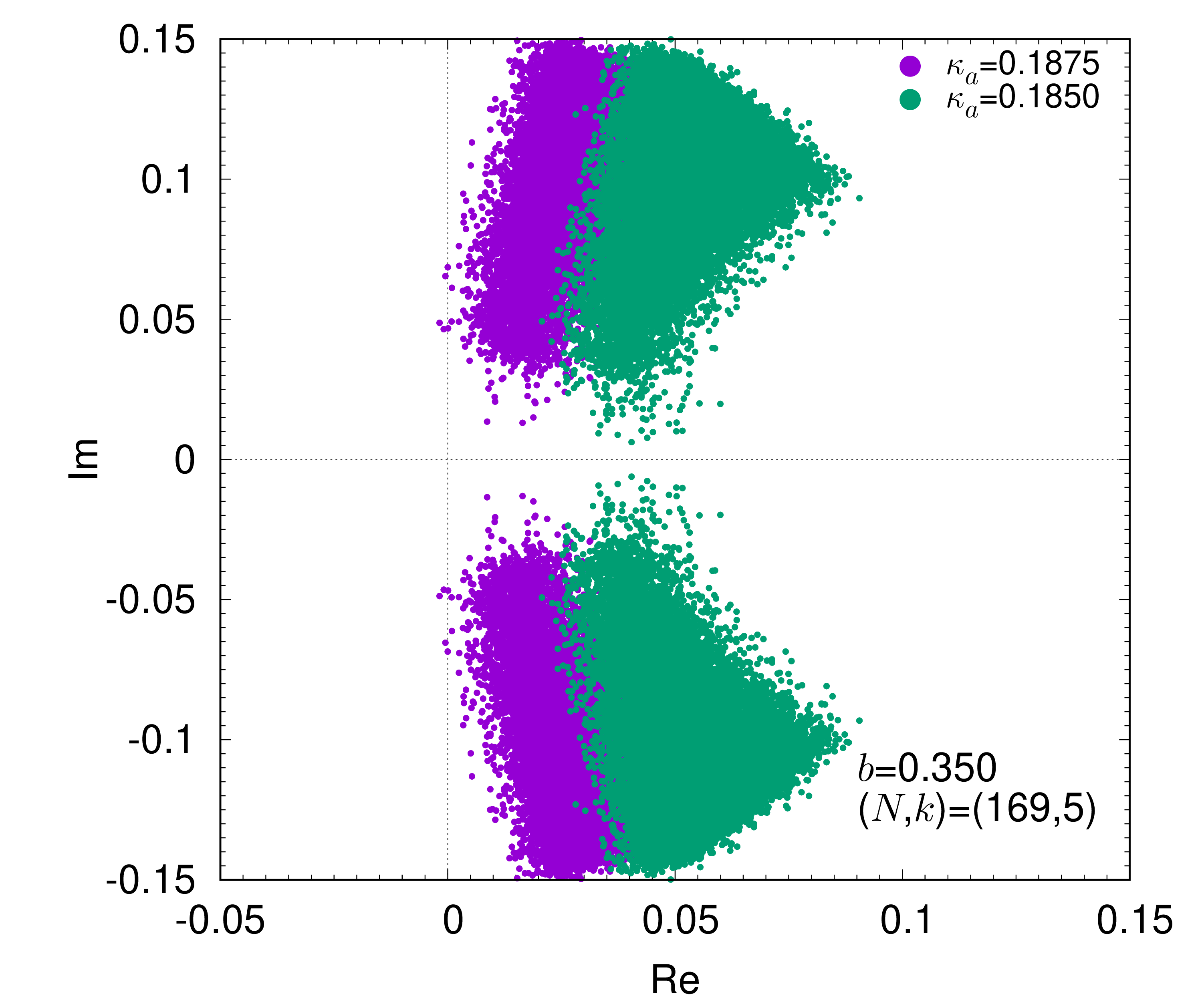}\\
    \includegraphics[clip,scale=\figscale]{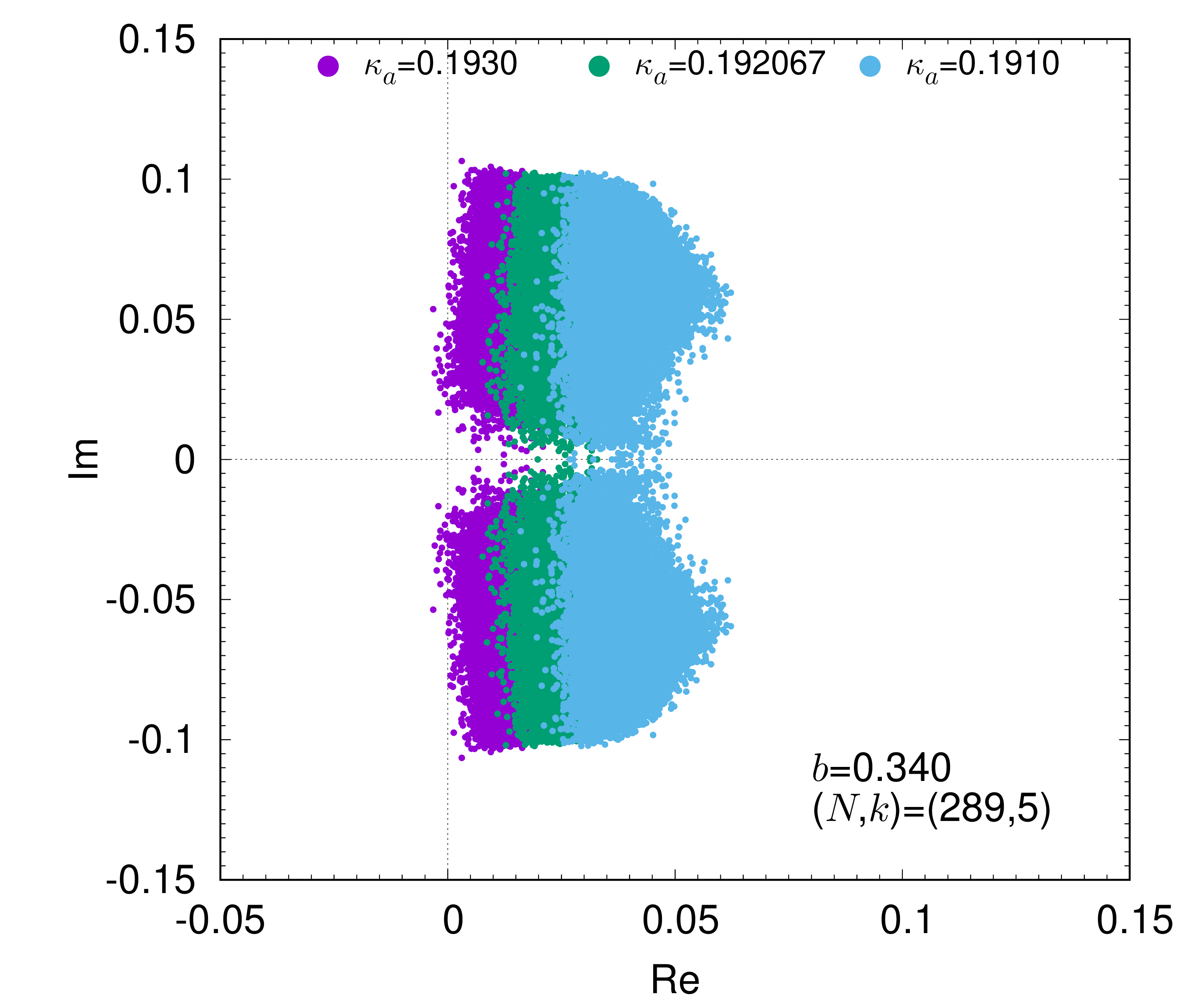}
    \includegraphics[clip,scale=\figscale]{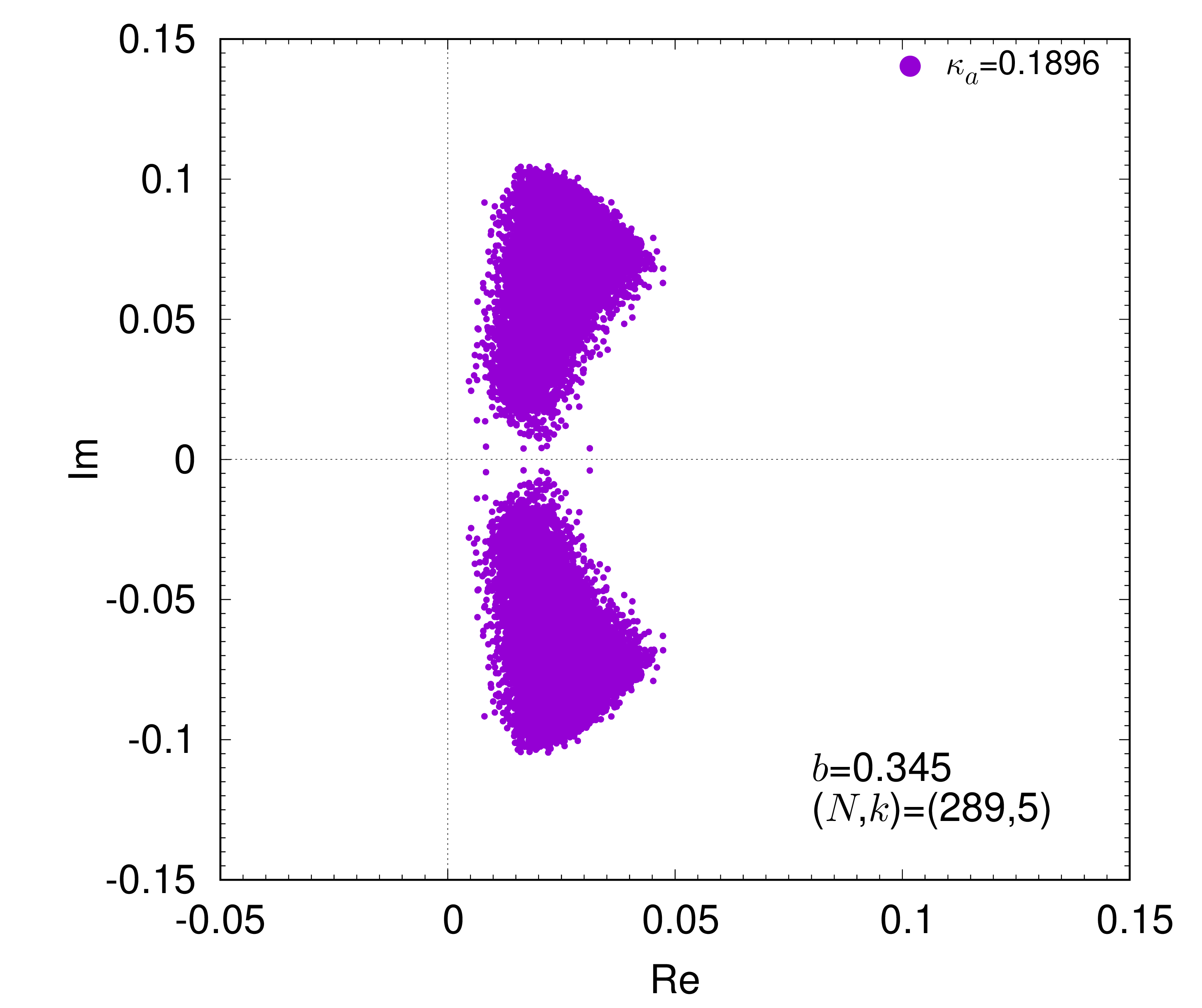}
    \includegraphics[clip,scale=\figscale]{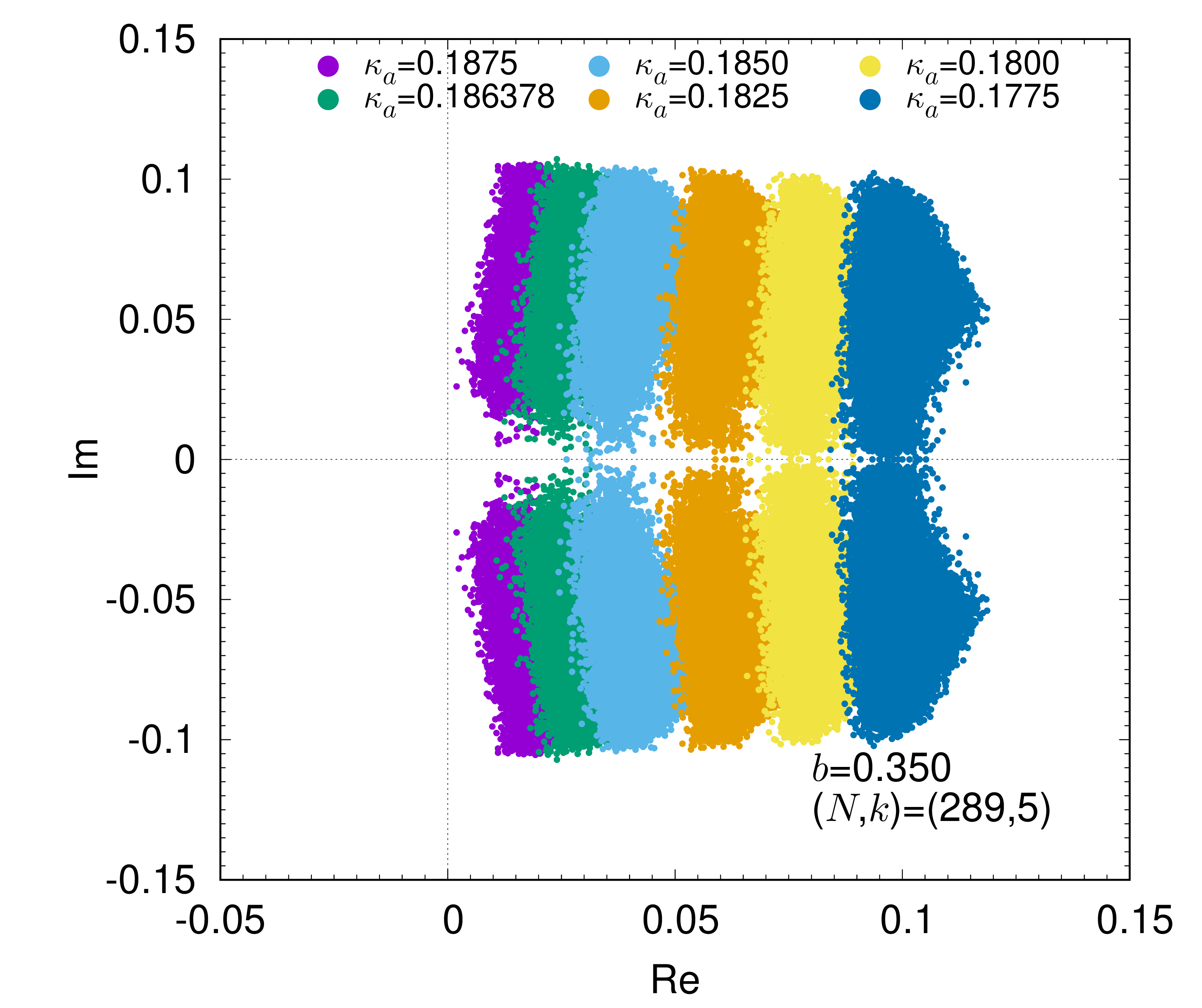}
    \includegraphics[clip,scale=\figscale]{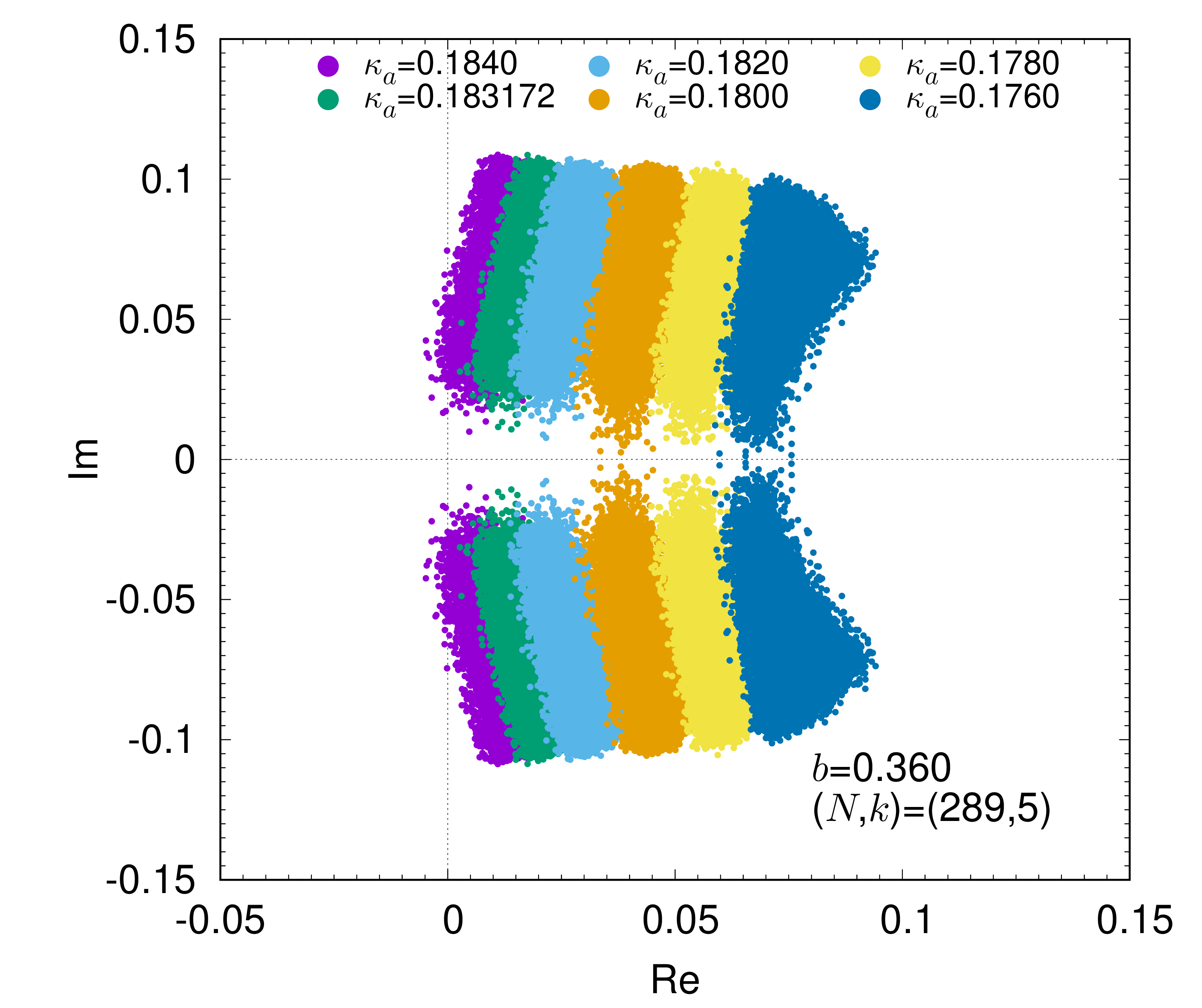}\\
    \includegraphics[clip,scale=\figscale]{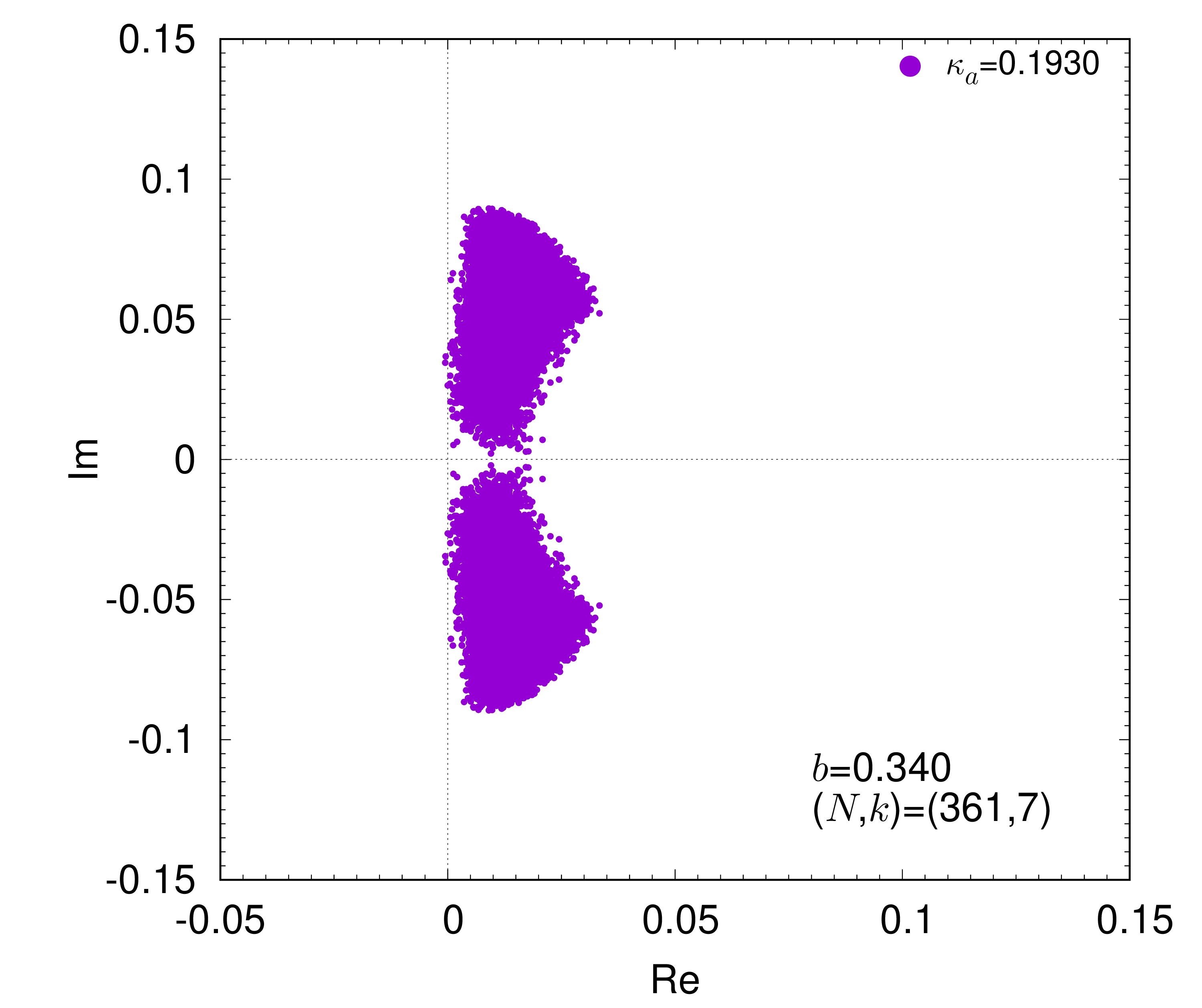}
    \includegraphics[clip,scale=\figscale]{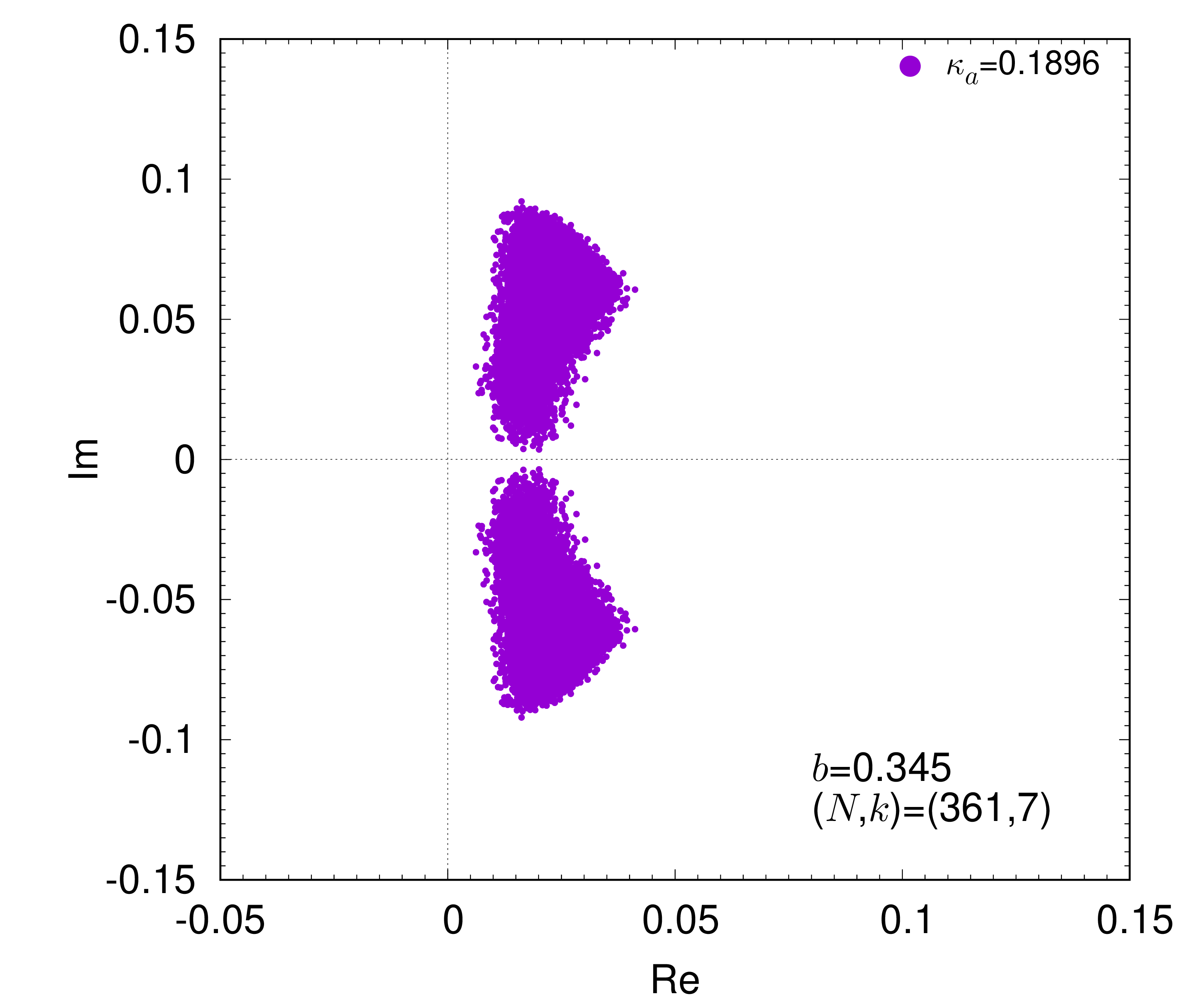}
    \includegraphics[clip,scale=\figscale]{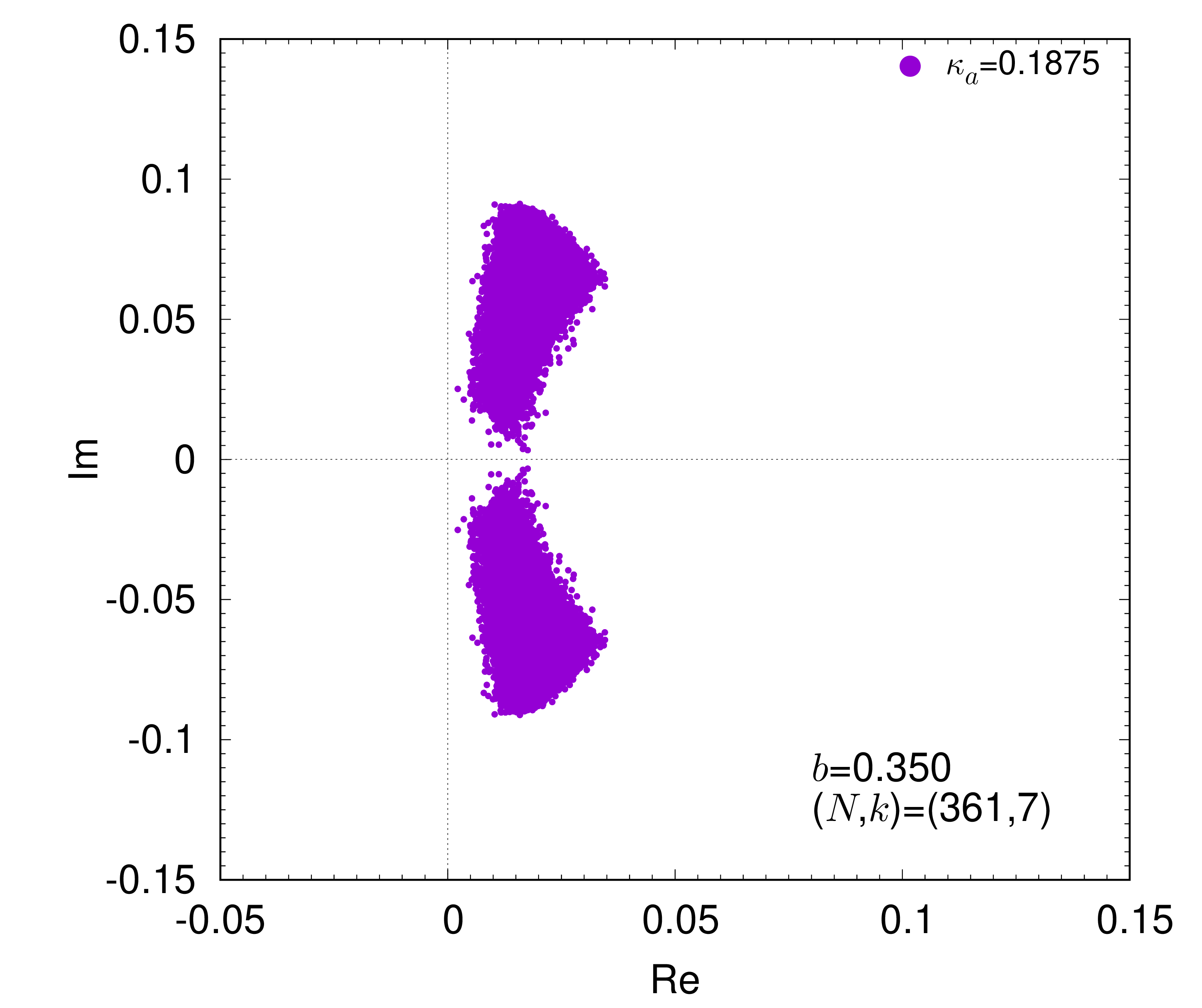}
    \caption{We display in the complex plane and for a subset of our simulation parameters at various values of $N$, the 100 eigenvalues of the Wilson-Dirac operator $D_W$ closest to $-0.1$. }
    \label{fig:complexeigen}
\end{figure}

\subsection{Eigenvalue spectrum and  the sign of the Pfaffian}\label{ss:pfaffian}

As mentioned in the previous section and the preceding subsection our
strategy has been to deal with the effect of the sign of the Pfaffian 
$\mathrm{sign}\qty(\mathrm{Pf}\qty(CD_W))$
in eq.~\eqref{eq:RHMC0} by means of the reweighting method.
In this context the expectation value of an observable $\expval{O}$ is evaluated with
\begin{equation}
\expval{\mathcal{O}} =\dfrac{\expval{\mathcal{O}\,\mathrm{sign}\qty(\mathrm{Pf}\qty(CD_W))}_{P(U)}}
      {\expval{   \mathrm{sign}\qty(\mathrm{Pf}\qty(CD_W))}_{P(U)}},
\end{equation}
where $\expval{\dots}_{P(U)}$ denotes the expectation value with the non-negative weight $P(U)$
which is statistically evaluated with the ensemble generated with the RHMC algorithm.

To compute the sign of the Pffafian for our configurations we make use
of the results of refs.~\cite{Wuilloud:2010wxi,Bergner:2011wf,Bergner:2011zp,Piemonte:2015,Bergner:2015adz},
according to  which, the sign  can be determined by counting the number of negative real 
eigenvalues of $D_W$. To perform this search, we carried out a complete
determination  of all the complex eigenvalues of $D_W$ near the origin of the complex plane, using the \texttt{ARPACK}  library~\cite{lehoucq1997arpack}. 
The shift invert mode with the shift parameter $\sigma=-0.1$, which
computes the extreme eigenvalues of $(D_W - \sigma)^{-1}$,
is used so that the eigenvalues having negative real part can be captured.
Figure~\ref{fig:complexeigen} shows, for all configurations, the 100 complex eigenvalue of $D_W$ 
closest to $z=-0.1$ in the complex plane.
Given the symmetry of $\qty(CD_W)^{t}=-\qty(CD_W)$, each point represents
a two-folded eigenvalue.

For heavy gluino masses we are not expecting flips of sign of the Pfaffian. Hence, our analysis focused on the lightest adjoint fermion masses at each $b$ and $N$.  We did not observe any negative-real eigenvalues in the spectrum. This absence was also checked for several heavier fermion masses. Therefore we conclude that the sign of the Pfaffian is always positive for the model parameters we employed, simplifying the reweighting method by validating the distribution using the absolute value of the Pfaffian. Similar results showing  that the negative sign of the Pfaffian are rare even with moderately light adjoint masses has been observed in lattice SUSY models~\cite{Demmouche:2010sf,Ali:2018dnd}.

    \section{SUSY restoration limit}
    \label{s:chirallimit}
         
 As mentioned in the introduction, $\mathcal{N}=1$ supersymmetry is broken by the lattice discretization and only appears as an emerging feature as one approaches the continuum limit and properly tunes the gluino mass to zero~\cite{Kaplan:1983sk,Curci:1986sm}.  
From a practical point of view, continuum and chiral (massless-gluino) limits are delicate to perform on the lattice. Nevertheless, while extrapolations to the continuum limit are quite standard in every lattice simulation, the methodology regarding how to explore the limit in which the gluino mass vanishes requires a special consideration, particularly when, as in our case, Wilson-Dirac fermions are employed to simulate the gluino. 
With Wilson fermions chiral symmetry is explicitly broken by an $\order{a}$ mass term. The consequence of this is that the bare mass of the fermion acquires an additive renormalization and a tuning of the hopping parameter $\kappa_a$ is required to reach the chiral limit. The way this tuning is done in QCD takes advantage of the fact that pions are pseudo-Goldstone bosons for the spontaneous breaking of chiral symmetry, and their mass is proportional to the square root of the renormalized quark mass. 
On the other hand, in $\mathcal{N}=1$ SUSY Yang-Mills, the chiral-symmetry breaking pattern is different, and no Goldstone excitation appears in the spectrum, one has instead an excitation analogous to the QCD $\eta'$. 
Several methods have been proposed in the literature to attain in this case the limit where SUSY is restored. One of them is to define a renormalized gluino mass using the supersymmetric Ward identities regularized on the lattice and tune this mass parameter to zero \cite{Ali:2018fbq}. Another possibility is to use the connected part of the adjoint-$\eta'$ correlator, this leads to a non-singlet adjoint-pion correlator that can be seen as composed of the gluino and a quenched Majorana fermion. On the basis of partially-quenched chiral perturbation theory~\cite{Munster:2014cja}, this adjoint-pion is a pseudo-Goldstone boson for the spontaneous breaking of the ``partially quenched" chiral-symmetry and can be tuned to vanishing mass to restore supersymmetry~\cite{Ali:2018dnd,Ali:2019agk}. 

In this work, the strategy used to determine the point of SUSY restoration consists of two very different methods. The first one makes use of the analysis of the eigenvalues of the Wilson-Dirac matrix, whose calculation has already been tackled in section~\ref{ss:pfaffian}. Let us   call $\lambda_{\rm min}^2$ the minimum eigenvalue of  $Q_W^2$. At 
infinite volume, we expect this quantity to vanish in the chiral limit. In our case this corresponds to it
vanishing at large $N$,  
as already observed in the context of Yang-Mills theories with $N_f=2$ Dirac flavors \cite{GarciaPerez:2015rda}. Hence, we can use its dependence on the fermion mass to define a proper massless gluino limit.

\renewcommand{\figscale}{0.65}
\begin{figure}[t]
    \centering
    \includegraphics[scale=\figscale,clip]{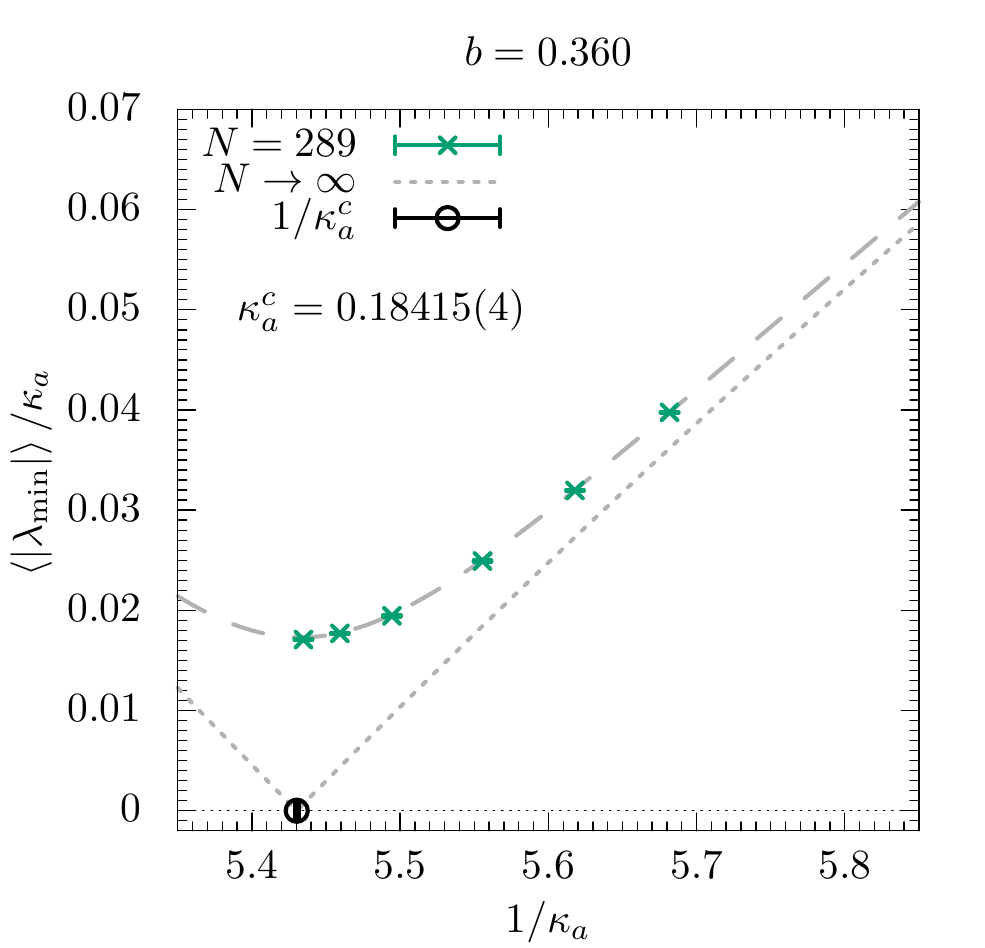}
    \includegraphics[scale=\figscale,clip]{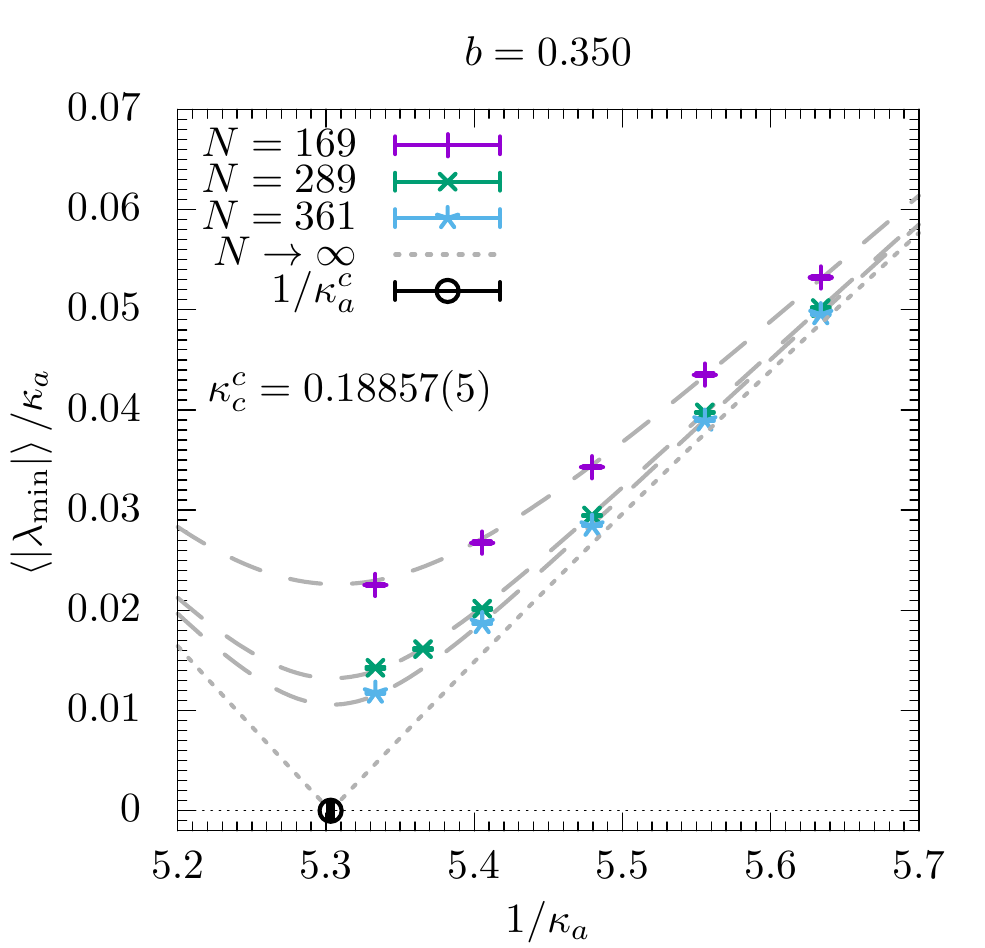}
    \includegraphics[scale=\figscale,clip]{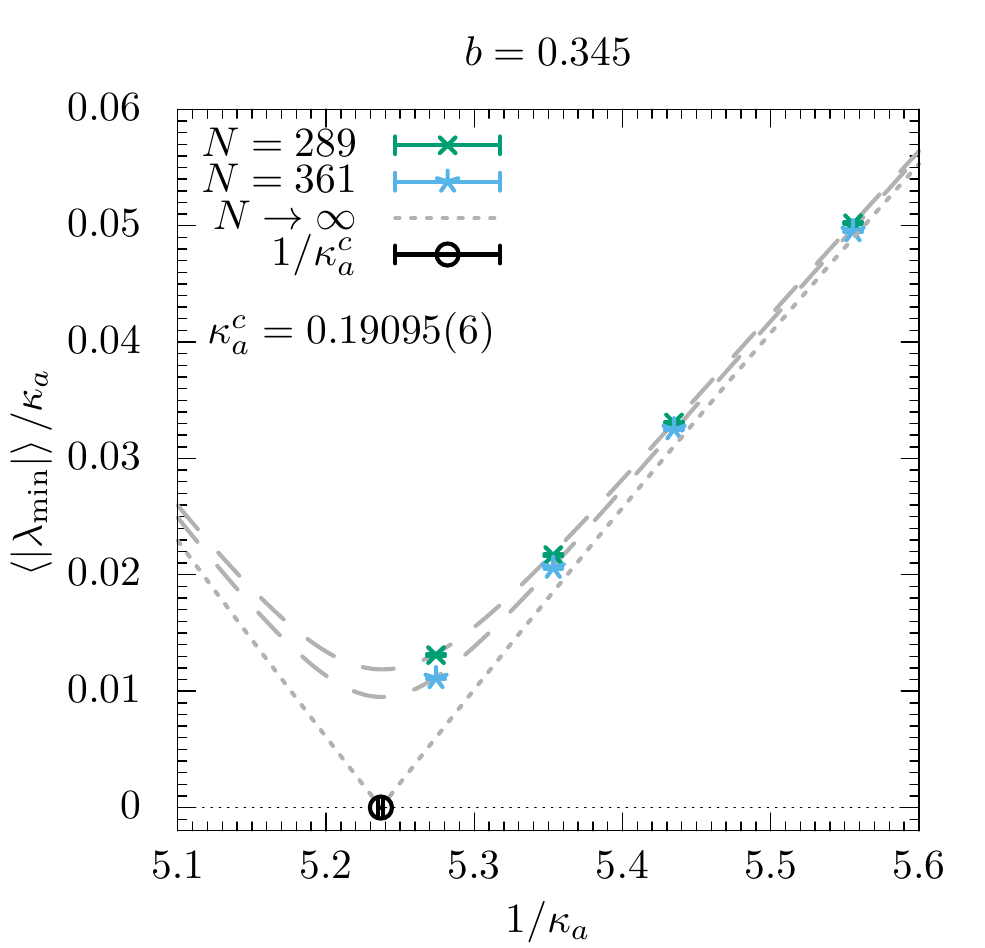}
    \includegraphics[scale=\figscale,clip]{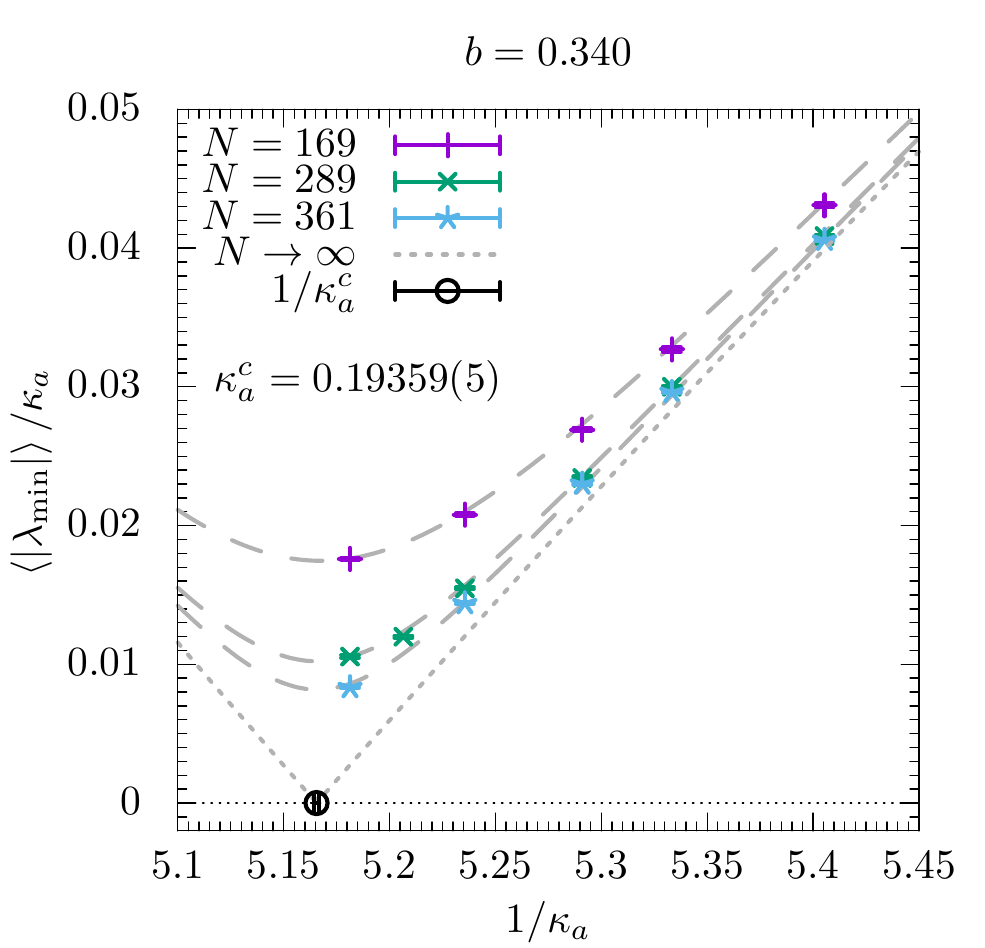}
    \caption{Dependence, for the 4 values of the bare coupling used in this work,  of $\expval{\qty|\lambda_{\mathrm{min}}|}/\kappa_{a}$ on  $1/(2\kappa_{a})$, where $\lambda^2_{\mathrm{min}}$ stands for the minimum eigenvalue of $Q_W^2$. The dashed lines in the various plots correspond to fits to eq.~\eqref{eq:LminFit}, while the dotted lines represent the infinite $N$ extrapolations, allowing to determine the value of the critical hopping parameter $\kappa_a^c(b)$. }
    \label{fig:LminVsKappa}
\end{figure}

The second one is in the same spirit of the aforementioned adjoint-pion mass tuning. Although in the target supersymmetric theory the gluino field is described by a single flavour, we make use of an additional ``quenched'' flavor which allows us to define a pseudoscalar $P=\bar{\Psi}\gamma_5\Psi$ and an axial current $A_\mu=\bar{\Psi}\gamma_\mu\gamma_5\Psi$. We can therefore define an analog of the PCAC mass as
\begin{equation}\label{eqn:mpcaccont}
	2m_\text{pcac} = \frac{\mel{0}{\partial_\mu A_\mu}{\pi}}{\mel{0}{P}{\pi}},
\end{equation}
where $\ket{\pi}$ stands for a generic state with the quantum numbers of the adjoint-pion. On the basis of partially-quenched chiral perturbation theory, we expect this mass to be directly proportional to the renormalized gluino mass and we tune it to zero to find the SUSY restoration limit.

\subsection{Determination from the eigenvalues of \texorpdfstring{$D_W \gamma_5$}{D\_W gamma\_5}}

As already mentioned, the first method that we have used to determine the point of SUSY restoration is based on  the eigenvalues of the Dirac matrix. In the continuum limit the minimum eigenvalue of the massive Dirac operator at infinite volume goes to zero linearly with the fermion mass. On the  lattice one expects a similar behaviour for the operator $D_W/\kappa_a$ which tends to the continuum one up to a renormalization factor. As explained earlier the lowest lying spectra of  $Q_W=D_W \gamma_5$  has been determined for all our configurations. This includes   $|\lambda_{\text{min}}|$, whose  average  value is listed in  table~\ref{tab:HMCAccAndLmin} for all our datasets.  Notice that this quantity has very small errors, making it a perfect observable for  the determination of the zero mass point. Indeed, the aforementioned continuum behaviour implies  that $\expval{\qty|\lambda_{\mathrm{min}}|}/\kappa_a$ should depend  linearly on $1/(2\kappa_a)$ and vanish at $\kappa_a=\kappa_a^c$, signalling the point of vanishing gluino mass. Since finite $N$ corrections amount to finite volume effects, this behaviour is only expected to hold  at infinite $N$. Building on this analogy we would expect  a dependence of the following form:
\begin{align}
  \qty(\dfrac{\expval{\qty|\lambda_{\mathrm{min}}|}}{\kappa_{a}})^2
  &=A \, \qty( \dfrac{1}{2 \kappa_{a}} - \dfrac{1}{2 \kappa_a^c})^2 + \delta(1/N^2).
    \label{eq:LminFit}
\end{align}
where the function $\delta(1/N^2)$ should vanish at large $N$. A correction of this type was observed earlier~\cite{GarciaPerez:2015rda}  for the large $N$ reduced model coupled to two flavours of adjoint fermions. In our case the  formula describes our data very well as
shown in figure~\ref{fig:LminVsKappa}. The dashed lines gives our  best fit with  $\delta(1/N^2)=B/N^2$ and the dotted line the extrapolation to infinite $N$.  The resulting critical hopping parameters $\kappa_a^c$ are presented in table~\ref{tab:kappa_a_c}. 

In the next subsection we will present an alternative method for determining the critical hopping values. 

\begin{table}
    \centering
    \begin{tabular}{c|c| c}
    \toprule
     b & $\kappa_a^c$ from  $\expval{\qty|\lambda_{\mathrm{min}}|}$ &
         $\kappa_a^c$ from  $m_{\text{pcac}}$ \\
     \midrule
     0.340 &  0.19359(5)& 0.19365(5)\\
     0.345 & 0.19095(6) & 0.19100(6)\\
     0.350 & 0.18857(5) & 0.18845(2)\\
     0.360 & 0.18415(4) & 0.18418(2)\\
    \bottomrule
    \end{tabular}
    \caption{Values of the adjoint critical hopping parameter $\kappa_a^c$ determined from the vanishing, as a function of $1/(2\kappa_a)$, of  $\expval{\qty|\lambda_{\mathrm{min}}|}$ or the $m_{\text{pcac}}$ mass.}
    \label{tab:kappa_a_c}
\end{table}

\subsection{Determination from the PCAC mass}

An alternative way of extracting the point of SUSY restoration is based on the vanishing of the non-singlet adjoint-pion correlator and the associated PCAC mass. Contrary to what happens in standard lattice simulations, and due to the form in which it is computed in the reduced setup, the signal-to-noise ratio in the adjoint-pion correlation function deteriorates significantly at large time~\cite{Gonzalez-Arroyo:2015bya}. For this reason, it turns out to be more convenient to do the tuning in terms of the PCAC mass which can be determined with very good precision, and this is the choice we have used in this work. As we will see below, the results obtained in this way turn out to be consistent with the less precise determination from the adjoint-pion mass.

\renewcommand{\figscale}{0.55}
\begin{figure}[t]
\centering
  \includegraphics[scale=\figscale]{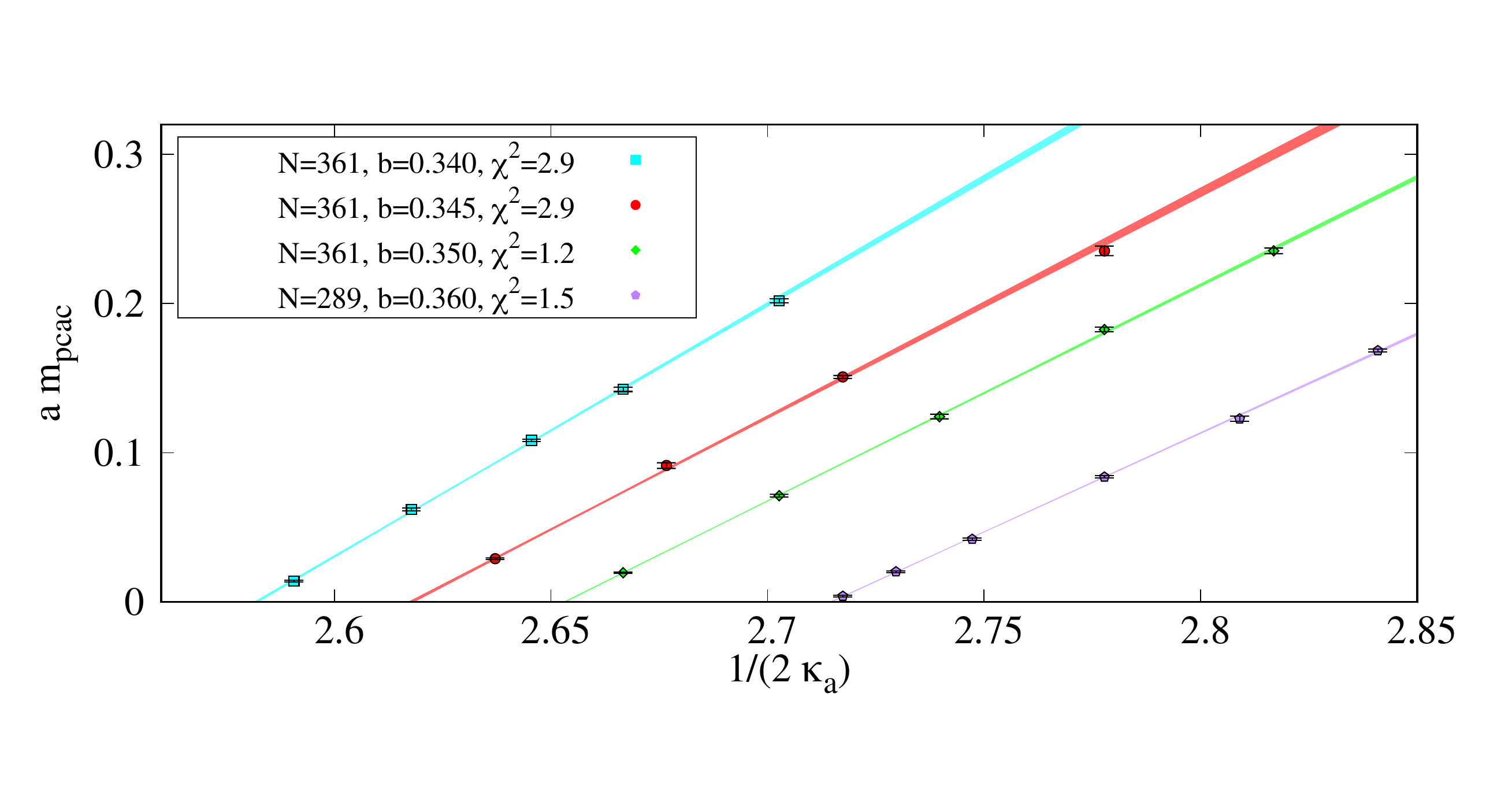}
\caption{Dependence of the PCAC adjoint fermion mass on $1/(2 \kappa_a)$ for all the values of the bare coupling $b$ and the largest value of $N$ within out set of simulations. The bands are linear fits at fixed value of $b$ used to determine the critical value of the adjoint hopping parameter $\kappa_a^c$, with the width indicating the error in the fit. The $\chi^2$ per degree of freedom of each of the fits is indicated in the legend. }
\label{fig:mpcac-adj}
\end{figure}

\begin{figure}[t] 
\centering
 \includegraphics[scale=\figscale]{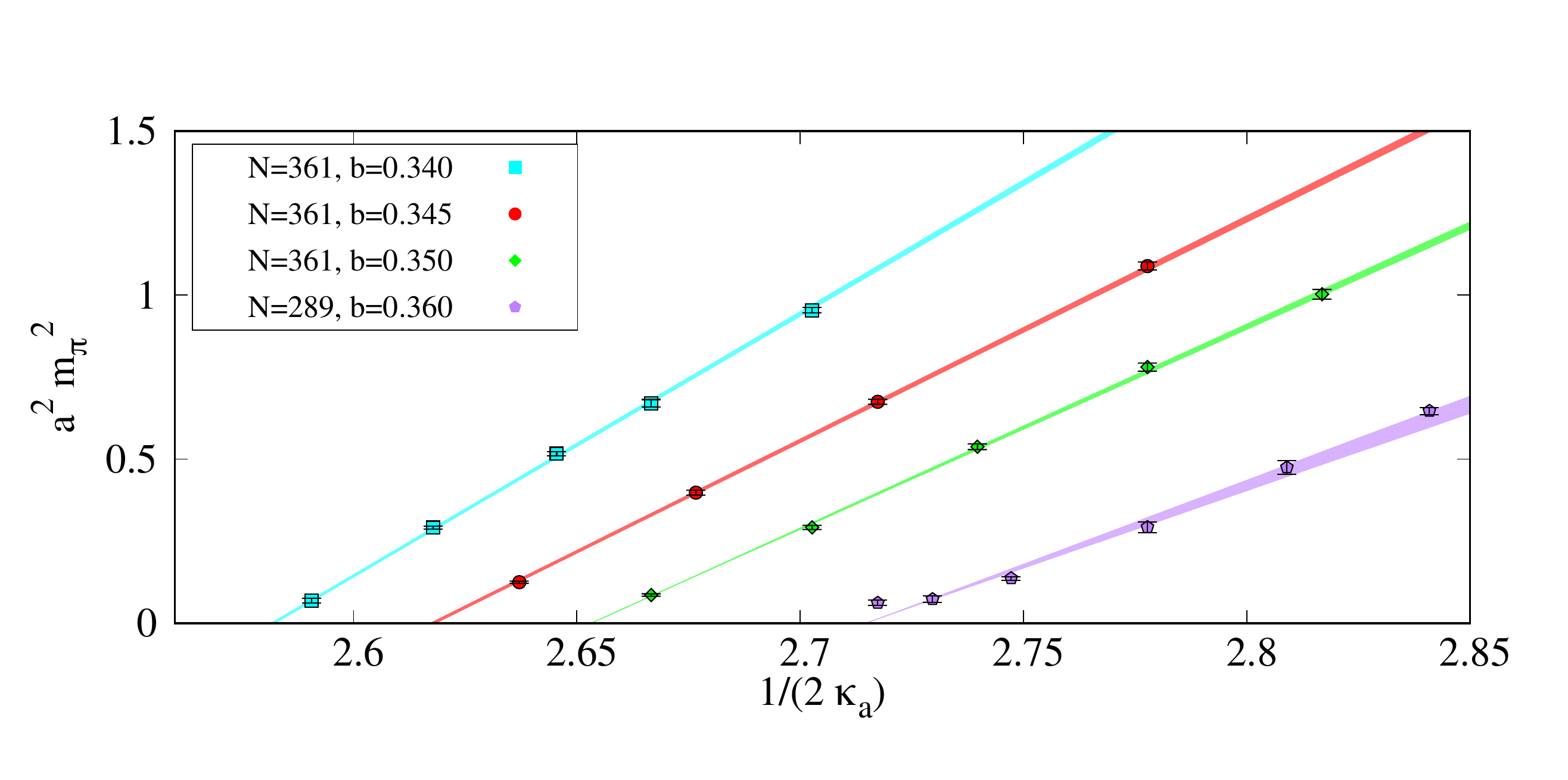}
\caption{Dependence of the adjoint pion mass on $1/(2 \kappa_a)$. The bands are fits to the data at fixed value of $b$ with the critical hopping parameter fixed to the one extracted from the vanishing of the PCAC mass, as detemined from figure~\ref{fig:mpcac-adj}.}
\label{fig:mpion-adj}
\end{figure}

In this and other sections we will be using correlators of meson operators made of fundamental quarks or adjoint gluinos. The methodology is
simple as it amounts to allowing the fermions to travel in a lattice of varying sizes, including infinite,  in the background field of the reduced model gauge field. This has been presented in
ref.~\cite{Gonzalez-Arroyo:2015bya} and used
extensively in ref.~\cite{Perez:2020vbn}. For other representations, like the adjoint, the
philosophy is the same (details will be given elsewhere ref.~\cite{AGA_mesons_new}). Some
specific features  about the implementation in this work are contained in appendix~\ref{app:meson_correlator}.

On the lattice, the PCAC mass can be determined from a discretized version of  eq.~\eqref{eqn:mpcaccont}. For that, we consider an ultralocal version of the axial vector current and an interpolating operator for the pion with optimized projection onto the pseudoscalar channel ground state, obtained as detailed in appendix~\ref{app:meson_correlator}. The PCAC mass is then determined by fitting to a constant the ratio of correlation functions: 
\begin{equation}\label{eq:mpcac}
    a m_\text{pcac} = \frac{\mathcal{C} (x_0+a; \gamma_0\gamma_5,\gamma_5^{\rm opt})-\mathcal{C} (x_0-a; \gamma_0\gamma_5,\gamma_5^{\rm opt}) }{4\, \mathcal{C}(x_0;\gamma_5,\gamma_5^{\rm opt} )},
\end{equation}
where $\mathcal{C}(x_0; A,B)$ stands for the two-point correlation function of the operators corresponding to channels $A$ and $B$.
The values of the PCAC masses extracted in this way are provided in table~\ref{tab:mpcac_adj}. We include in the table results for our datasets with $N=289$ and 361. The general agreement, within errors, of the results for the  two different values of $N$ gives an indication of the small influence of finite $N$ effects on this quantity.

In the chiral limit  the  PCAC mass is expected to be proportional to the fermion mass. This implies that close to this limit $m_{\text{pcac} }$ should  depend linearly on $1/\kappa_a$ and vanish at the critical value. Indeed,  in figure~\ref{fig:mpcac-adj} we display  our results, for the datasets with larger value of $N$, together with a fit to a  linear function of $1/(2 \kappa_a)$. We see that the straight line provides a very good qualitative description of the data. Given the small errors, an additional quadratic term is needed in some cases to get a fit with $\chi^2$ per degree of freedom of the order of one, if we include also the largest masses. In any case, for the purpose of determining the point of vanishing of the PCAC mass both fits give identical results.

\begin{table}
\centering
    \begin{tabular}{c c c c c c }
    \toprule
    $b$ & $\kappa_a$ & $am_{\text{pcac}} (N\!=\!289)$ &  $am_{\text{pcac}} (N\!=\!361)$ &$am_\pi (N\!=\!289)$ &  $am_\pi (N\!=\!361)$\\
     \midrule
0.340& 0.1850&   0.1957(21)& 0.2018(14) & 0.970 (6)& 0.977 (4)\\
0.340& 0.1875&   0.1426(14)& 0.1424(15) & 0.814 (7)& 0.819 (7)\\
0.340& 0.1890&   0.1041(12)& 0.1083 (7) & 0.699 (5)& 0.719 (4)\\
0.340& 0.1910&   0.0603 (9)& 0.0619 (9) & 0.534 (6)& 0.540 (5) \\
0.340& 0.192067& 0.0344 (8)&  --        & 0.397 (4)& -- \\
0.340& 0.1930&   0.0120 (3)& 0.0138 (6) & 0.258 (6)& 0.263(13)\\
\midrule
0.345& 0.1800&   0.2390(60)& 0.2352(32) &1.040(17)& 1.043 (6)  \\ 
0.345& 0.1840&   0.1477(22)& 0.1507(11) &0.814(16)& 0.821 (5) \\ 
0.345& 0.1868&   0.0901 (7)& 0.0913(18) &0.630 (7)& 0.631 (6) \\   
0.345& 0.1896&   0.0275 (5)& 0.0289 (5) &0.349 (6)& 0.353 (5) \\
\midrule
0.350& 0.1775&   0.2363(29)& 0.2353(19) &1.013(12)& 1.001 (7) \\
0.350& 0.1800&   0.1805(25)& 0.1826(16) &0.878 (9)& 0.883 (7)  \\
0.350& 0.1825&   0.1242(24)& 0.1242(15) &0.719(13)& 0.733 (6) \\
0.350& 0.1850&   0.0733(14)& 0.0712(10) &0.534 (6)& 0.540 (6) \\
0.350& 0.1864&   0.0427 (9)& --         &0.433 (7)& --\\
0.350& 0.1875&   0.0192 (7)& 0.0195 (4) &0.291 (8)& 0.293 (6)  \\
\midrule
0.360& 0.1760&  0.1684 (9)& -- &0.804(7)& --\\
0.360& 0.1780&  0.1227(17)& -- &0.689(15)& --\\
0.360& 0.1800&  0.0837 (8)& -- &0.541(15)& --\\ 
0.360& 0.1820&  0.0420 (7)& -- &0.370 (7)& --\\
0.360& 0.1832&  0.0202 (6)& -- &0.270(19)& --\\
0.360& 0.1840&  0.0038 (5)& -- &0.249(17)& --\\
    \bottomrule
    \end{tabular}
    \caption{Values of the PCAC and adjoint-pion masses determined for all values of the bare coupling and adjoint hopping parameter in our set of simulations with $N=289$ and $N=361$.}
    \label{tab:mpcac_adj}
\end{table}

Our final results for the critical value of $\kappa_a$, corresponding to the largest value of $N$ simulated in each case, are collected in table~\ref{tab:kappa_a_c}. The values obtained are in remarkable agreement with the determination based on the eigenvalues of the Dirac operator. In the rest of this work, the vanishing of the PCAC mass will be used as the criteria to tune all other quantities to the zero gluino mass limit.

Finally, to check for the consistency of our approach, we have as well determined the adjoint-pion mass and analyzed its dependence on the hopping parameter $\kappa_a$. Results are collected in table~\ref{tab:mpcac_adj}.  
Figure~\ref{fig:mpion-adj} displays $m_\pi^2$ vs $1/(2 \kappa_a)$.  The coloured lines shown in the plot correspond to linear fits where, for each value of the bare coupling $b$, the critical value of the hopping parameter is fixed  to the one determined from the PCAC mass; only the slope is left as a free parameter. Linearity works very well in all cases, giving $\chi^2$ per degree of freedom of the order of one, except on the smallest lattice with bare coupling $b=0.36$, where a deviation is observed for the lighter masses (the two lightest ones are excluded from the fit). This difference can be due to finite size effects, which come out more pronounced for the pion mass than for the PCAC mass. For this reason it is also preferable to use the latter in order to determine the point of SUSY restoration.

%%% Local Variables:
%%% mode: japanese-latex
%%% TeX-master: t
%%% End:
    
    \section{Scale setting: choosing the observables}
    \label{s:scalesetting}
    % \section{Scale setting}

In this section we will present our results on the  scale setting for our
lattice model. By extrapolation to the massless gluino limit we will
later achieve our main goal of determining the scale for the $\mathcal{N} = 1$ Supersymmetric gauge theory.

Scale setting  is a standard and rather crucial step one has to face in lattice
gauge theories. Different procedures start by defining what observable
is going to be used as a unit of energy. On the lattice this
observable will appear as a dimensionless quantity given by the
product of the continuum quantity times the lattice spacing. Hence, by
measuring this particular lattice observable we are actually measuring
the lattice spacing in units of the inverse of that physical quantity.
It is obvious that there are infinitely many ways to fix the scale
corresponding to the infinitely many observables that one can choose.
In theories with a massive spectrum one could simply take one of the
masses as unit. For QCD or Yang-Mills theory the lowest glueball mass, the
vector meson  mass  or the square root of the string tension  are typical
units of this kind.  However, in practice making the right choice of a
unit is a crucial one. A unit must be of the appropriate size to the
objects to be measured, it should be precise, easy to compute and as
insensitive as possible to  typical statistical or systematic errors.
From that point of view the natural type of units mentioned earlier might not
be the optimal,  since they involve taking asymptotic limits. Other
possible family of choices involve dimensionless monotonous functions of an
energy scale. The unit of energy is taken to be the value at which  the function
takes a particular numerical value. To this class belong some of the most common
scales used nowadays by the lattice community such as  the Sommer scale ~\cite{Sommer:1993ce},
based on the quark-antiquark potential, and those relying on
the use of the gradient flow such as $t_0$~\cite{Luscher:2010iy} or
$w_0$~\cite{Borsanyi:2012zs}. For a review of these and other  common
methods to fix the scale,  the interested reader can consult ref.~\cite{Aoki:2021kgd}.

For theories like pure Yang-Mills or $\mathcal{N}=1$ SUSY Yang-Mills with
a single relevant operator the ratio of two scales is just a constant,
given by the  ratio of the corresponding units. On the lattice this is
only true close enough to the critical point and the phenomenon is
called scaling.

In this work we will study three different methods to fix the scale.
This will allow us to measure the lattice spacing in each of the three
units, and in addition to  test whether our data shows scaling
at the values of $b$ at which we are simulating. Obviously, scaling is
expected to work better for the larger values of $b$.  In making our
choices of units we have taken into account three factors. First of all
the precision of the corresponding observable. Second its
insensitivity to the gluino mass. And third the insensitivity to
finite $N$ effects.  The latter,  as mentioned in
Section~\ref{s:reducedmodels},  are equivalent to finite volume effects, and  represent
a very important source of systematic errors.

Here we will mention briefly what the methods are about and in the
next three subsections develop in more detail what are the
corresponding observables. The technical details of its implementation
on the lattice as well as the final results will be covered in the
following section.

The first method  is a modification of gradient flow based methods devised to reduce finite volume effects in the determinations of $t_0$ and $w_0$,
equal in spirit to the one introduced for periodic boundary conditions in~\cite{Fodor:2014cpa}, but making use in this case of twisted boundary conditions
that have well known advantages from the point of view of perturbation theory. The second, a variant of which has been already used in large $N$ pure
Yang-Mills theory~\cite{\refst}, is analogous to the Sommer scale but involves Wilson loops of fixed aspect  ratio. The advantage of this is that the
extrapolation to asymptotically large time needed to compute the quark-antiquark potential is no longer required.
Finally, the third method involves using the spectra of mesons made of
quarks in the  fundamental representation to set the scale. In the large
$N$ limit fundamental  fermion loops are suppressed and the quenched approximation is exact.
The computation of the meson spectrum can therefore be determined at no additional
cost on our set of dynamical adjoint fermion configurations. These
meson masses depend on an  additional scale, which is the  fundamental
quark mass,  but in the zero mass limit this dependence disappears
and the mass of the lightest  non-singlet vector meson becomes a natural scale for SUSY Yang-Mills.

We emphasize that the methods presented in the following subsections
can also be used at finite $N$ and for other gauge theories and are
particularly well suited when finite size effects are an important
source of concern.

\subsection{Gradient flow observables}
\label{s:general-flow}

The gradient flow~\cite{Narayanan:2006rf,Lohmayer:2011si,Luscher:2009eq,Luscher:2010iy} is a smoothing technique that has received much
attention in recent years. The idea is to replace the original gauge fields $A_\mu(x)$ by a set of flow time-dependent fields obtained by solving the gradient flow equations: $\partial_t B_\mu (x;t) = D_\nu G_{\mu\nu}(x;t)$ (with $B_\mu (x;0) =A_\mu(x)$) and leading to an effective smearing of the  gauge fields over a length scale $\sqrt{8t}$. The 
flow time, having physical dimensions of a length squared, induces natural candidates for scale setting; one just has to find a dimensionless, flow time dependent, quantity and  determine the flow time at which it equals a particular pre-fixed value. The most common choices used for this purpose are based on the quantity:
\begin{equation}\label{flowedenergydensity}
    \Phi(t) = \expval{\frac{t^2 E(t)}{N}},
\end{equation}
where $E(t)$ stands for the \textit{flowed energy density}:
\begin{equation}\label{energydensity}
    \expval{E(t)} = \frac{1}{2}\expval{\Tr G_{\mu\nu}(x,t)G_{\mu\nu}(x,t)}.
\end{equation}
 Particular examples are the ones obtained by solving either of the two following implicit equations:
\begin{align}
    \eval{\Phi(t)}_{t=t_s } &= s,
\label{tc}\\
    \eval{t \dv{t}\Phi(t)}_{t=w_s^2} &= s,
\label{wc}
\end{align}
where $s=0.1$ corresponds to the standard choices in the literature denoted by $t_0$~\cite{Luscher:2010iy}  and $w_0$~\cite{Borsanyi:2012zs}~\footnote{
Note that these definitions differ from the standard ones used in lattice QCD simulations due to the extra, $N$-dependent, factor, required to have a well 
defined large $N$ limit of these quantities.}.

All these considerations take place in infinite volume. However, numerical simulations are implemented on a $L$-site lattice
of finite physical size $l=a L$, a fact that leads to finite size effects in the determination of the scale whenever the flow radius extends over 
a significant fraction of the box size. Our proposal, analogous to the one in ref.~\cite{Fodor:2014cpa},  is to take an alternative version of the observable $\Phi(t)$ with the correct infinite
volume limit but with reduced finite size effects. The idea makes use of the fact that in infinite volume one can define a renormalized \textit{gradient flow} (GF) coupling constant proportional to $\Phi(t)$~\cite{Luscher:2010iy}: 
\be
\lambda_\GF (\mu=1/\sqrt{8t}) = \frac{1} {\mathcal{K}(N)}\,  \Phi(t),
\label{gf-coupling}
\ee
with a flow-time independent proportionality constant $\mathcal{K}(N)$
fixed by imposing that the GF coupling equals the bare one at leading order in perturbation theory.
 On a finite torus of size $l$ with twisted boundary conditions, this proportionality factor, denoted from now on by $\mathcal{N}(c(t),N)$, acquires a volume and flow time dependence parameterized in terms of the dimensionless variable $c(t)\equiv \sqrt{8 t /N l^2}$, for further details see appendix~\ref{ap:flow-imp}. We then define the finite volume quantity:
\be
\hat \lambda (t,l,N) = \frac{1}{\mathcal{N}(c(t),N)}\, \,  \Phi(t,l,N),
\label{improved-gf-coupling}
\ee
 which tends to the gradient flow coupling in the infinite volume limit, and, by analogy, a modified flowed energy density:
\begin{equation}
    \hat {\Phi}(t,l,N) = \, \mathcal{K}(N) \hat \lambda(t,l,N).
    \label{improved-phi-cont}
\end{equation}
As we will see in subsection~\ref{s:tekflow}, finite volume (finite $N$ in our set-up) effects are considerably reduced with this choice. Our scale definitions are then obtained by replacing $\Phi$ by  $\hat{\Phi}$ in eqs. \eqref{tc} and \eqref{wc}, which at infinite volume coincide with the standard ones.

Let us finally mention that the
twisted finite volume normalization factor, $\mathcal{N}(c(t),N)$, has been computed and used before for a series of step scaling studies~\cite{Ramos:2014kla,GarciaPerez:2014azn,Bribian:2021cmg}; 
 explicit formulas are given in 
appendix~\ref{ap:flow-imp}, where we also give some more details on the perturbative expansion and the relevant references.
This procedure also accounts for correcting lattice artefacts at leading order if the normalization factor is computed instead in lattice perturbation
theory.
In subsection~\ref{s:tekflow} we will discuss the specific lattice implementation of this method and the methodology applied to extract the scale from the implicit eqs.~\eqref{tc} and \eqref{wc}.

\subsection{Wilson loops}
\label{s:general-creutz}

We move now onto the discussion of the second alternative to fix the scale, based on Wilson loops. As already mentioned, a particular version of this procedure was used to set the scale in large $N$ pure Yang-Mills  theory~\cite{\refst}. Standard   
physical quantities such as the string tension or
glueball masses are difficult  observables to measure as they involve taking asymptotic limits. This makes them  more affected by corrections and systematic uncertainties. It is in this spirit that Sommer scale~\cite{Sommer:1993ce}, which is based on the quark-antiquark potential but not at infinite separation,  is superior. However, the Sommer scale
still involves the study of loops that are asymptotically long in
time. Our proposal is  based on fixed aspect ratio Wilson
loops and hence it involves no limit. 

Ultimately all gluon observables are based on Wilson loop expectation values. However, the Wilson loops themselves are UV divergent quantities and thus not suitable observables. Our proposal is based upon the following observable function
\begin{equation}
    F(r,r') = -\pdv{\log \mathcal{W}(r,r')}{r}{r'},
    \label{F-of-r}
\end{equation}
where $\mathcal{W}(r,r')$ are expectation values for rectangular
$r'\times r$ Wilson loops. The double derivative gets rid of perimeter and corner singularities
and one gets a  well-defined continuum quantity depending on two scales.
We can reduce it to a single scale by fixing the aspect ratio $r'/r$.
Different choices give different definitions.  Notice that Sommer scale
involves the limit $r\gg r'$.  Here we will restrict
ourselves to the opposite limit given by symmetric  loops $r=r'$ (a restriction taken after the
derivative is evaluated).

Indeed we claim that $F(r,r)$ is a very interesting physical quantity for SU($N$) Yang-Mills theory, which  has been computed in
ref.~\cite{\refst} both for finite and infinite $N$. Obviously the
string tension is given by
\begin{equation}
\sigma = \lim_{r \longrightarrow \infty} F(r,r).
\end{equation}
However, it is better to fix the scale in a different way. For that
purpose one notices that $F(r,r)$ has dimensions of energy square, so we consider the dimensionless observable $G(r)=r^2 F(r,r)$. A physical scale $\bar{r}(f_0)$ can be defined, \'{a}-la Sommer,  as follows
\begin{equation}
   G(\bar{r}(f_0))= \eval{r^2 F(r,r)}_{r=\bar{r}} = f_0,
    \end{equation}
where $f_0$ is some numerical value that can be chosen arbitrarily.
This is essentially the method used in ref.~\cite{\refst} to fix the
scale with $f_0=1.65$. Here we will consider a variant of the method that makes use of the gradient flow and results advantageous from the point of view of the lattice implementation.

One can use the gradient flow to
construct  flow-time dependent Wilson loops $ \mathcal{W}_t (r,r')$
and use them to define a function $F_t (r,r)$, in analogy to eq.~\eqref{F-of-r}. Our new proposal is to use the flowed functions at non-zero flow time to set the scale. In order to define a function of a   single energy scale in the problem, we fix the flow smearing radius to be proportional to  the loop size:
$\sqrt{8 t}=\sqrt{8\bar{t}(r,s)}\equiv s r$. This amounts to defining Wilson loops having  {\em fat} edges with thickness proportional to the size of the loop. This choice defines a dimensionless function of a single length  $r$
\begin{equation}\label{Ghatdef}
\hat{G}(r;s)=r^2 F_{\bar{t}(r,s)}(r,r).
\end{equation}
Applying now the, by now well-known, prescription we can define a physical scale  $\bar{r}(f_0,s)$ as follows
\begin{equation}\label{creutzcond}
   \hat{G}(\bar r(f_0,s);s)=  f_0,
    \end{equation}
Any choice of  $f_0$ and $s$ defines a different scale, but they should all be proportional to each other.  
The concrete choice of $s$ and $f_0$ is dictated by practical reasons of accessibility and insensitivity of the corresponding lattice observable to different error sources. From that viewpoint the choice used previously $\bar{r}(1.65,0)$ is not the most adequate here.

\subsection{Fundamental meson spectrum}
\label{s:general-mesons}

In the large $N$ limit, fermions in the fundamental representation play a completely different role than adjoint ones.  If the limit is taken á-la 't Hooft, sending $N$ to infinity while keeping fixed the number of fermion flavours $N_f$, fundamental-quark loops are suppressed and the quenched approximation is exact. As already mentioned, these fundamental quarks give rise to a meson spectrum, that can be used to determine an additional scale for the $\mathcal{N}=1$ SUSY Yang-Mills theory at large $N$.  
Even though the fundamental spectrum introduces an additional scale, the fundamental fermion mass, 
this scale is removed in the chiral limit  (not to be confused with the massless gluino limit described in previous sections). It is in this particular limit when the masses of non-Goldstone fundamental mesons can be considered a natural scale of the SUSY theory.
For the purpose of this paper we have selected the lightest vector meson state to determine this additional scale. 
In subsection~\ref{s:mesons} we will present our results for the determination of the lattice spacing using this method.

     \section{Scale setting: lattice determination of the scale}
    \label{s:results}
    
In this section, we are presenting the results we obtained by implementing on the lattice the three different scale setting methods described in the previous one. 

\subsection{Setting the scale with the flow}
    \label{s:tekflow}
    We will firstly present our results for determining the scale using the improved version of the flow discussed in subsection~\ref{s:general-flow}.
We start by addressing the lattice implementation and by analyzing its efficiency in reducing finite 
size effects; as we will see the method works remarkably well inside a properly defined scaling window.
The final analysis leading to the determination of the lattice spacing $a(b,\kappa_a)$ in units of the flow scale
is presented in subsection~\ref{s:analysis-flow}.  

\subsubsection{Methodology}

So far our discussion of the flow observables in subsection~\ref{s:general-flow} has been in the continuum.
A generalization to the lattice is rather straightforward, one has to select a discretization of the energy density 
and of the flow equations. We adopt the clover version of the field strength which, in our one-site reduced lattice,
leads to a discretized version of eq.~\eqref{energydensity} given by:
\begin{equation}
    \hat{E} = -\frac{1}{128}   \sum_{\mu,\nu}
    \Tr\left [  z_{\mu\nu} \left (U_\nu U_\mu U^\dagger_\nu U^\dagger_\mu
    + U_\mu U^\dagger_\nu U^\dagger_\mu U_\nu
    + U^\dagger_\nu U^\dagger_\mu U_\nu U_\mu
    + U^\dagger_\mu U_\nu U_\mu U^\dagger_\mu \right )
    - \text{h.c.}\right]^2.
\label{clover}
\end{equation}
As for the flow,  the flow time is discretized in units of the lattice spacing as: $t=Ta^2$, where we will from now on use capital letters 
to denote lattice, dimensionless quantities. We employ the so-called Wilson flow~\cite{Luscher:2010iy} and integrate the discretized flow
equations by using a 3rd order Runge-Kutta integrator with constant time interval $\Delta T = 0.03125$. 
We have checked that changing the time step does not produce any sizeable difference in the integrated flowed observable, resulting only in a different computational cost for different lattice spacings.

In terms of $\hat{E}(T)$, the naive dimensionless flowed energy density  can be estimated on the lattice from:
\begin{equation}
    \expval{\frac{T^2 \, \hat{E}(T)}{N}}.
\label{naive-phi}
\end{equation}
The lattice equivalent of the normalization factor $\mathcal{N}(c(t),N)$,  corresponding to a concrete  discretized definition of the energy density,  is also easily determined by a
tree level calculation in lattice perturbation theory.
For our choice of the clover discretization and
for the one-point lattice, an explicit expression has been computed in ref.~\cite{GarciaPerez:2014azn} and is provided in Appendix~\ref{ap:flow-imp}.
One advantage of using the lattice determined instead of the continuum norm is that one corrects finite lattice artefacts on top of finite size effects at tree level.

With all this, our final formula for the discretized version of the flow is given by:
\be
\hat {\Phi}_L(T,b,N) = \frac{3}{128 \pi^2 \mathcal{N}_L(\sqrt{8T/N},N)}  \expval{\frac{T^2 \, \hat{E}(T)}{N}},
\label{improved-phi}
\ee
with $\mathcal{N}_L(x,N)$ given by eq.~\eqref{lattnorm}.

\begin{figure}[t]
\centering
\begin{subfigure}{.5\textwidth}
  \centering
\raggedleft
  \includegraphics[width=1.\linewidth]{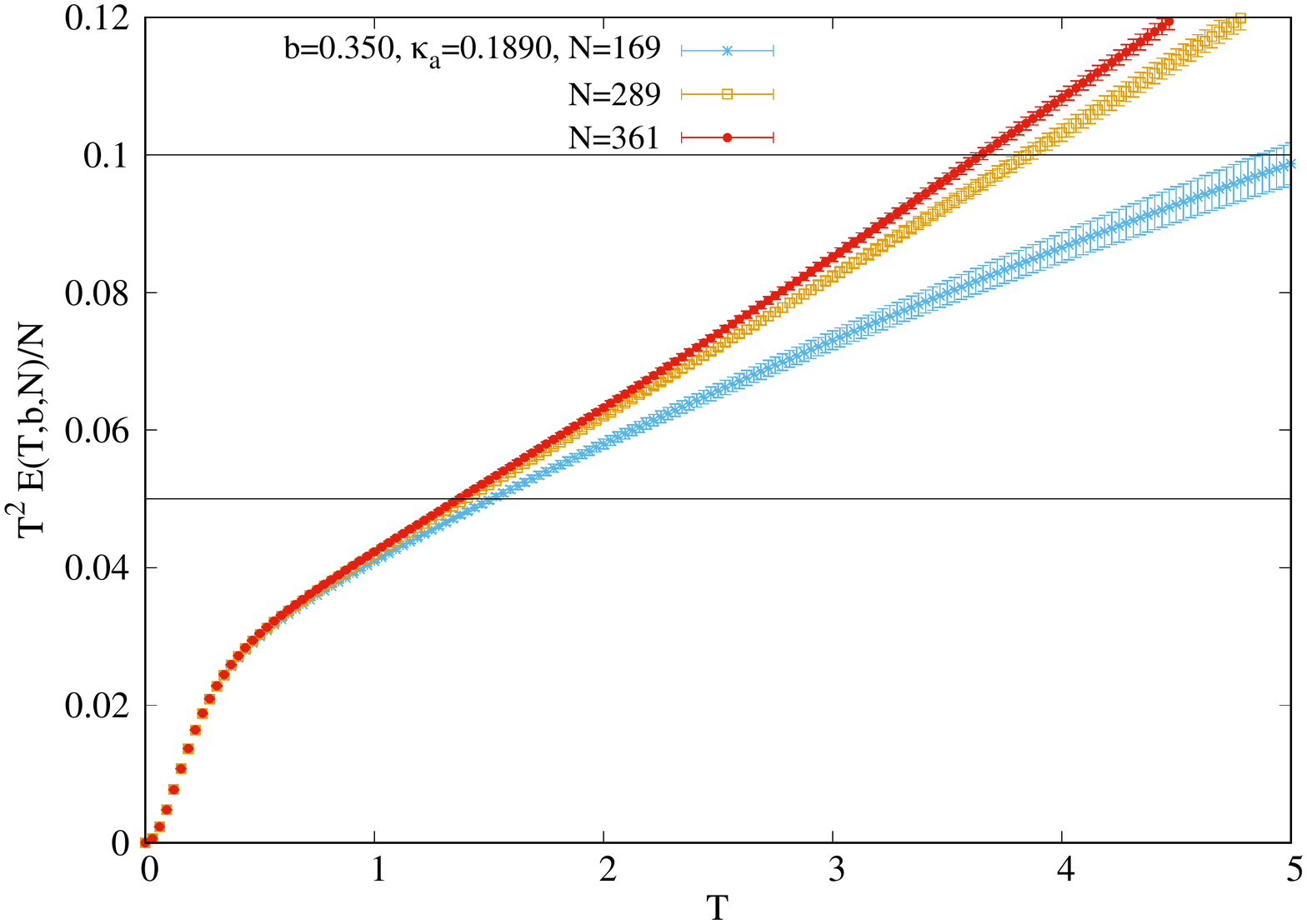}
\captionsetup{width=.9\textwidth}
\caption{Naive flow (eq.~\eqref{naive-phi}).}
\label{fig:finiteN_naive}
\end{subfigure}%
\begin{subfigure}{.5\textwidth}
  \centering
\raggedleft
 \includegraphics[width=1.\linewidth]{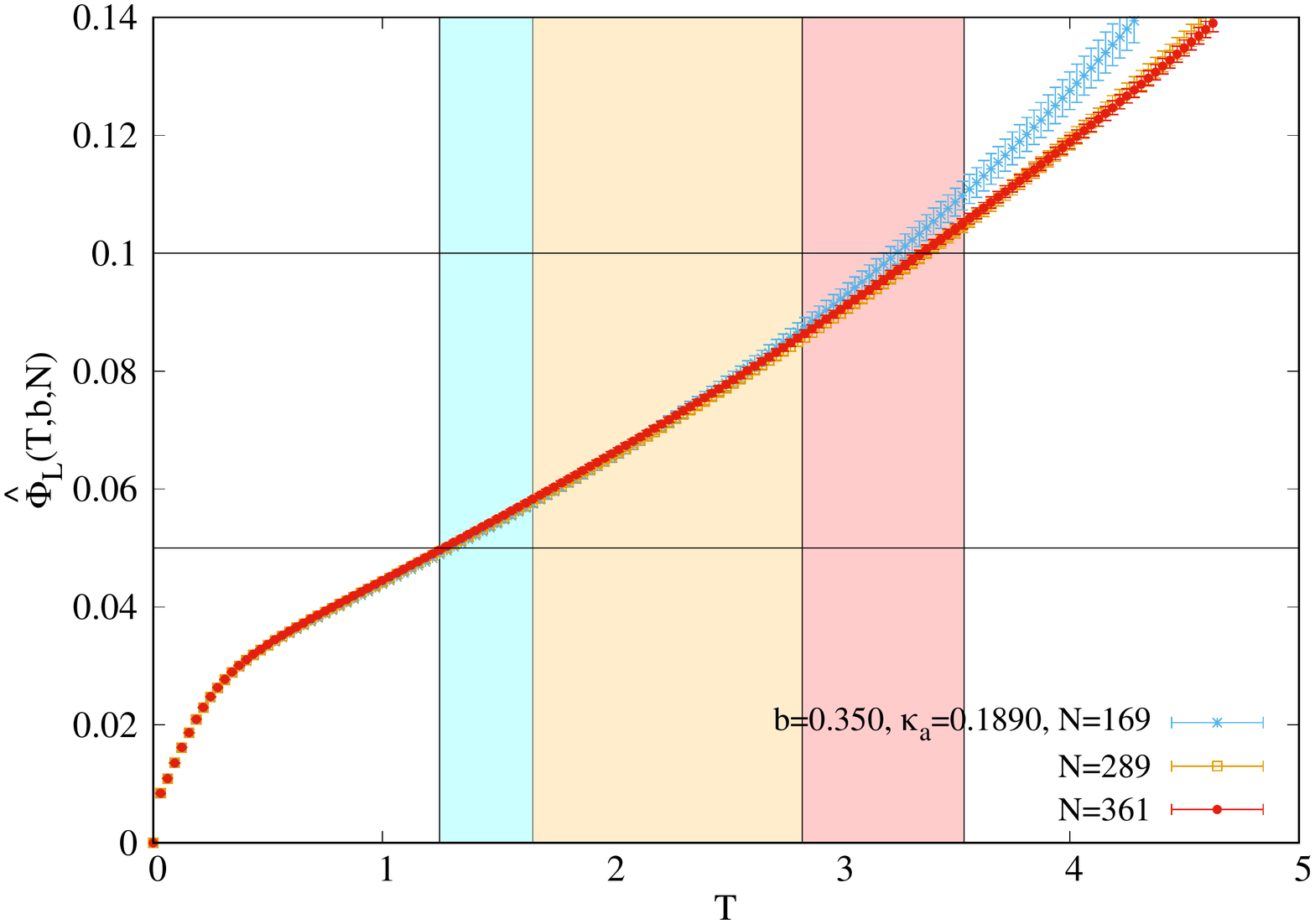}
\captionsetup{width=.9\textwidth}
\caption{Improved flow (eq.~\eqref{improved-phi}).}
\label{fig:finiteN_imp}
\end{subfigure}
\caption{Finite $N$ effects on the dimensionless flowed energy density. The coloured bands in the right plot represent the scaling windows for each value of $N$, c.f. $ T \in [ 1.25, \gamma^2 N/8]$, with $\gamma=0.28$.}
\end{figure}

In order to illustrate the kind of improvement 
attained with the use of eq.~\eqref{improved-phi} we present in figure~\ref{fig:finiteN_naive} the dependence on $T$ of the naive
expression  for three different values of $N$, corresponding to $N=169$, 289 and 361.
As can be observed, finite $N$  corrections are a sizable source of systematics. 
The plot of figure~\ref{fig:finiteN_imp} displays instead the flowed energy density obtained using the improved observable $\hat \Phi$. In this case, the window in which the three curves collapse to a single one extends over a much larger range.  
Generically, this window is set by the ratio $c(t)= \sqrt{8T/N}$, determining the fraction of the box  occupied by the smearing radius. 
An empirical observation is that the correction is efficient over a
scaling window given by:
\be
  T \in [ 1.25, \gamma^2 N/8], \text{ with } \gamma\lesssim 0.3,
\label{fit-window}
\ee
where the lower end of the interval is set to have under control the remaining lattice artefacts.

\subsubsection{Results}
\label{s:analysis-flow}

Before presenting our results, there is an additional consideration that has to be made. To have a precise determination of the scale it is important that the value of $T_s\equiv t_s/a$ appearing in the lattice counterparts of eqs.~\eqref{tc} and~\eqref{wc} falls well within the scaling window given by eq.~\eqref{fit-window} and can therefore be reached by interpolation. With our set of parameters this is best attained by choosing $s=0.05$ instead of the commonly used value of $s=0.1$; the corresponding lattice scale will be denoted by $T_1$ from now on. However, on our smaller lattices, particularly those corresponding to $b=0.36$, even this value leads to a scale that can only be reached by extrapolation. 
We address this issue by simultaneously fitting our data, for all values of $b$ and hopping parameter $\kappa_a$, to a single universal curve depending on $T/T_1(b,N,\kappa_a)$. Our combined data covers a window that runs from the   perturbative small flow-time region up to values around $T_0/T_1$. In playing this game we are neglecting violations of universality that may come from lattice artefacts, remnant finite size effects or a dependence of the flow on the gluino mass. As we will see below, these assumptions are well satisfied by our results. 
 
In appendix~\ref{app:t1_all_flow} we give full details of the form of the universal fitting functional used to describe the flow, here let us just indicate that we have selected it so as to describe appropriately the perturbative domain for small values of $T/T_1$. For that one can use that the infinite volume flow defines a renormalized coupling constant and the small $T/T_1$ regime is therefore well described by using the two-loop perturbative $\beta$-function.  As we will see below, this two-loop function describes very well the time dependence of the flow in a remarkably large window of flow times.
 
 \begin{figure}[t]
\centering
    \hspace*{-0.8cm}
     \includegraphics[width=\linewidth+0.8cm]{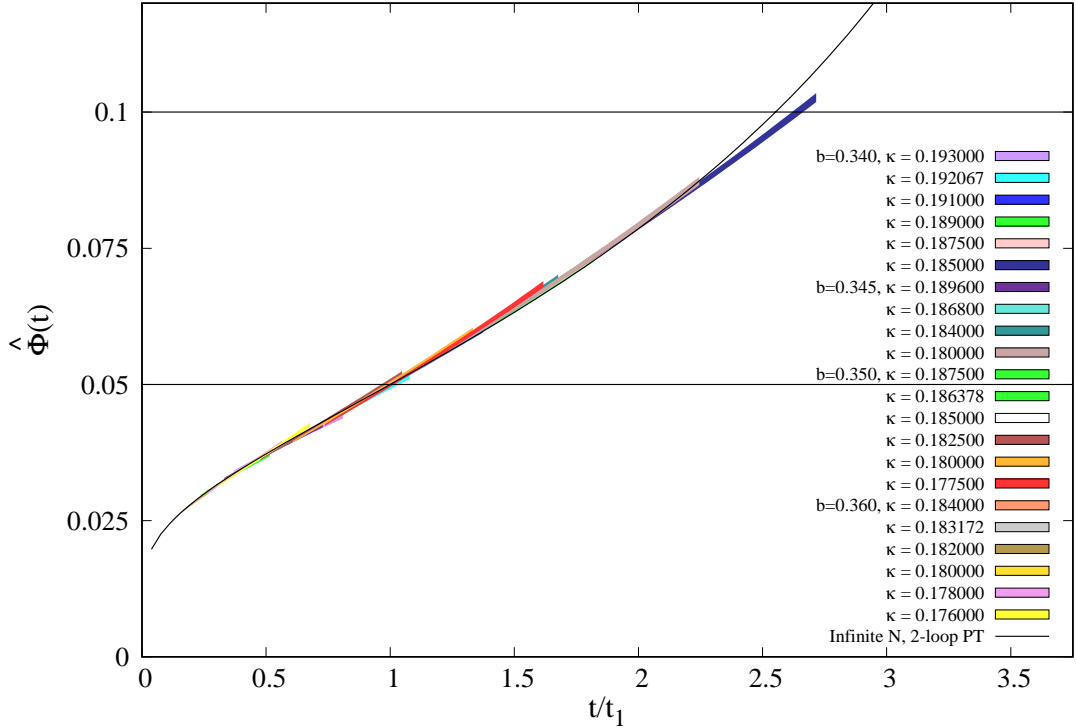}
\caption{We display $\hat \Phi(t)$ as a function of $t/t_1$ for all our datasets with $N=289$. The width of the coloured bands indicates the error on the different quantities. The black line on the plot corresponds to the two-loop infinite $N$ (infinite volume) perturbative prediction for the flow-time dependence of $\hat \Phi(t)$. }
\label{fig:t1_all_flow}
\end{figure}

To serve the purpose of illustrating how well universality holds,  we display in figure~\ref{fig:t1_all_flow} the dependence of $\hat \Phi(T)$ on $T/T_1$ for $N=289$. 
The different data displayed in the plot correspond to different values of $b$ and $\kappa_a$ restricted to the scaling window eq.~\eqref{fit-window} with $\gamma=0.28$.  
The values of $T_1$ have been obtained from a universal fit to the data  as the one described by eq.~\eqref{fit-pert} with  three parameters of the $\beta$-function, in addition to the two universal ones corresponding to $\mathcal{N}=1$ Supersymmetric Yang-Mills, and gives a $\chi^2$ per degree of freedom of 1.2 (we obtain $\chi^2/\# {\rm dof}=3.1$ for the datasets with $N=361$). As becomes evident from the plot, the advantage of the joint fit is that it allows to constrain the time dependence of the flow in a region of scales much larger than the actual fitting window and permits to determine $T_1$ even when it falls outside it. Finally, we also display for comparison the prediction of two loop perturbation theory at infinite $N$, given by the black line in the plot, which, as mentioned previously,  describes quite well our results in a large window of flow times.

Let us now present our results. In order to have an additional check on the effective reduction of finite $N$ effects, we have fitted separately the $N=289$ and 361 data sets, obtaining compatible results within errors. Our final values for $T_1$ are given in table~\ref{tab:6scales}, the first quoted error is statistical and the second systematic. The latter is determined so as to cover various determinations of the scale corresponding to different fitting functional and ranges as detailed in appendix~\ref{app:t1_all_flow}.  

We end this section by discussing the relation of the scale $t_1$ to those more common in the literature, as $t_0$ or $w_0$.  We have already mentioned that in most of our simulations $t_0$ falls out of the scaling window. Nevertheless, and under the assumption of scaling, we can use the universal fitting functional describing the flow to obtain a determination of the ratio  $R=\sqrt{T_0/T_1}$. The result is
$R=1.624(50)$ and $R=1.631(70)$ for $N=361$ and 289 respectively. This is in perfect agreement with the ratios obtained at the few cases where we can determine $T_0$ directly by interpolation.  The error quoted in all cases covers for the systematics in the fitting functional and fitting ranges following the same procedure used to determine $t_1$.

Finally, we have also determined the scales $w_1$ and $w_0$ derived by solving the implicit equation eq.~\eqref{wc} with $s=0.05$ and $s=0.1$ respectively. The strategy to determine these scales is very similar to the one used for $t_1$. We rely on the universality of the flow  and fit $t d \hat \phi(t)/dt$ as a function of $t/a^2$ simultaneously for all our datasets within the scaling window corresponding to $\gamma=0.28$. In this case we use a degree-two polynomial fit, with the systematic error obtained by varying the fitting range.  The resulting scales can be compared to $t_1$. 
Excluding the sets at $b=0.36$, which have very large systematic errors and give nevertheless results consistent within errors, and restricting to the cases with $m_{\text{pcac}}\sqrt{8t_1}>0.3$, a fit of the dimensionless ratio of scales to a constant gives: $w_1/\sqrt{8t_1}= 0.4535(49) $ and $w_0/\sqrt{8t_1}=0.586(10) $ with $\chi^2$ per degree of freedom equal to 0.24 and 0.15 respectively.

 We collect  our final results for the ratio of $\sqrt{8 t_0}$, $w_0$ and $w_1$ to $\sqrt{8t_1}$ in table~\ref{tab:flow_scales}. These ratios can be used to convert all the results given in this section, in particular the values of the lattice spacing as a function of the bare coupling and gluino mass, to the other more standard units used in the literature.  

\begin{table}[t]
    \centering
    \begin{tabular}{ccc}
    \toprule
         $\sqrt{t_0/t_1}$ &   $w_1/\sqrt{8 t_1}$&$w_0/\sqrt{8 t_1}$   \\
    \midrule
        1.627(50) & 0.4535(49) & 0.586(10) \\
    \bottomrule
    \end{tabular}
    \caption{Different scales determined from the flow expressed in units of $\sqrt{8t_1}$. Scales $t_1(t_0)$ and $w_1(w_0)$ are derived respectively from eqs.~\eqref{tc} and~\eqref{wc} setting $s=0.05(0.1)$. }
    \label{tab:flow_scales}
\end{table}

\subsection{Setting the scale with Wilson loops (Creutz ratios) }
    \label{s:creutz}
    
In this subsection we focus on the second class of observables chosen to set the scale. These are the logarithm of the Wilson loop expectation values  or rather its derivative $F(r,l)$. As usual we have to look for  lattice counterparts of these observables.  These turn to be very  well-known lattice quantities, the Creutz ratios, defined as follows:  
    \begin{equation}\label{creutzratio}
            \chi(R,R') = -\log
	    \frac{W(R+0.5,R'+0.5)W(R-0.5,R'-0.5)}{W(R+0.5,R'-0.5)W(R-0.5,R'+0.5)}.
     \end{equation}
In this formula $W(R,R')$ is the lattice Wilson loop evaluated for a rectangle of
size $r\times r'$, where $r=R a$ and $r'=R'a$ are integer multiples of the  lattice spacing $a$. Thus, in our definition of  $\chi(R,R')$, the
arguments take half-integer values. Taylor expanding the Wilson loops  we see that 
 \begin{equation}
  \chi(R,R') = a^2 F(Ra, R'a) - \frac{a^4}{12}   \frac{\partial^4\log
  \mathcal{W}(r,r')}{\partial r^3 \partial r'} + \ldots.
\end{equation}
Thus, in the continuum limit one has 
\begin{equation}\label{dimlesscreutz}
 G_L(R) \equiv R^2\chi(R,R) \xrightarrow{a\rightarrow 0} r^2 F(r,r) +
    \order{\frac{a^2}{r^2}}.
    \end{equation}
Notice that the first term is universal, being independent of the lattice bare coupling, once 
$r$ is measured in units of an implicitly defined $\bar{r}$.

A practical problem which appears when implementing this method is that Creutz ratios  $\chi(R,R')$ 
are very noisy quantities for $R$ and $R'$ large.  To solve this problem one can use, as mentioned in the previous section,  the gradient flowed (or APE smeared) equivalents of these observables: $\mathcal{W}_t(r',r)$,  $F_t(r,r)$ and $G_t(r)$. The lattice counterparts employ the same methodology as in the previous subsection allowing the definition of a lattice equivalent to $\hat{G}(r;s)$ in eq.~\eqref{Ghatdef}: $\hat{G}_L(R;s)$. In this case, the flow time in lattice units is fixed in terms of $R$ as follows: $\sqrt{8 T}= s R$. Although, this function is well-defined in the limit $s \to 0$, performing this extrapolation as in ref.~\cite{\refst} is unnecessary. For an effective statistical error reduction it is enough to keep $s$ larger than $0.1$ and smaller than $1$. Notice that intuitively our lattice observables are square Creutz ratios obtained from  {\em fat} links of thickness proportional to its edge length.  

Having defined the lattice observables to be used, we will now describe the process leading to the determination of the lattice spacing in units of $\bar{r}(f_0,s)$, c.f. eq.~\eqref{creutzcond}. This involves a collection of technical steps that we list below.
\begin{itemize} 
\item For each configuration of our simulation parameters ($b$, $\kappa_a$, $N$) we evolve the lattice link variables using the same discretized flow that was  explained earlier. 
\item At each  flow time $T$ we compute the Wilson loops and Creutz
ratios for $R=R'=$1.5, 2.5, 3.5, 4.5 and 5.5, and determine from them 
the corresponding value of $\hat{G}_L(R;s)$ and its error by averaging over configurations having the same simulation parameters.
\item We investigated the $N$ dependence of these values for those
cases in which we have at least two values of $N$. As expected the
sensitivity depends on the quantity $R/\sqrt{N}$ giving the ratio of the
loop size to the effective lattice size. This is indeed what happens
when computing  these quantities in perturbation theory at finite
volume. To keep the finite volume correction smaller than 1-2\%, one
should set $R/\sqrt{N}\le 0.25$. In  pure Yang-Mills theory we could
reach $N=841$ and $N=1369$ which allowed us to go up to 
$R_{\mathrm{max}}=7.5$ and $R_{\mathrm{max}}=9.5$ respectively. 
Here our largest value of $R_{\mathrm{max}}=5.5$ gives
ratios of 0.29, 0.32 and 0.49 for $N=361$, 289 and 169 respectively,
which are larger than 0.25. Indeed, for $\hat{G}_L(5.5;s)$ the difference
between the value at $N=289$ and $N=361$ can reach up to 10\%. Thus, in
order to process the results we first extrapolate to $N=\infty$ using the  $N=289$ and 361 data and assuming a
$1/N^2$ dependence as predicted by  perturbation theory for large enough $N$. In practice this limits our determination of the scale to the cases in which there is $N=361$ data available. 

\begin{figure}[t]
\centering
    \hspace*{-1.cm}
     \includegraphics[width=\linewidth+1cm]{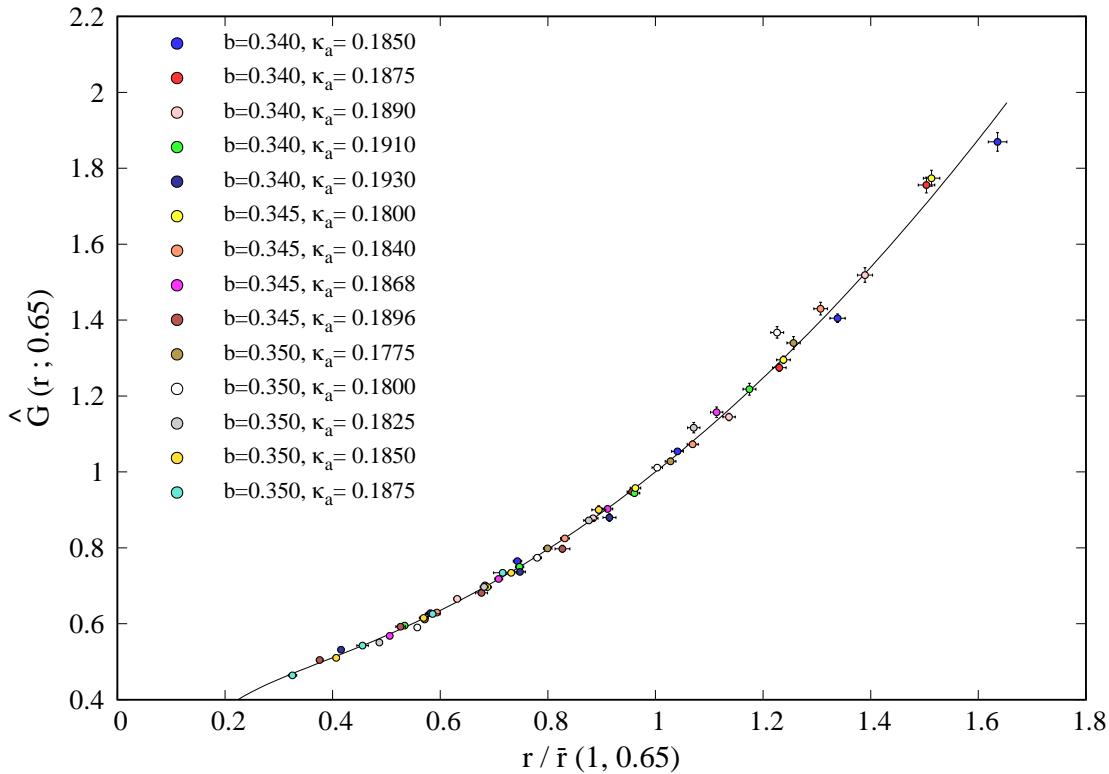}
\caption{For all our $N=361$ data samples, we display $\hat{G}(r;0.65)$ as a function of $r/\bar{r}(1,0.65)$, where $\bar{r}(1,0.65)$ is the scale determined using eq.~\eqref{creutzcond} setting $f_0=1$ and $s=0.65$. The black line is obtained by doing a joint fit of all the data with $R>1.5$ to eq.~\eqref{creutz_params}, as described in the text. }
\label{fig:universal_function}
\end{figure}

\item  To extract the value of $a/\bar{r}(f_0,s)$ one should deal with two additional issues. The first is that 
the values of $r$ obtained on the lattice are multiples of the lattice spacing. Determining the value $\bar{r}$ at which 
$\hat{G}(\bar{r};s)=f_0$ must be done by interpolation. The second is that $\hat{G}_L$ differs from the continuum function $\hat{G}$ by terms of order $a^2$
as described in eq.~\eqref{dimlesscreutz}. One can deal with both issues simultaneously by a method that gives a more robust determination of the scale. It involves a simultaneous fit to all our data points with $2.5\le R\le 5.5$ by a function 
\begin{equation}\label{creutz_params}
    G_L(R,s) = \alpha\qty(\frac{R}{\bar{R}})^2 + 2\gamma +
    4\qty(\frac{\bar{R}}{R})^2\qty(\frac{f_0-\alpha-2\gamma}{4} + \frac{\delta}{\bar{R}^2}),
\end{equation}
where $\alpha$, $\gamma$ and $\delta$ only depend on the value of $s$ and $f_0$, and $1/\bar{R}=a(b,\kappa_a)/\bar{r}(f_0,s)$ expresses the lattice spacing in $\bar{r}(f_0,s)$ units. Thus, the fitted data  contains 4 values of $R$ for each of the 14 total simulation parameters, and the fit parameters are the 14 values of $\bar{R}$ and the three additional parameters $\alpha$, $\gamma$ and $\delta$. The rationale behind the parameterization is given by the flowed equivalent to eq.~\eqref{dimlesscreutz}.  The universal function  $F_{\bar{t}(r;s)}(r,r)$ is well described by a second order polynomial in $(\bar{r}(f_0,s)/r)^2$ forced to be equal to $f_0$ for $\bar{r}(f_0,s)/r=1$. This parameterization is inspired by the results at $s=0$ in which $\alpha$ is given by the string tension $\sigma\bar{r}^2(f_0,s)$ and  $\gamma$ by a L\"uscher-type term. 
The parameter $\delta $ is introduced to account for the $a^2$ correction appearing in eq.~\eqref{dimlesscreutz}. 
\end{itemize}

The procedure can be performed for various values of $s$ and $f_0$ and the results should be compatible up to a change in the unit. In particular we chose two values of $f_0$ (0.65 and 1) and two values of $s$ (0.5 and 0.65) to check consistency. The results for different $f_0$ are perfectly compatible since they involve fitting the same data points. The data  just predicts the ratio $\bar{r}(1,0.65)/\bar{r}(0.65,0.65)=1.611$  and  $\bar{r}(1,0.50)/\bar{r}(0.65,0.50)=2.045$. On the other hand a change in $s$ involves data at different flow times so that the comparison serves to check independence of this choice. If we fit the ratio of scales to a constant we get perfect compatibility with a  constant value of $\bar{r}(1,0.65)/\bar{r}(1,0.50)=1.120(6)$.

Finally, we will express our lattice spacing in units of $\bar{r}(1,0.65)$ which are  the ones affected by smaller errors. The results are given in table~\ref{tab:6scales}. Notice that  the final values come from global fit to the data which assumes  scaling. Hence, the errors do include a part associated to the amount of  scaling violation present in  our data. A visual determination of how well our data satisfies scaling can be obtained by plotting the best fit to the continuum function $\hat{G}(r;0.65)$. This is given in figure~\ref{fig:universal_function}. Together with the function we plot all our data points after subtraction of the lattice artefact $\delta$ term. The horizontal errors come from the errors in the determination of the scales $\bar{R}$. The overall agreement is very good.

\subsection{Setting the scale with fundamental meson spectrum}
    \label{s:mesons}
    \newcommand{\kappaf}{\kappa_{f}}

In this section we will explain the most basic details of the lattice implementation of the  determination of the meson spectrum of fundamental quarks in this theory. More technical aspects will be collected in appendix~\ref{app:meson_correlator}.

Fundamental  quarks are quenched if the number of flavours over the number of colours tends to zero so that meson masses can be considered observables of the gauge-gluino theory. From that point of view the methodology applied to the determination of the fundamental-meson spectroscopy does not differ from the one applied for the large $N$  pure gauge twisted reduced models  in previous publications~\cite{Gonzalez-Arroyo:2015bya,Perez:2020vbn}. We refer the reader to these publications for derivations and a more detailed explanation. In any case, as mentioned previously,  the main idea is quite simple. We  let the fermions propagate in an extended lattice on the background of the reduced model gauge field. This background field is periodic up to a twist and repeats itself after $a\sqrt{N}$ periods in each direction. Effectively it amounts to having gauge fields defined on a periodic lattice of size $(\sqrt{N})^4$. The fermion fields can live in a much bigger lattice including infinite. For fundamental quark fields it is natural to impose that they propagate on a lattice of the same size $(\sqrt{N})^4$, which considerably simplifies the formulas. For practical reasons in computing correlators in time it is useful to duplicate the lattice temporal length $L_0=2 \sqrt{N}$.

To achieve our final goal we have first to compute the correlation functions of bilinear quark operators. Here we restrict to the pseudoscalar, axial and vector channels. The next step is to determine the mass of the lightest state having the corresponding quantum numbers. This is done for various values of the fundamental quark mass. Then we extrapolate those masses to the fundamental chiral limit. The pion should behave as a Goldstone boson and the PCAC relation should hold as we approach this limit. The procedure is repeated for all our gauge couplings and gluino masses, and the resulting vector meson mass in that chiral limit is precisely the observable that we use to set the scale of the theory. All these steps will be spelled out and the results presented in the following susubsections.     
\begin{figure}
    \centering
        \hspace*{-1.cm}
     \includegraphics[width=\linewidth+2cm]{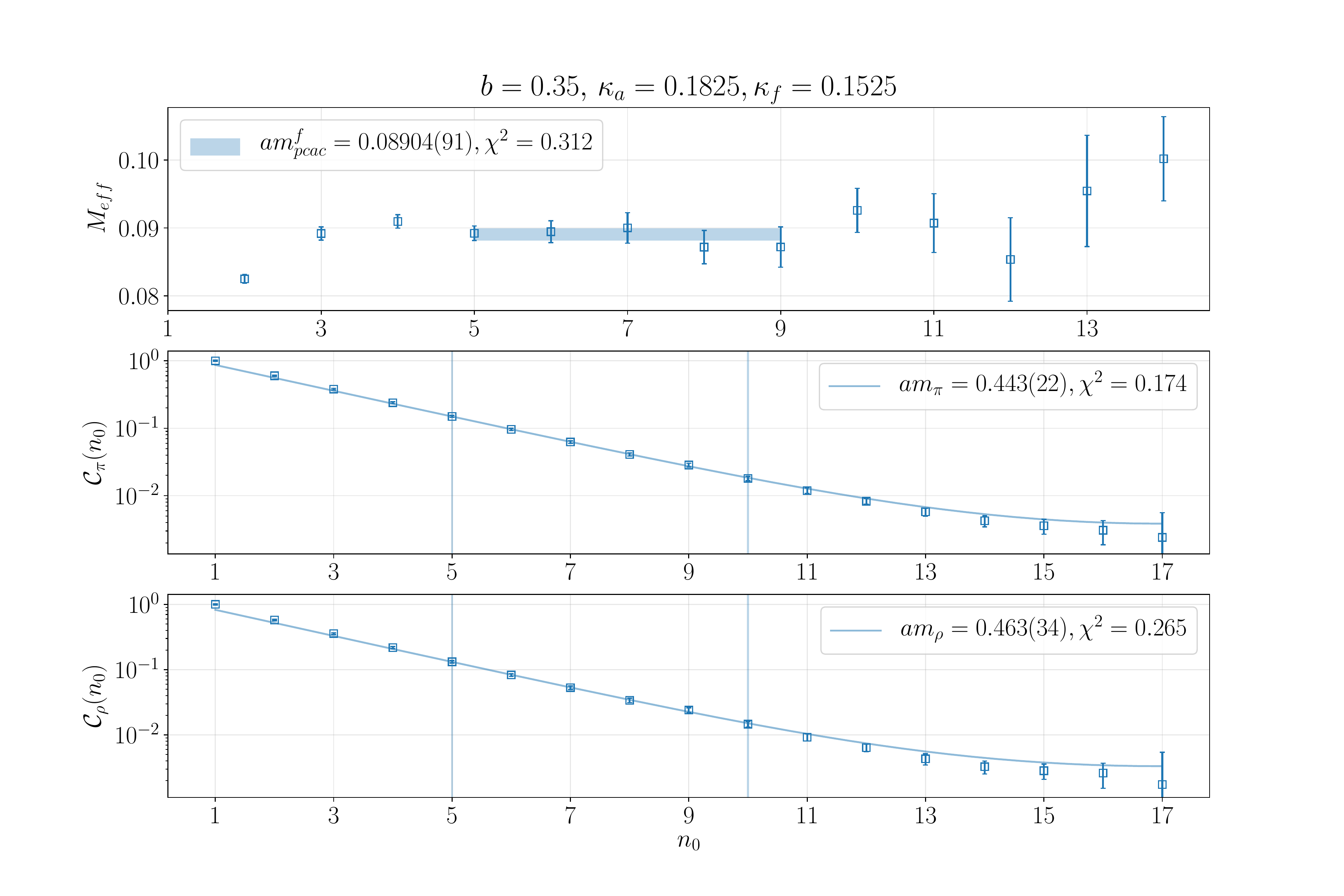}
    \caption{Stack plot of the correlators in this analysis for one example case. The first one shows the signal for $am^f_\text{pcac}$ as an effective mass. The blue band is the result of the fit, the length corresponds to the time slices used to fit the mass, while the width correspond to the error. The second and the third, displays the pion and the rho correlator signal, respectively. The blue line is the result of the fit performed in the region between the vertical lines.}
    \label{fig:corr_example}
\end{figure}

\subsubsection{Methodology}\label{sssec:methodology}

The meson correlators in time of fundamental quarks are expectation values 
\begin{equation}
{\cal C}_{A B}(n_0)= \langle \mathbf{O}_A(0)   \mathbf{O}_B^\dagger(n_0) \rangle,
\end{equation}
where $\mathbf{O}_{A}$ and $\mathbf{O}_{B}$ are gauge invariant bilinear quark operators projected to zero spatial momentum. As explained previously, for the purpose of this paper we restrict ourselves to pseudoscalar and vector-meson operators, as for example: $\sum_{\vec{n}}\mathbf{\overline{\Psi}}(n_0,\vec{x}) \, \gamma_5 \, \mathbf{\Psi}(n_0,\vec{n}) $ and $\sum_{\vec{n}}\mathbf{\overline{\Psi}}(n_0,\vec{n}) \, \gamma^\mu \, \mathbf{\Psi}(n_0,\vec{n}) $. However, there are infinitely many operators with the same quantum numbers and this is an essential advantage that is used by us and other researchers. In practice we will be considering spatially non-local operators obtained by applying two well-known algorithms: the Wuppertal smearing of fermion bilinears~\cite{Gusken:1989ad,Gusken:1989qx,Bali:2016lva} and the three-dimensional APE smearing~\cite{APE:1987ehd}. The explicit expressions have been used in previous papers and are recalled in appendix~\ref{app:meson_correlator}. The advantage of using these operators is that their coupling to the lowest mass state in each channel is enhanced with respect to excited states. In any case, it is useful to consider linear combinations of these operators designed to optimize the coupling to the ground state in each channel relative to other low mass states. This variational procedure is by now quite standard and goes by the name GEVP~\cite{Michael:1985ne,Luscher:1990ck}. 
As an example of the outcome of the method, we display in figure~\ref{fig:corr_example} the correlators of the optimal operators in the pseudoscalar and vector channels for the case $b=0.35$, $\kappa_a=0.1825$, $\kappa_{f}=0.1525$, with $\kappa_f$ the fundamental hopping parameter. The time-dependence of these correlators shows a clear exponential decay from which the mass 
of the  $\pi$-meson (pseudoscalar) and  $\rho$-meson (vector) can be extracted. The numbers obtained and the $\chi^2$ of the fits to an exponential are also displayed on the figure. We also show the signal that allows one to extract the fundamental PCAC mass $m^f_\text{pcac}$, defined as in eq.~\eqref{eq:mpcac}. 

Summing up, the procedure to obtain the final results is the one described in the example. It amounts to determining the optimal operator and then to fit the corresponding correlator in a certain interval to an exponential (rather to a hyperbolic cosine function, taking into account the periodic nature of the temporal direction).  There are of course some specific details which reflect the selection of the operator and the choice of fitting interval that affect the final numerical value for the mass. In the end, variations of these types which are of similar statistical significance are accounted for  as a systematic error of the determination. The specific details of the GEVP method  that we have used will be collected in appendix~\ref{app:meson_correlator}.  Here we will comment briefly on the choice of the fitting interval. The main points to be taken into account in this selection are finite volume effects, lattice artefacts and the contamination of excited states.  Typically, finite-size effects are more relevant close to the chiral limit, as the pion mass goes to zero and its Compton wavelength becomes comparable to the effective volume. To avoid too severe effects our selected values of the hopping parameter should stay sufficiently far from the chiral limit. In particular in our data the lightest cases still had $m_\pi a \sqrt{N}\sim 3.4$. 
Nonetheless, finite-size effects may still reflect in the appearance of a constant term in the correlator arising from the propagation of quarks wrapping around the finite extent of the lattice~\cite{Umeda:2007hy}. Although this effect disappears in the large volume (large $N$) limit, we have observed that in some cases the addition of a small constant to the hyperbolic cosine was required to obtain a good fit.
Finally, we should comment about the lower limit of the fitting interval. A smaller time separation implies a smaller relative error at the expense of a larger contamination of excited states. A balance is then necessary, our choice has been to fix the lower limit of the interval in physical units setting it equal to $R\sqrt{8t_1}$. We have taken in all cases $R=0.95$ except for $b=0.36$ for which, to increase the signal-to-noise ratio,  a value $R=0.8$ was preferred. As for the upper limit of the interval, it was set in all cases so as to have at least 6 lattice points in the fitting range.

\begin{table}[H]
    \centering
    \small
    \begin{tabular}{ccllll}
        \toprule
            $b$   &    $\kappa_a$  &  $\kappa_f$ & $am^f_\text{pcac}$ &  $am_\pi$   & $am_\rho$    \\
        \midrule
            \multirow{15}*{0.34} & \multirow{4}*{0.189}   &      0.15500 &   0.09711(88)  &    0.507(12)   &    0.549(20)  \\  
                                 &                        &      0.15700 &   0.05755(89)  &    0.381(17)   &    0.434(31)  \\   
                                 &                        &      0.15800 &   0.0392(11)   &   0.350(14)    &    0.421(27)  \\    
                                 &                        &      0.15920 &  0.0149(11)   &   0.248(39)    &    0.336(56)  \\
                                 \cline{2-6}   
                                 & \multirow{4}*{0.191}   &      0.15250 &  0.1253(11)   &   0.545(16)    &    0.570(21)  \\
                                 &                        &      0.15500 &   0.07464(76)  &    0.423(17)   &    0.450(29)  \\   
                                 &                        &      0.15580 &  0.05929(86)  &    0.370(14)   &    0.416(22)  \\   
                                 &                        &      0.15700 &   0.03601(89)  &    0.345(33)   &    0.373(54)  \\
                                 \cline{2-6}   
                                 & \multirow{3}*{0.192}   &      0.15380 &  0.08564(87)  &    0.430(20)   &    0.458(27)  \\
                                 &                        &      0.15550 &  0.05090(88)  &    0.368(21)   &    0.405(32)  \\
                                 &                        &      0.15650 &  0.03086(75)  &    0.336(21)   &    0.379(34)  \\
                                 \cline{2-6}    
                                 & \multirow{4}*{0.193}   &      0.15240 &  0.1000(12)   &   0.443(21)   &    0.450(30)   \\
                                 &                        &      0.15380 &  0.06932(92)  &    0.318(24)   &    0.317(30)  \\   
                                 &                        &      0.15530 &  0.03904(60)  &   0.301(21)    &    0.299(27)  \\   
                                 &                        &      0.15630 &  0.01982(96)  &   0.288(34)    &    0.283(46)  \\
                                 \midrule  
            \multirow{22}*{0.35} & \multirow{4}*{0.1775}  &      0.15000 &    0.1730(11)   &   0.701(16)    &    0.741(22)  \\   
                                 &                        &      0.15250 &  0.11939(90)  &    0.558(13)   &    0.599(20)  \\   
                                 &                        &      0.15500 &   0.06896(73)  &    0.425(15)   &    0.480(27)  \\   
                                 &                        &      0.15625 & 0.04288(45)  &    0.328(18)   &    0.382(28)  \\
                                 \cline{2-6}   
                                 & \multirow{4 }*{0.18 }  &      0.15000 &    0.15806(93)  &    0.608(12)   &    0.626(15)  \\
                                 &                        &      0.15250 &  0.10532(64)  &    0.488(13)   &    0.514(19)  \\   
                                 &                        &      0.15500 &   0.05327(78)  &    0.343(14)   &    0.377(24)  \\   
                                 &                        &      0.15625 & 0.02986(60)  &    0.323(18)   &    0.409(33)  \\
                                 \cline{2-6}   
                                 & \multirow{5 }*{0.1825} &      0.14700 &   0.2093(16)   &   0.719(18)    &    0.731(21)  \\
                                 &                        &      0.15000 &    0.1457(13)   &   0.560(19)    &    0.582(25)  \\   
                                 &                        &      0.15250 &  0.08904(91)  &    0.443(22)   &    0.463(34)  \\   
                                 &                        &      0.15500 &   0.03765(92)  &    0.306(30)   &    0.303(50)  \\   
                                 &                        &      0.15580 &  0.02157(81)  &    0.314(22)   &    0.319(43)  \\
                                 \cline{2-6}    
                                 & \multirow{5 }*{0.185}  &      0.14930 &  0.1410(19)   &   0.533(26)    &    0.537(33)  \\
                                 &                        &      0.15100 &   0.1025(13)   &   0.430(24)    &    0.432(32)  \\   
                                 &                        &      0.15250 &  0.0704(12)   &   0.351(25)    &    0.346(34)  \\   
                                 &                        &      0.15380 &  0.0428(10)   &   0.288(30)    &    0.277(42)  \\   
                                 &                        &      0.15500 &   0.0176(11)   &   0.259(42)    &    0.262(78)  \\
                                 \cline{2-6}  
                                 & \multirow{5 }*{0.1875} &      0.14180 & 0.1538(23)   &   0.496(32)    &    0.495(39)  \\
                                 &                        &      0.15000  & 0.1106(17)   &   0.447(28)    &    0.452(35)  \\   
                                 &                        &      0.15100 &  0.0882(14)   &   0.397(29)    &    0.401(36)  \\   
                                 &                        &      0.15250 & 0.0547(12)   &   0.323(34)    &    0.323(45)  \\
                                 &                        &      0.15380 & 0.0258(11)   &   0.253(52)    &    0.249(68)  \\
                                 \midrule  
            \multirow{3}*{0.36}  & \multirow{3 }*{0.1831} &      0.14750 &  0.1194(20)   &   0.413(28)    &    0.406(33)  \\   
                                 &                        &      0.15000 &    0.0630(14)   &   0.282(39)    &    0.282(46)  \\   
                                 &                        &      0.15100 &   0.0430(12)   &   0.289(55)    &    0.282(67)  \\   
        \bottomrule
    \end{tabular}
    \caption{The values of $m^f_\text{pcac}$, $m_\pi$ and $m_\rho$ for fermions in the fundamental representation and for each $(b,\kappa_a,\kappa_f)$ value at $N=289$ are given. The fitting procedure is the one described in sec.~\ref{sssec:methodology}, from which we obtained $\chi^2$ per degrees of freedom in general smaller than one. }
    \label{tab:fundmeson}
\end{table}

\subsubsection{Results}

The results we obtained for the PCAC mass, the pion mass and the vector meson mass of fundamental fermions are reported in lattice units in table~\ref{tab:fundmeson}. 
We use these results to explore the (fundamental) chiral limit of the theory and extract an alternative scale to the ones presented in subsections~\ref{s:tekflow} and~\ref{s:creutz}. 

\begin{figure}[t]
    \centering
    \hspace*{-1.cm}
     \includegraphics[width=\linewidth+2cm]{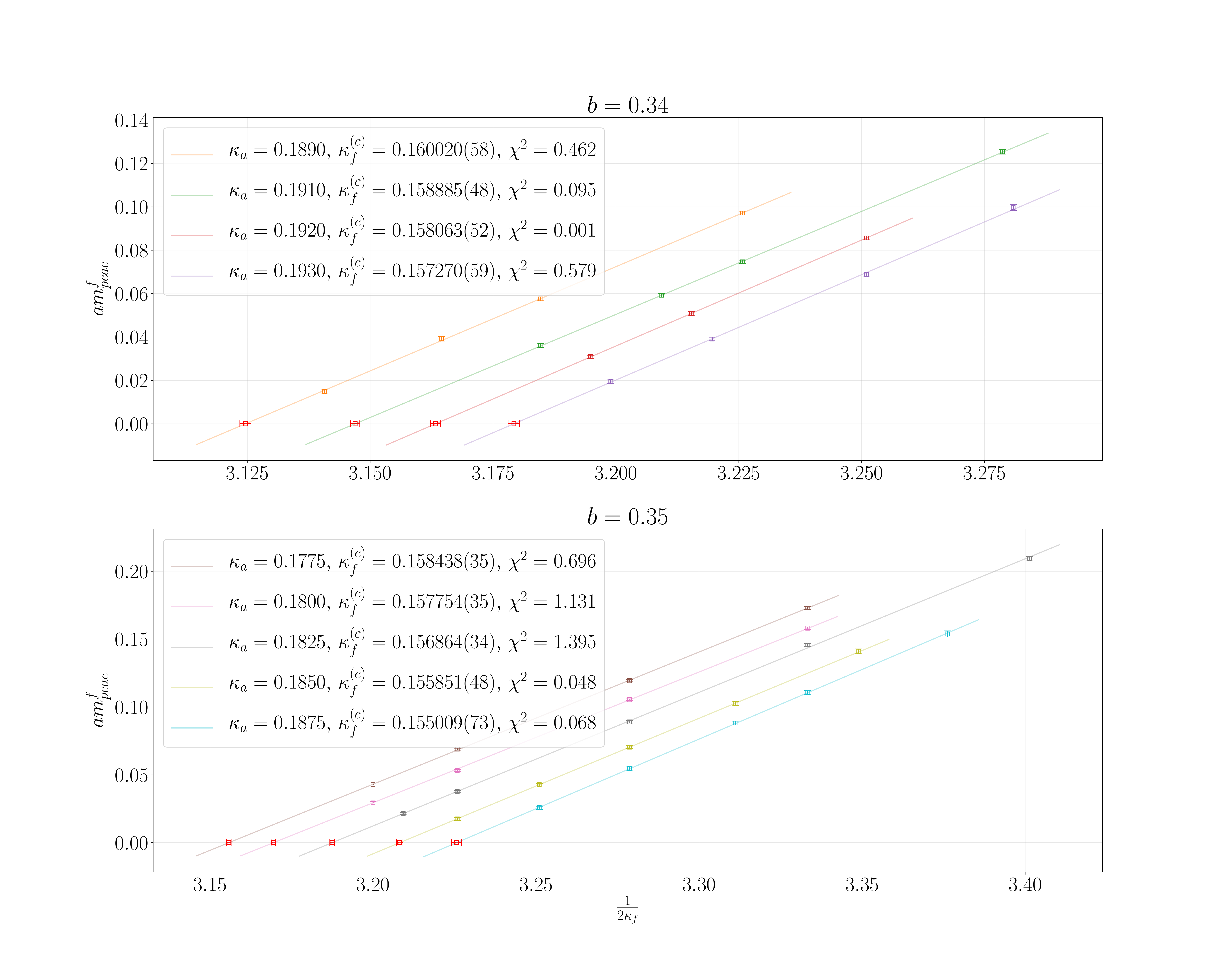}
    \caption{Extrapolation of $am^f_\text{pcac}$ to 0. In the labels we report the corresponding value of $\kappa_{f}^{(c)}$  extracted from the fit and the corresponding $\chi^2$ per degrees of freedom. Errors are calculated using standard jackknife techniques. }
    \label{fig:mpcac_vs_k}
\end{figure}

The first check we do is to analyze the dependence of the PCAC mass on the fundamental hopping parameter $\kappa_{f}$.  In order to determine the critical hopping parameter where the fundamental fermions become massless we follow the same strategy as for gluinos, i.e. we analyze the dependence of the PCAC mass on $\kappa_{f}$ and determine the critical hopping parameter as the point where the fundamental PCAC mass vanishes.
In figure~\ref{fig:mpcac_vs_k} we plot $am^f_{\text{pcac}}$ as a function of $1/(2\kappa_{f})$. 
Performing separate linear fits for each value of $(b,\kappa_a)$, we extracted in each case the critical value of the hopping parameter $\kappa_{f}^{(c)}$. As signalled by the $\chi^2$ per degree of freedom reported in the legend of the plot, the observed linear dependence is very good and confirms what one would expect from chiral symmetry restoration in the limit of vanishing quark masses for Wilson fermions.

        \begin{table}[t]
    \centering
    \begin{tabular}{clcccc}
        \toprule
            $b$              &    $\kappa_a$   &   $\frac{\sqrt{8t_1}}{a}$ (N=289)&   $\frac{\sqrt{8t_1}}{a}$ (N=361) & $\frac{\bar{r}(1,0.65)}{a}$  & $\frac{1}{am_\rho^\chi}$ \\
        \midrule
        \multirow{6}*{0.34}  & 0.185           &  2.878(2)(46)  & 2.883(2)(58)  & 3.362(36)   & -                \\
                             & 0.1875          &  3.209(3)(68)  & 3.188(3)(90)  & 3.658(37)   & -               \\                   
                             & 0.189           &  3.514(3)(65)  & 3.488(3)(88)  & 3.959(40)   &   3.10(13)    \\                 
                             & 0.191           &  4.049(4)(37)  & 4.042(4)(58)  & 4.682(48)   &   3.53(17)    \\                 
                             & 0.192067        &  4.568(6)(58)  &     -           &      -     &   3.54(28)    \\                    
                             & 0.193           &  5.273(7)(161) & 5.244(8)(175) & 6.014(78)   &   4.86(51)    \\   
        \midrule
        \multirow{4}*{0.345} & 0.18            & 3.166(3)(74)  & 3.145(2)(93)    & 3.635(37)   &  -              \\
                             & 0.184           & 3.664(4)(77)  & 3.645(3)(98)    & 4.208(41)   &   -             \\                   
                             & 0.1868          & 4.294(5)(40)  & 4.274(5)(55)    & 4.939(51)   &    -            \\                 
                             & 0.1896          & 5.559(8)(154) & 5.614(8)(201)   & 6.649(111)  &     -           \\                 
        \midrule
        \multirow{6}*{0.35}  & 0.1775          &  3.730(3)(97)   & 3.737(3)(105)  & 4.377(43)  &  3.08(14)     \\
                             & 0.18            &  4.109(4)(75)   & 4.003(7)(109)  & 4.485(43)  &  3.65(17)     \\                   
                             & 0.1825          &  4.634(6)(70)   & 4.516(5)(64)   & 5.135(56)  &  3.94(26)     \\                 
                             & 0.1850          &  5.364(8)(144)  & 5.323(7)(65)   & 6.187(108) &  5.06(37)     \\                 
                             & 0.186378        &  5.909(9)(192)  &  -              &     -     &       -         \\                    
                             & 0.1875          &  6.608(11)(249) & 6.582(12)(287) & 7.681(180) &  5.34(71)     \\  
        \midrule
        \multirow{6}*{0.36}  & 0.1760          &  5.77(1)(32)    &  -              &         -   &  -              \\
                             & 0.1780          &  6.32(1)(31)    &   -             &        -    &   -             \\                   
                             & 0.18            &  7.05(1)(31)    &    -            &       -     &  -    \\                 
                             & 0.1820          &  7.85(2)(35)    &     -           &      -      &  -     \\                 
                             & 0.183172        &  8.68(2)(44)    &      -          &     -       &  6.99(87)    \\                    
                             & 0.184           &  9.16(2)(52)    &       -         &    -        &     -           \\  
        \bottomrule
    \end{tabular}
    \caption{Values of the inverse lattice spacing in units of the three different scales determined in this work. For flow related scales, the first error is statistical, the second is systematic.}
     \label{tab:6scales}
\end{table}

We finally come to the determination of the scale based on the vector meson mass. 
To determine the chiral extrapolation of this quantity we perform a global fit of $am_\rho$ as a linear function of $am^f_\text{pcac}$ for each value of $(b,\kappa_a)$, by imposing a common slope and extracting $am_\rho^\chi(b,\kappa_a)$ as the intercept at vanishing PCAC mass. We obtain a $\chi^2$ per degree of freedom of 0.56.

The final values of the inverse lattice spacing in units of $m_\rho^\chi$ obtained in this way are reported in our summary table~\ref{tab:6scales}. As we will discuss in the next section, the ratio between this scale and $\sqrt{8t_1} $ turns out to be quite close to one.

\begin{figure}[t]
    \centering
     \includegraphics[width=\linewidth]{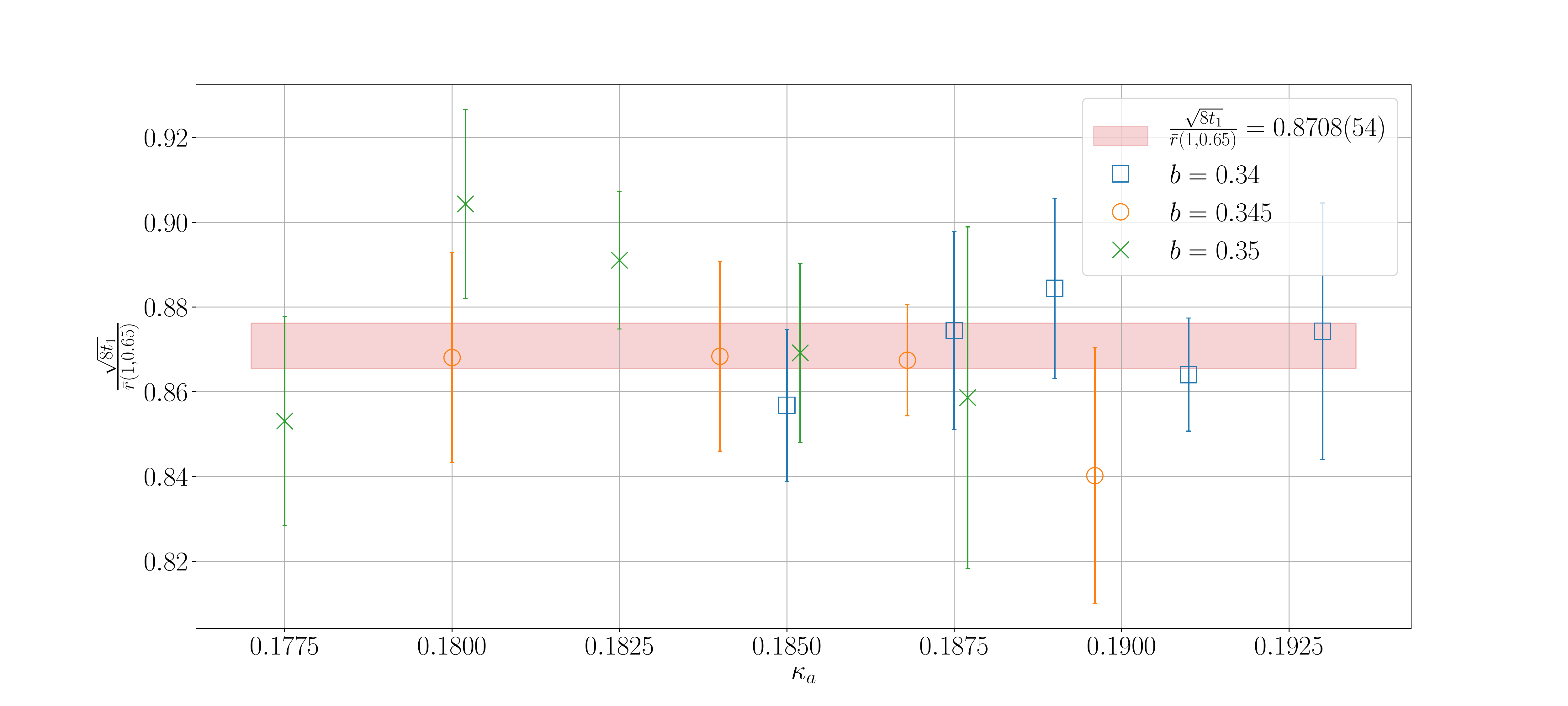}
    \includegraphics[width=\linewidth]{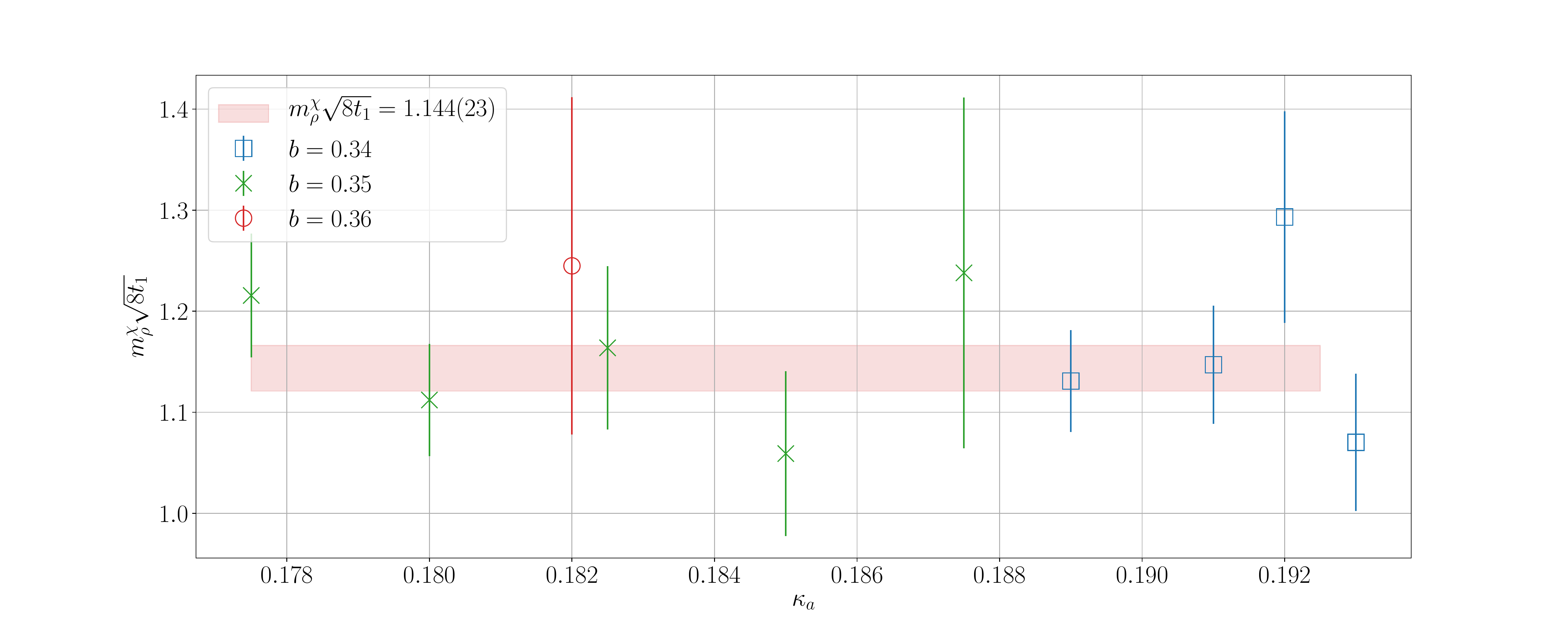}
    \caption{Comparison of scales as a function of the adjoint hopping parameter $\kappa_a$ for different gauge couplings. In the top panel it is displayed the ratio $\sqrt{8t_1}/\bar{r}(1,0.65)$, in the bottom one $m_\rho^\chi\sqrt{8t_1}$. The red bands represent the average weighted over the errors, while their width represent the statistical uncertainty over the average ratio. When needed, points corresponding to the same $\kappa_a$ have been slightly shifted in the x-axes to avoid overlapping.}
    \label{fig:ratio}
\end{figure}

\subsection{Scale comparison}
    \label{s:comparison}

We dedicate this last part of the section to the comparison of the results obtained with the three different scale setting methods which are summarized in table~\ref{tab:6scales}. In each method we have determined the lattice spacing in terms of  a different unit:  $a/\sqrt{8t_1}$, $a/\bar{r}$ and $am_\rho^\chi$. Although each of these quantities changes considerably when we change $b$ (the inverse lattice 't Hooft coupling) and the hopping parameter $\kappa_a$ (related to the gluino mass), scaling dictates that the ratio should stay constant and be given by the ratio of the corresponding units of energy. In   figure~\ref{fig:ratio} we display the two independent ratios 
$\sqrt{8t_1}/\bar{r}(1,0.65)$ and $m_\rho^\chi\sqrt{8t_1}$  for all the cases in which it is available. For the case of $a/\sqrt{8t_1}$ we have used an average of the results of $N=289$ and $N=361$ with errors that are dominated by the systematic ones. The results are compatible with being a constant within errors. From the best fit we estimate that the conversion factor between the two units $\sqrt{8t_1}$ and $\bar{r}(1,0.65)$ is $0.8708(54)$, while between  $\sqrt{8t_1}$ and $1/m_\rho^\chi$ it is $1.144(23)$. 

\begin{figure}[t]
    \centering
     \includegraphics[width=\linewidth]{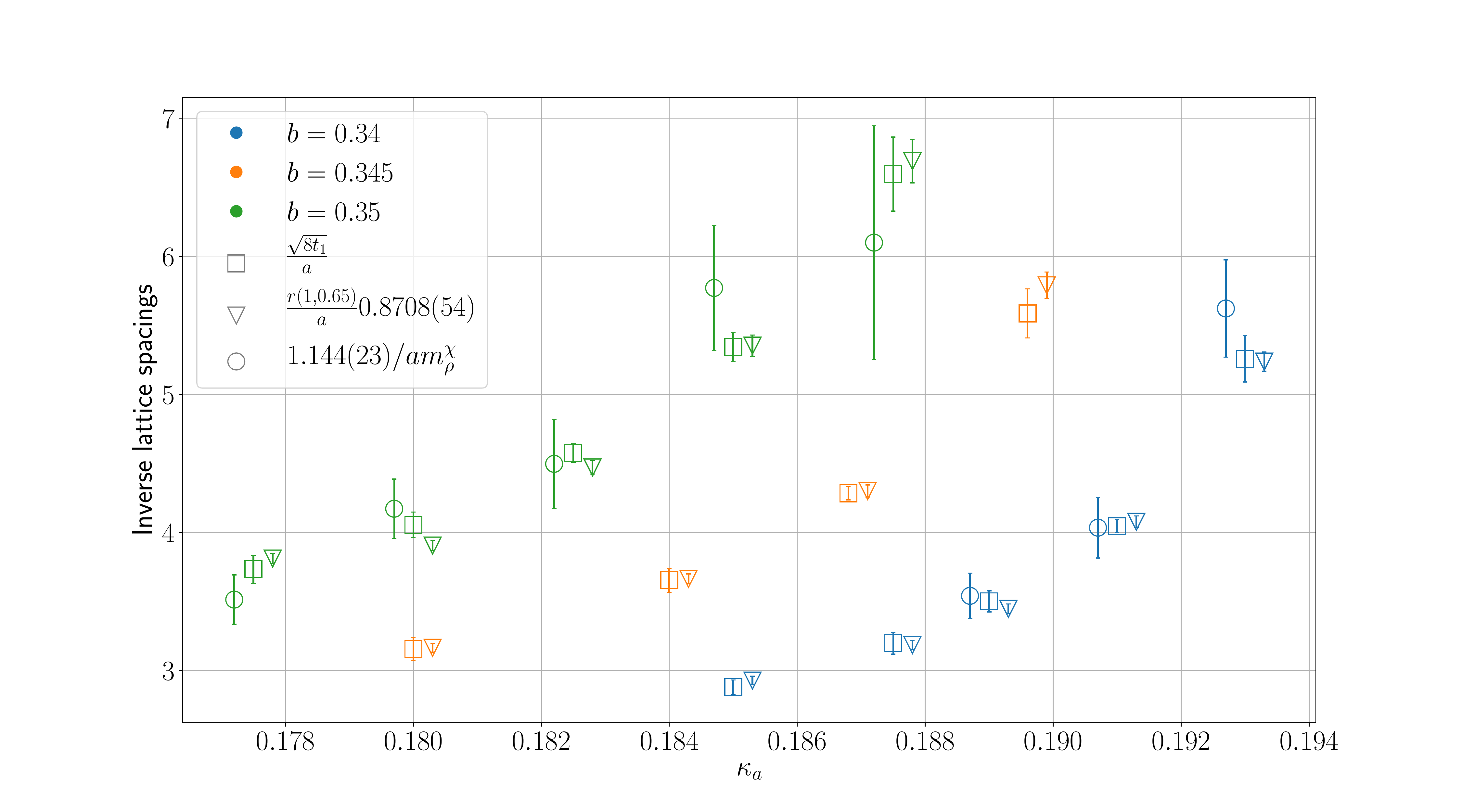}
    \caption{The inverse lattice spacing for the different values of $(b,\kappa_a)$. The quantities $\bar{r}(1,0.65)/a$ and $1/am_\rho^\chi$ have been rescaled with the conversion factor indicated in the legend to match the value of $\sqrt{8t_1}/a$.}
    \label{fig:comparison}
\end{figure}
\begin{figure}[t]
    \centering
     \includegraphics[width=\linewidth]{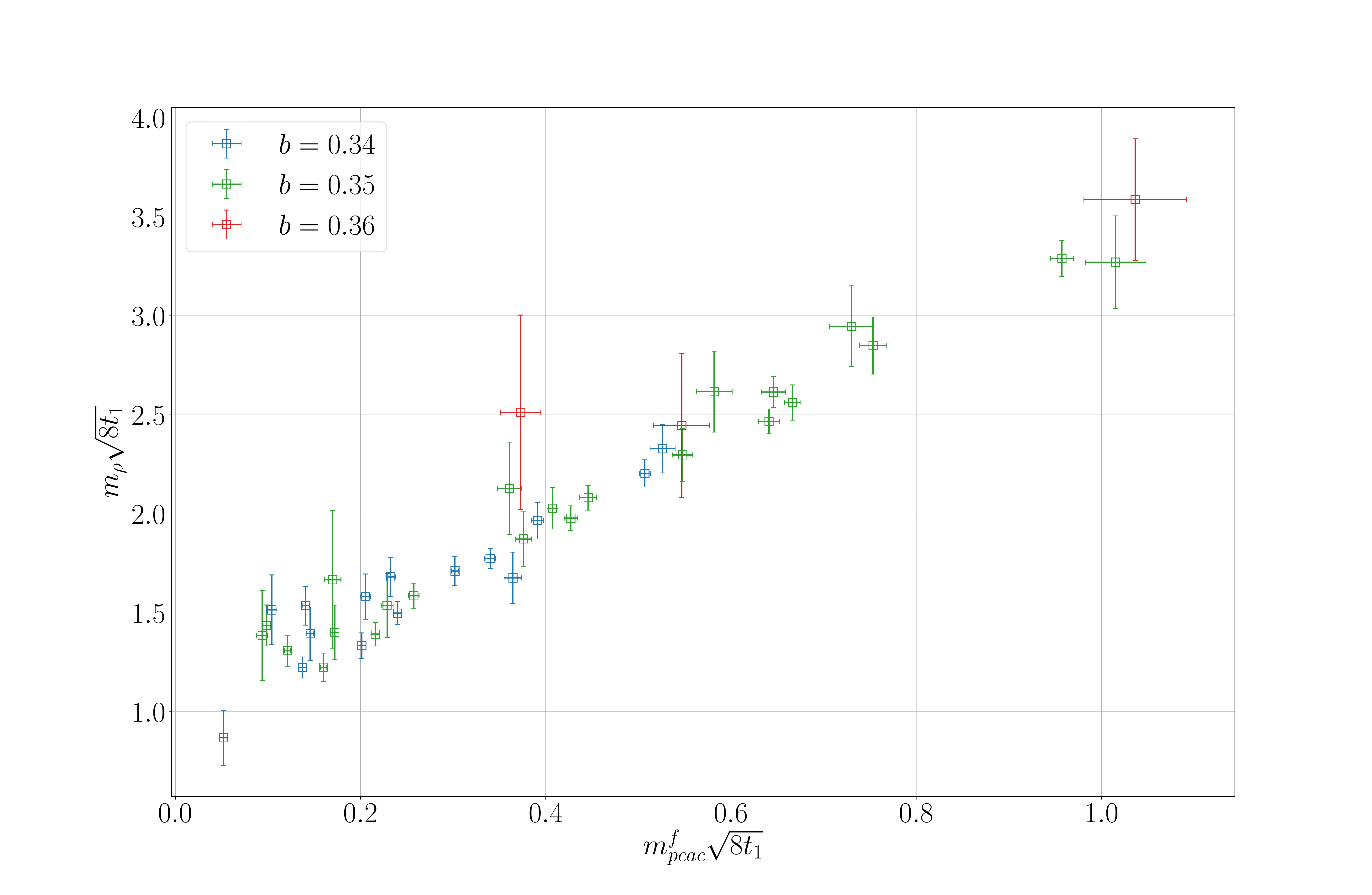}
    \caption{Dependence of $m_\rho\sqrt{8t_1}$ on $m^f_\text{pcac} \sqrt{8t_1}$ for all the data samples included in table~\ref{tab:fundmeson}. }
    \label{fig:mrho_vs_mpcac}
\end{figure}

A more visual impression of  how scaling works can be seen in figure~\ref{fig:comparison}
where the three lattice spacing determinations of the inverse of the lattice spacing are displayed side by side after applying the conversion factors determined earlier. All datasets for which $\bar{r}(1, 0.65)$ was available are displayed. The figure shows how the three scales change considerably within all the datasets following the same trend in a consistent way. One can also notice the relative size of the errors of the three determinations of the lattice spacing. It looks as if the scale  based on the Creutz ratios is the  most precise, but the $N$ dependence had to be corrected for and systematic errors might be underestimated. Furthermore, our flow-based  scale is the one that has been determined for all our datasets and hence, it will be used for the determination of the scale in the Supersymmetric limit to be presented in the next section. 

Concerning  the rho mass, although  also  consistent,  it is much less precise than the other two. Indeed, the scaling behaviour also holds for the  meson masses built from massive fundamental quarks. This can  be seen in figure~\ref{fig:mrho_vs_mpcac} in which the rho mass in $1/\sqrt{8 t_1}$ units is plotted against the fundamental PCAC mass in the same units. Although with large errors all the data points seem to follow the same trend.

    \section{Final results and Conclusions}
    \label{s:conclusion}
        \begin{figure}[t]
    \centering
     \hspace*{-.5cm}
     \includegraphics[width=\linewidth+1cm]{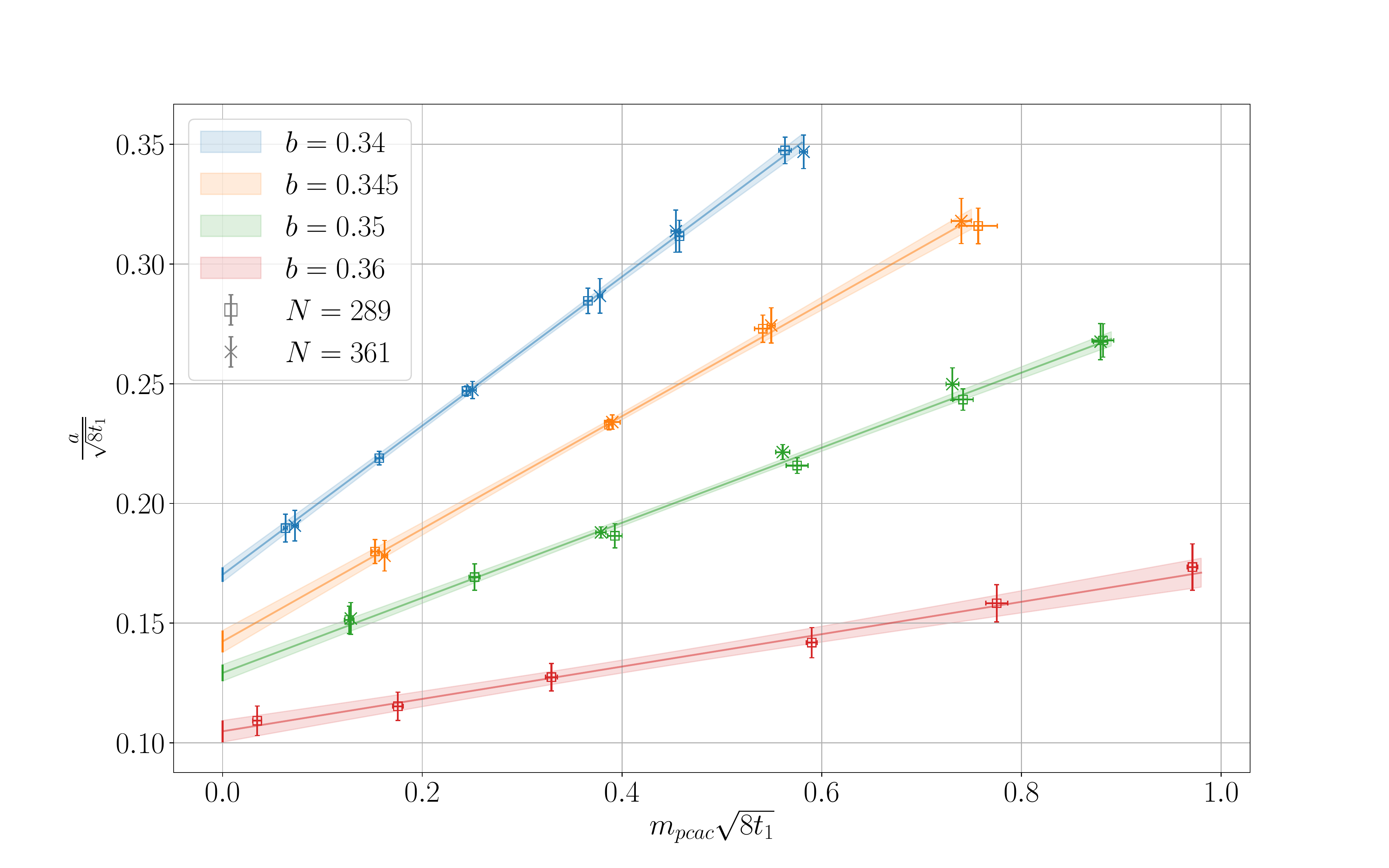}
    \caption{The lattice spacing expressed in units of the gradient flow scale $\sqrt{8t_1}$ as a function of the adjoint PCAC mass in physical units. 
    Different markers represent a different number of colours $N$ (volume). The straight lines represent the extrapolation of the lattice spacing to the massless-gluino limit, obtained  through a joint fit to all the points for each $b$. The error bars on the $y$-coordinate are dominated by systematics -- see table~\ref{tab:6scales} -- while those on the horizontal axis take only into account the errors on the bare PCAC mass.}
    \label{fig:chirallimit}
\end{figure}

\subsection{The lattice spacing and the \texorpdfstring{$\beta$}{beta}-function in the SUSY limit}

After having completed the determination of the scale with good precision for our massive gluino values, 
 we will here attempt  achieving   our main goal of determining the scale for the supersymmetric theory. This will be done  by  extrapolation of the scale to the massless-gluino limit. Given that we have used different scale setting observables and units which are mutually compatible, we will here focus on the flow unit $\sqrt{8t_1}$ since  it covers all our simulation points and is both  precise and relatively insensitive to finite $N$ corrections.  Hence, we extrapolate the lattice spacing $a$ in those units to the limit of vanishing $m_\text{pcac}$ always expressed in physical units. The resulting plot is shown in figure~\ref{fig:chirallimit}.
In the plot, different markers were used to display the points belonging to $N=289$ and $N=361$. By looking at the plot, it is visible by eye that the points corresponding to different markers are compatible within errors, showing that we were able to control finite-volume effects. 
The points are well fitted by a straight line and no higher-order polynomial terms are necessary to perform the extrapolation, which is remarkable, taking into account the wide range of $m_\text{pcac}$ values covered. 
\begin{table}[t]
    \centering
    \begin{tabular}{lcccl}
    \toprule
        $b$   &  $\frac{a^\chi}{\sqrt{8t_1}}$ &    $\frac{a^\chi}{\sqrt{8t_0}}$ & $\frac{a^\chi}{w_0}$ & $P(b)$  \\
    \midrule
        0.34  & 0.1702(31) & 0.1046(37) & 0.2904(72) & 0.5620(3) \\
        0.345 & 0.1423(46) & 0.0875(39) & 0.2428(89) & 0.57027(15) \\
        0.35  & 0.1292(35) & 0.0794(33) & 0.2205(71) & 0.57810(15) \\
        0.36  & 0.1048(45) & 0.0644(34) & 0.1788(83) & 0.5934(1) \\
    \bottomrule
    \end{tabular}
    \caption{Lattice spacing in units of $\sqrt{8t_1}$ for the supersymmetric theory. For comparison with other authors we also convert to $\sqrt{8t_0}$ and $w_0$ units using the conversion factor determined in the previous section table~\ref{tab:flow_scales}. The last column displays the plaquette expectation value extrapolated to the massless gluino limit.}
    \label{tab:achi}
\end{table}

The values of the scale extrapolated to the massless-gluino limit are reported in table~\ref{tab:achi} for different values of the gauge coupling $b$. Having these values greatly simplifies future studies in both selecting the parameters in which to simulate and in expressing the results in physical units.

Our results provide the value of the lattice spacing as a function of the inverse lattice 't Hooft coupling $\lambda$. This is precisely the dependence that follows from the $\beta$-function of the theory as given by the following formula: 
\begin{equation}\label{eqn:loga}
   \log{a(\lambda)} + \log\Lambda = -\int^\lambda \frac{\dd x}{\beta(x)},
\end{equation}
where $\Lambda$ is just an integration constant. Thus, by assuming some functional form for the $\beta$-function, integrating it and comparing it with our data values we can determine the parameters appearing in this functional form. This $\beta$-function is the one describing the bare coupling and depends on the discretization procedure, and on the  definition of this bare coupling. However,  perturbation theory predicts the leading behaviour of the function up to next to leading order. For Yang-Mills theory coupled to $N_f$ flavours of adjoint Dirac fermions this gives
\begin{equation}\label{eqn:ptbeta}
    \beta(\lambda) = - b_0 \lambda^2 - b_1 \lambda^3 + \order{\lambda^4} = -\dv{\lambda}{\log{a}},
\end{equation}
with the coefficients  $b_0$ and $b_1$  given by
\begin{align}
    b_0 &= \frac{1}{8\pi^2}\frac{11-4N_f}{3} \label{eqn:b0pietro}, \\
    b_1 &= \frac{1}{128\pi^4}\frac{34-32N_f}{3} \label{eqn:b1pietro}.
\end{align}
There is one particular scheme in which the $\beta$-function is known to all orders. This is the so-called Novikov-Shifman-Vainshtein-Zakharov (NSVZ) $\beta$-function~\cite{Novikov:1983uc}:
\begin{equation}
    \beta(\lambda) = -\frac{b_0\lambda^2}{1-\frac{b_1}{b_0}\lambda}.
\end{equation}
This functional form is particularly well-suited for performing the integration of the inverse of the $\beta$-function which is what is needed for eq.~\eqref{eqn:loga}. One gets 
\begin{equation}\label{eqn:logapar}
    -\log{a} = \frac{1}{b_0\lambda} + \frac{b_1}{b_0^2}\log{\lambda} + \log\Lambda.
\end{equation}
Thus, although we are certainly not in NSVZ scheme, this formula incorporates nicely the universal part of the $\beta$-function and provides a suitable parametrization which allows adding extra terms proportional to the coupling and to higher powers of it.

Now let us apply these ideas to our data. It is well-known that for lattice QCD the naive lattice coupling $\sfrac{1}{b}$ is not particularly well-suited for comparison with the perturbative predictions of scaling. Different authors have proposed improved couplings that do behave much better in this respect. We will then consider these same definitions for the Supersymmetric Yang-Mills theory. One possible choice is the one given in ref.~\cite{Allton:2007py}:
\begin{equation}
    \lambda_I = \frac{1}{b P(b)} \label{eqn:allton},
\end{equation}
where $P(b)$ represents the average value of the plaquette extrapolated to the massless-gluino limit, whose values are reported in table~\ref{tab:achi}. 
Fitting the NSVZ $\beta$-function (with $N_f=\sfrac{1}{2}$) to our 4 data points gives a chi-square per degree of freedom $\chi^2/\#\mathrm{dof}=1.42$, and the single fitted parameter is $(\sqrt{8t_1}\Lambda)^{-1}=45.2(7)$. If we modify the fit to include a higher order term in the $\beta$-function the fit gives a worse $\chi^2/\#\mathrm{dof}$ and the additional parameter comes out compatible with zero. We conclude that our data  does not have enough sensitivity to determine modifications to the NSVZ  $\beta$-function. However, our data does have sensitivity to the leading coefficients of the $\beta$-function. A two-parameter fit leaving $N_f$ and $\Lambda$ free having $\chi^2/\#\mathrm{dof}=1.38$, gives $N_f=0.31(20)$ ($24\pi^2 b_0=9.76(80)$) to be compared with the perturbative result $N_f=0.5$ ($24\pi^2 b_0=11-4N_f=9$).

\begin{figure}
    \centering
    \hspace*{-1.cm}
     \includegraphics[width=\linewidth+2cm]{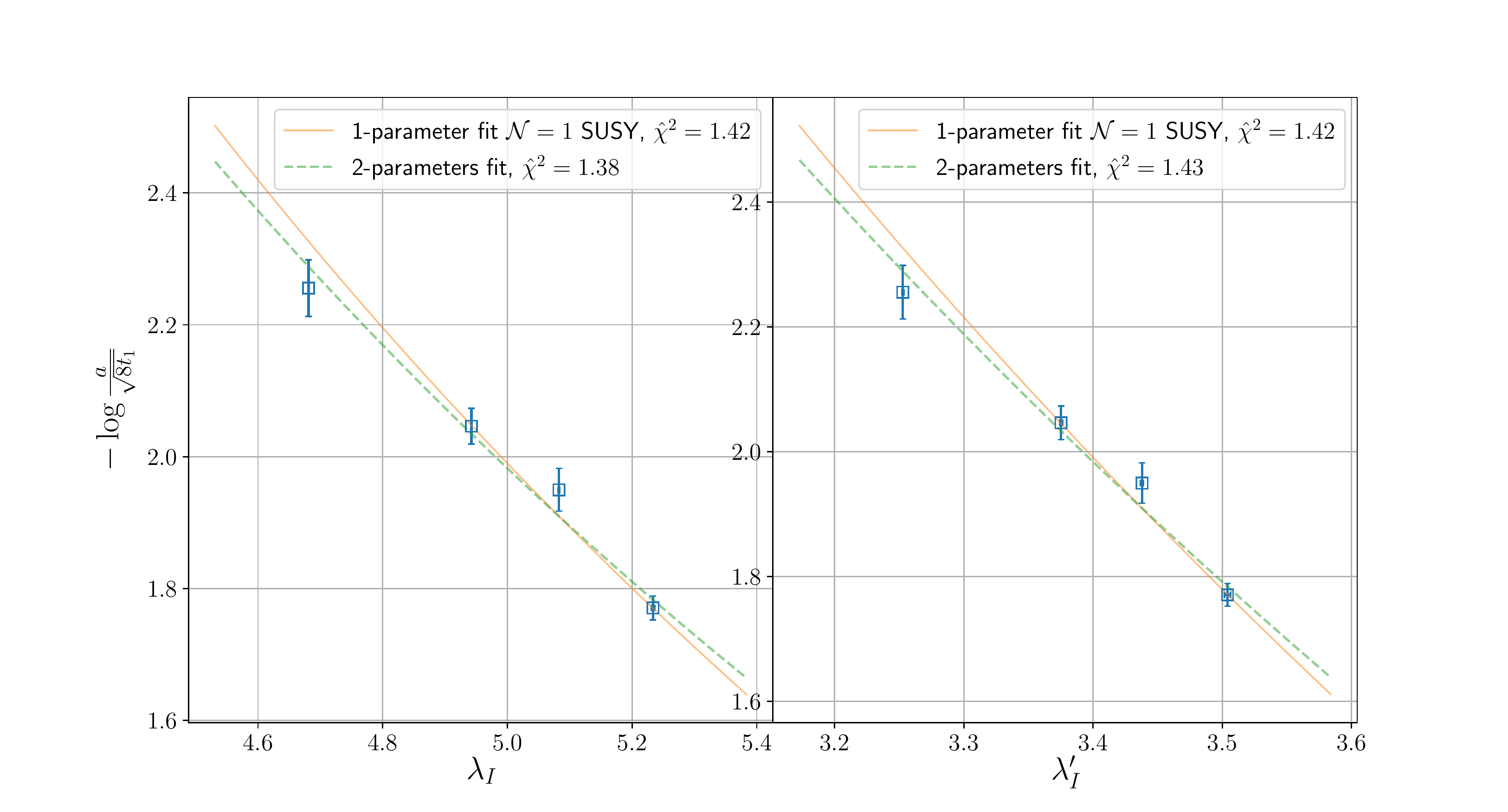}
    \caption{Dependence of the logarithm of the lattice spacing as a function of the improved coupling $\lambda_I$ defined in eq.~\eqref{eqn:allton} (left panel), and $\lambda_I'$ defined in \eqref{eqn:other} (right panel). 
    Points have been fitted with eq.~\eqref{eqn:logapar} leaving only $\Lambda$ free (solid orange line), and with the same analytical form leaving also $N_f$ as a free parameter (dashed green line).
    }
    \label{fig:betafunction}
\end{figure}

One can repeat the procedure with another choice of improved coupling constant, like the one proposed in refs.~\cite{Martinelli:1980tb,Edwards:1997xf}:
\begin{equation}
    \lambda'_I = 8(1-P(b)).
    \label{eqn:other}
\end{equation}
 The one-parameter fit to the NSVZ $\beta$-function gives  $\chi^2/\#\mathrm{dof}=1.42$ with a best fit $(\sqrt{8t_1}\Lambda)^{-1}=472(7)$. Again a two-parameter fit leaving also $N_f$ free gives $N_f=0.30(22)$ ($24\pi^2 b_0=9.78(87)$), completely compatible with that obtained for the other improved coupling. In figure~\ref{fig:betafunction} we display side-by-side the  logarithm of the lattice spacing as a function of both improved couplings together with the lines corresponding the fits described before.

In summary, we emphasize that the behaviour of the scale in the range explored in our study is certainly not far and even compatible with the dependence predicted by perturbation theory. The data also shows the tendency expected by the addition of an adjoint Majorana fermion since $N_f$ comes out larger than the value 0 corresponding to the pure gauge theory and not incompatible with the value 0.5, expected for the SUSY Yang-Mills theory at asymptotic small values of the coupling. This implies that the leading perturbative coefficient agrees within a $10\%$ with the expected value $b_0=9/(24\pi^2)$.

\subsection{Summary}
The purpose of this paper was to study the large $N$ limit of the $\mathcal{N}=1$ SUSY Yang-Mills theory on the lattice. More precisely our goal was the  determination of the dependence of the lattice spacing on the bare coupling constant, a necessary first step in connecting lattice to continuum results. The paper includes dealing with many aspects of a technical nature, although some of them have been moved to appendices to facilitate readability. Also with this idea in mind we consider appropriate to add this short final summary to point to our main results. 

We have used redundancy to ensure the robustness of our final conclusions. Hence, we have computed the lattice spacing using three different and rather independent observables. First we did this for all our different simulation runs having various values of the lattice coupling and the gluino mass. The results are collected in table~\ref{tab:6scales}. Since each observable expresses the lattice spacing in its associated unit,  compatibility demands that the values are proportional to each other with a proportionality constant given by the ratio of units. This is actually happening as can be deduced from the table or more visually from figure~\ref{fig:comparison}. The proportionality is in principle only exact in the continuum limit, but our results show that this holds for our lattice sample within errors. 

The last step is to extrapolate the results to the massless gluino limit giving the desired value of the lattice spacing for each bare lattice coupling (table~\ref{tab:achi}). Remarkably, the resulting dependence of the beta function for two common improved lattice couplings comes close to the prediction of perturbation theory (figure~\ref{fig:betafunction}). This also shows that the range of bare couplings that we have explored is within the precocious scaling regime.

    \section*{Acknowledgments}
  
This work is partially supported by grant PGC2018-094857-B-I00 funded by MCIN/AEI/ 10.13039/501100011033 and by “ERDF A way of making Europe”, and by the Spanish Research Agency (Agencia Estatal de Investigación) through grants IFT Centro de Excelencia Severo Ochoa  SEV-2016-0597 and  No CEX2020-001007-S, funded by MCIN/AEI/10.13039
/501100011033. We also acknowledge support from the project H2020-MSCAITN-2018-813942 (EuroPLEx) and the EU Horizon 2020 research and innovation programme,\linebreak  
STRONG-2020 project, under grant agreement No 824093. M.O. is supported by JSPS KAKENHI Grant Numbers 21K03576.
K.-I.I. is supported by MEXT as ``Program for Promoting Researches on the Supercomputer Fugaku''
(Simulation for basic science: from fundamental laws of particles to creation of nuclei, JPMXP1020200105) and JICFuS.
This work used computational resources of SX-ACE (Osaka U.) and Oakbridge-CX (U. of Tokyo)
through the HPCI System Research Project (Project ID: hp220011, hp210027, hp200027, hp190004) and Subsystem B of ITO system (Kyushu U.).
We acknowledge the use of the Hydra cluster at IFT and HPC resources at CESGA (Supercomputing Centre of Galicia).

    \appendix

\section{Simulation Parameters and Statistics}
\label{app:simulation}

In this appendix we collect extra information regarding the
simulation in addition to the one described in
section~\ref{s:generationofconfs}. In particular, the full set of model
parameters and the number of configurations generated for them are
collected in  table~\ref{tab:ParameterAndStatistics}. The hyper-parameters of the RHMC algorithm and the statistics for each ensemble
are also included in the table.
Each configuration is separated with five trajectories with a
trajectory length of $\tau= 1$.
The number of the MD time steps in the generalized multiple-time step
integrator is denoted by $N_{\mathrm{G}}$ for the gauge action,
and $N_{\mathrm{UV}}$ ($N_{\mathrm{IR}}$) for the UV (IR) part of the
pseudo-fermion action, respectively.
Table~\ref{tab:HMCAccAndLmin} shows the HMC acceptance rate, 
the expectation value of the plaquette, and
the expectation value $\qty|\lambda_{\mathrm{min}}|$, with $\lambda_{\mathrm{min}}^2$ the minimum eigenvalue
 of $Q_W^2=(\gamma_5 D_W)^2$.

The choice of
hopping parameters have been performed  having in mind the ultimate goal
of taking the massless limit. Thus, our $\kappa_{a}$ values
approach the critical value from below. On the opposite edge, for small
hopping parameter the theory becomes equivalent to the TEK model.   
This model exhibits a first order phase transition at a certain
$b_c \sim 0.346$ for $(N,k)=(169,5)$ and below this point one enters a strong coupling region of
the model. Indeed,  this phase transition extends into a   first order phase transition line 
in the current model in the $b$-$\kappa_{a}$ plane.
Hence, we first surveyed the phase diagram for the weak coupling region from which the proper continuum limit can be approached,
and the model parameters in the table are chosen to stay in the weak coupling region.

The RHMC algorithm program was primarily written in Fortran 2003/2008 language and uses OpenMP thread parallelization. 
To speed up the generation of configurations for the cases with $N=361$, the pseudo-fermion force computation in the MD evolution 
is offloaded to the corresponding GPU kernel written in the CUDA language.
To further speed up the configuration generation we made several replica with the same parameter set 
but with different random number sequence so that several computational resources are concurrently 
available on  multi-node systems. We discarded the first 200--500 trajectories in each replica for the thermalization, 
where the seed configurations are chosen in such a way that the system remains in the weak coupling region avoiding the phase transition.

\begin{center}
\begin{longtable}{cccccc}
\hline
      $(N,k)$ &
      $b$ &
      $\kappa_{a}$ &
      $(N_R, N_{\mathrm{split}}, a, b)$ &
      $(N_{\mathrm{G}},N_{\mathrm{UV}},N_{\mathrm{IR}})$ &
      {\# of configs} \\ 
      \hline%\midrule
  (361,7) & 0.350 &  0.1775000 & ( 12, 5, 0.001700, 4.630) & ( 200, 100, 90) & 600 \\
          &       &  0.1800000 & ( 12, 5, 0.000700, 4.700) & ( 200, 100, 90) & 600 \\
          &       &  0.1825000 & ( 14, 5, 0.000500, 4.780) & ( 200, 100, 90) & 600 \\
          &       &  0.1850000 & ( 12, 6, 0.000200, 4.890) & ( 200, 100, 90) & 607 \\
          &       &  0.1875000 & ( 14, 6, 0.000020, 4.960) & ( 200, 100, 90) & 600 \\
%===
          & 0.345 &  0.1800000 & ( 10, 5, 0.002000, 4.645) & ( 200, 100, 90) & 600 \\
          &       &  0.1840000 & ( 10, 5, 0.000800, 4.780) & ( 200, 100, 90) & 712 \\
          &       &  0.1868000 & ( 10, 6, 0.000300, 4.875) & ( 200, 100, 90) & 600 \\
          &       &  0.1896000 & ( 12, 6, 0.000030, 4.980) & ( 200, 100, 90) & 600 \\
%===
          & 0.340 &  0.1850000 & ( 14, 6, 0.000700, 4.790) & ( 200, 100, 90) & 640 \\
          &       &  0.1875000 & ( 14, 6, 0.000400, 4.860) & ( 200, 100, 90) & 638 \\
          &       &  0.1890000 & ( 14, 6, 0.000300, 4.910) & ( 200, 100, 90) & 604 \\
          &       &  0.1910000 & ( 12, 6, 0.000080, 4.990) & ( 200, 100, 90) & 660 \\
          &       &  0.1930000 & ( 12, 6, 0.000005, 5.070) & ( 200, 100, 90) & 625 \\
\hline%\midrule
%===
  (289,5) & 0.360 &  0.1780000 & ( 12, 5, 0.000500, 5.050) & ( 200, 80, 64) & 516 \\
          &       &  0.1800000 & ( 12, 5, 0.000200, 4.800) & ( 200, 80, 64) & 800 \\
          &       &  0.1760000 & ( 12, 5, 0.000800, 4.700) & ( 200, 80, 64) & 600 \\
          &       &  0.1820000 & ( 12, 5, 0.000080, 4.850) & ( 200, 80, 64) & 720 \\
          &       &  0.1831720 & ( 12, 5, 0.000100, 4.863) & ( 200, 80, 64) & 600 \\
          &       &  0.1840000 & ( 12, 5, 0.000020, 4.900) & ( 200, 80, 64) & 800 \\
%===
          & 0.350 &  0.1800000 & ( 12, 5, 0.001000, 4.700) & ( 200, 80, 64) & 680 \\
          &       &  0.1850000 & ( 12, 5, 0.000150, 4.900) & ( 200, 80, 64) & 600 \\
          &       &  0.1775000 & ( 12, 5, 0.001900, 4.620) & ( 200, 80, 64) & 600 \\
          &       &  0.1825000 & ( 12, 5, 0.000500, 4.800) & ( 200, 80, 64) & 600 \\
          &       &  0.1863780 & ( 12, 5, 0.000100, 4.905) & ( 200, 80, 64) & 600 \\
          &       &  0.1875000 & ( 12, 5, 0.000040, 5.000) & ( 200, 80, 64) & 680 \\
%===
          & 0.345 &  0.1800000 & ( 12, 5, 0.002000, 4.650) & ( 200, 80, 64) & 624 \\
          &       &  0.1840000 & ( 10, 5, 0.000800, 4.785) & ( 200, 80, 64) & 600 \\
          &       &  0.1868000 & ( 12, 5, 0.000300, 4.880) & ( 200, 80, 64) & 600 \\
          &       &  0.1896000 & ( 12, 5, 0.000050, 4.980) & ( 200, 80, 64) & 600 \\
%===
          & 0.340 &  0.1850000 & ( 10, 5, 0.000900, 4.780) & ( 200, 80, 64) & 610 \\
          &       &  0.1875000 & ( 10, 5, 0.000400, 4.860) & ( 200, 80, 64) & 600 \\
          &       &  0.1890000 & ( 12, 5, 0.000200, 4.920) & ( 200, 80, 64) & 608 \\
          &       &  0.1910000 & ( 12, 5, 0.000080, 4.990) & ( 200, 80, 64) & 600 \\
          &       &  0.1920670 & ( 12, 5, 0.000050, 5.030) & ( 200, 80, 64) & 610 \\
          &       &  0.1930000 & ( 12, 5, 0.000005, 5.070) & ( 200, 80, 64) & 630 \\
\hline%\midrule
%===
  (169,5) & 0.350 &  0.1775000 & ( 10, 3, 0.001000, 4.700) & ( 100, 80, 40) & 600 \\
          &       &  0.1800000 & ( 10, 3, 0.000800, 4.750) & ( 100, 80, 40) & 600 \\
          &       &  0.1825000 & ( 10, 3, 0.000400, 4.820) & ( 100, 80, 40) & 600 \\
          &       &  0.1850000 & ( 10, 3, 0.000150, 4.890) & ( 100, 80, 40) & 700 \\
          &       &  0.1875000 & ( 10, 3, 0.000080, 4.980) & ( 100, 80, 40) & 600 \\
%===
          & 0.340 &  0.1850000 & ( 10, 3, 0.000900, 4.780) & ( 100, 80, 40) & 600 \\
          &       &  0.1875000 & ( 10, 3, 0.000300, 4.870) & ( 100, 80, 40) & 600 \\
          &       &  0.1890000 & ( 10, 3, 0.000250, 4.930) & ( 100, 80, 40) & 610 \\
          &       &  0.1910000 & ( 10, 3, 0.000050, 5.000) & ( 100, 80, 40) & 610 \\
          &       &  0.1930000 & ( 10, 3, 0.000020, 5.075) & ( 100, 80, 40) & 600 \\
\hline%      \bottomrule  
%    \end{tabular}
    \caption{Data sample: For each value of $N$ and twist factor $k$, we list the bare coupling $b$, the adjoint hopping parameter $\kappa_a$,  the set of RHMC hyper-parameters used in the generation of configurations and the final number of configurations for each ensemble.}
    \label{tab:ParameterAndStatistics}
%\end{table}
\end{longtable}
\end{center}

\begin{center}
\begin{longtable}{cccccc}
%\begin{table}[H]
%    \centering
%    \small
%    \begin{tabular}{cccccc}
\hline %      \toprule
      $(N,k)$ &
      $b$ &
      $\kappa_{a}$ &
      {HMC acc.} & $\expval{P}$ &
      $\expval{\qty|\lambda_{\mathrm{min}}|}$ \\ \hline % \midrule
(361, 7) & 0.350 & 0.1775000 &  0.7723(98)  &   0.561689(73)  &   0.049554(65)  \\
         &       & 0.1800000 &  0.766(10)   &   0.564610(70)  &   0.038933(51)  \\
         &       & 0.1825000 &  0.7475(92)  &   0.568020(55)  &   0.028445(42)  \\
         &       & 0.1850000 &  0.7400(86)  &   0.571769(42)  &   0.018722(79)  \\
         &       & 0.1875000 &  0.745(11)   &   0.576334(91)  &   0.011781(77)  \\
         & 0.345 & 0.1800000 &  0.726(13)   &   0.55072(10)   &   0.049524(85)  \\
         &       & 0.1840000 &  0.7340(80)  &   0.556355(81)  &   0.032503(57)  \\
         &       & 0.1868000 &  0.7137(96)  &   0.561253(71)  &   0.020603(59)  \\
         &       & 0.1896000 &  0.733(11)   &   0.567106(91)  &   0.011117(97)  \\
         & 0.340 & 0.1850000 &  0.7530(85)  &   0.541414(99)  &   0.040589(71)  \\
         &       & 0.1875000 &  0.7402(91)  &   0.546237(83)  &   0.029619(64)  \\
         &       & 0.1890000 &  0.720(11)   &   0.549479(88)  &   0.023011(64)  \\
         &       & 0.1910000 &  0.7248(81)  &   0.55419(10)   &   0.014393(64)  \\
         &       & 0.1930000 &  0.7341(85)  &   0.560145(74)  &   0.008346(73)  \\
\hline % \midrule
%===
(289, 5) & 0.360 & 0.1760000 &  0.8060(78)  &   0.583233(62)  &   0.039755(62)  \\
         &       & 0.1780000 &  0.814(10)   &   0.585296(89)  &   0.031968(77)  \\
         &       & 0.1800000 &  0.8057(75)  &   0.587629(60)  &   0.024920(65)  \\
         &       & 0.1820000 &  0.795(11)   &   0.590255(64)  &   0.019453(67)  \\
         &       & 0.1831720 &  0.7820(85)  &   0.591912(75)  &   0.017702(68)  \\
         &       & 0.1840000 &  0.8023(63)  &   0.593188(51)  &   0.017090(80)  \\
         & 0.350 & 0.1775000 &  0.8040(97)  &   0.561674(93)  &   0.050168(76)  \\
         &       & 0.1800000 &  0.7726(76)  &   0.564575(72)  &   0.039760(56)  \\
         &       & 0.1825000 &  0.769(13)   &   0.568034(85)  &   0.029459(77)  \\
         &       & 0.1850000 &  0.7843(79)  &   0.57191(10)   &   0.020167(95)  \\
         &       & 0.1863780 &  0.776(11)   &   0.57421(10)   &   0.016132(79)  \\
         &       & 0.1875000 &  0.7956(74)  &   0.576383(73)  &   0.01427(11)   \\
         & 0.345 & 0.1800000 &  0.7917(93)  &   0.55045(12)   &   0.050209(83)  \\
         &       & 0.1840000 &  0.7477(81)  &   0.55636(10)   &   0.033087(77)  \\
         &       & 0.1868000 &  0.785(10)   &   0.56126(11)   &   0.021689(91)  \\
         &       & 0.1896000 &  0.7403(99)  &   0.567256(95)  &   0.013094(99)  \\
         & 0.340 & 0.1850000 &  0.7141(84)  &   0.541611(88)  &   0.040868(72)  \\
         &       & 0.1875000 &  0.752(10)   &   0.546422(99)  &   0.029998(74)  \\
         &       & 0.1890000 &  0.7638(88)  &   0.54981(14)   &   0.02343(10)   \\
         &       & 0.1910000 &  0.7347(89)  &   0.554363(99)  &   0.015480(69)  \\
         &       & 0.1920670 &  0.7793(93)  &   0.557248(99)  &   0.011992(77)  \\
         &       & 0.1930000 &  0.7803(86)  &   0.56039(14)   &   0.010557(92)  \\
\hline % \midrule
%====
(169, 5) & 0.350 & 0.1775000 &  0.832(10)   &   0.56191(18)   &   0.05322(12)   \\
         &       & 0.1800000 &  0.8357(72)  &   0.56479(16)   &   0.04353(10)   \\
         &       & 0.1825000 &  0.845(10)   &   0.56834(15)   &   0.034286(94)  \\
         &       & 0.1850000 &  0.8503(68)  &   0.57194(14)   &   0.02676(11)   \\
         &       & 0.1875000 &  0.8333(86)  &   0.57680(12)   &   0.022537(83)  \\
         & 0.340 & 0.1850000 &  0.8060(78)  &   0.54165(23)   &   0.04309(15)   \\
         &       & 0.1875000 &  0.8367(85)  &   0.54666(23)   &   0.03270(17)   \\
         &       & 0.1890000 &  0.8292(80)  &   0.55007(19)   &   0.02691(11)   \\
         &       & 0.1910000 &  0.8085(91)  &   0.55495(17)   &   0.020784(92)  \\
         &       & 0.1930000 &  0.8447(74)  &   0.56003(28)   &   0.017592(78)  \\
\hline %      \bottomrule  
%    \end{tabular}
    \caption{For each value of $N$,  $b$ and $\kappa_a$ we list the HMC acceptance rate, the expectation value of the plaquette $\expval{P}$, and $\expval{\qty|\lambda_{\mathrm{min}}|}$, with $\lambda^2_{\mathrm{min}}$ the lowest eigenvalue of $Q_W^2= (D_W \gamma_5)^2$.}
    \label{tab:HMCAccAndLmin}
%\end{table}
\end{longtable}
\end{center}

%%% Local Variables:
%%% mode: japanese-latex
%%% TeX-master: t
%%% End:

    \section{Finite size effects in the gradient flow}
\label{ap:flow-imp}

 In this appendix we summarize the basic ingredients for the derivation of the improved expression of the flow presented in
eq.~\eqref{improved-phi} and provide also explicit formulas for computing the normalization constant $\mathcal{N}_L (x,N)$ on the lattice.

The perturbative expansion of the infinite volume flow in terms of the $\MS$ 't Hooft coupling 
at scale $\mu=1/\sqrt{8t}$ is given by~\cite{Luscher:2010iy}:
\be
    \Phi_\infty (t,N) =  \mathcal{K}(N)  \lambda_\MS(\mu) \left( 1 + c_1 \lambda_\MS (\mu)  \right)\Big |_{\mu=1/\sqrt{8t}},
    \label{eq:pt-flow-inf}
\ee 
where:
\be
\mathcal{K}(N) = \frac{3 (N^2 -1)}{128 \pi^2 N^2}.
\label{eq:norm-inf}
\ee
where $N$ stands for the number of colours.
This expression is used to define the \textit{gradient flow} (GF) coupling constant~\cite{Luscher:2010iy}: 
\be
\lambda_\GF (\mu=1/\sqrt{8t}) \equiv \frac{ \Phi_\infty(t) } {\mathcal{K}(N)}=\lambda_\MS(\mu) \left( 1 + c_1 \lambda_\MS (\mu)  \right) ,
\ee
showing that $c_1$ stands for the finite one-loop renormalization constant that relates the $\Lambda$ parameters in the two schemes. 

One can derive an analogous expression for the finite volume flow. In our case, on a box of size $l^4$ with twisted boundary conditions as the ones used in this work, one obtains at second order in the coupling~\cite{Bribian:2019ybc}:
\begin{equation}
    \Phi(t,l, N) = \mathcal{N} \left(c(t), N\right)  \lambda_\MS(\mu) \left \{1 + c_1 \lambda_\MS(\mu) + \mathcal{C}\left (c(t), N \right) \lambda_\MS(\mu) \right  \}|_{\mu=1/\sqrt{8t}},
\end{equation}
where finite volume effects depend on the dimensionless variable:
\be
c(t)=\frac{\sqrt{8 t}}{\tl}, \quad \tl\equiv \sqrt{N}l,
\ee
 and are encoded in the functions $\mathcal{N} (c(t), N)$ and $\mathcal{C}(c(t), N )$.  
The former has a simple expression in terms of Jacobi $\theta_3$ functions 
and reads:
\be
\mathcal{N} (x,N) = \frac{3x^2}{128} \left ( \theta_3^4(0,i \pi x)- \theta_3^4(0,i \pi N x) \right),
\ee
with:
\be
\theta_3(0,ix) =  \sum_{m\in\mathbb{Z}} e^{-\pi x m^2}  \equiv \frac{1}{\sqrt{x}}  \sum_{m\in\mathbb{Z}} e^{-\pi m^2/x}
.
\ee
In the large volume limit, taken by sending $l$ to infinity at fixed flow time, i.e. by sending $c(t)$ to zero,  this normalization factor tends to the infinite volume one, c.f. eq.~\eqref{eq:norm-inf}.
The function $\mathcal{C}(c(t), N )$ goes to zero in this limit and the infinite volume expansion of the flow presented above is recovered.
Let us also mention that, on account of volume independence, the same expressions can be obtained for $SU(\infty)$ 
by taking the $c(t)$ to zero limit in a different way, i.e. by sending $N$ to infinity at fixed torus size $l$ (in the particular case of the one-point lattice, $l=a$).  In that limit, the explicit $N$ dependence of the normalization $\mathcal{K}( N )$ factor disappears.

This exercise indicates a simple way to correct the flow at leading order in the coupling. 
We introduce the quantity
\begin{equation}
    \hat {\lambda}(t,l,N) = \frac{1}{\mathcal{N}(c(t),N)} {\Phi}(t,l,N),
\end{equation}
which, at leading order in the coupling, has the correct  perturbative expansion
and we define a modified flowed energy density given by:
\be
\hat {\Phi}(t,l,N) = \mathcal{K}(N) \hat {\lambda}(t,l,N). \ee 
This removes finite size effects at tree level. At second order in $\lambda$, the remnant volume dependence  comes from the $c(t)$ dependence of the function $\mathcal{C}\left (c(t), N \right)$, which has not yet been computed in perturbation theory for the case envisaged in this work (with a dynamical Majorana fermion and the symmetric twist). As discussed in section~\ref{s:analysis-flow}, within the numerical accuracy we have achieved, this correction is very small for values of $c(t)\lesssim 0.3$.

On the lattice, an analogous correction, which in addition takes into account lattice artefacts, is derived by computing the tree level factor $\mathcal{N}(c(t),N)$ in lattice perturbation theory. For our choice of the clover discretization of the energy density and 
for the one-point lattice~\cite{GarciaPerez:2014azn}:
\be
\mathcal{N}_L(x,N) = \frac{x^4}{128} \sum_{\mu \ne \nu} \sum'_q e^{- \frac{N x^2}{4} \widehat q^2} 
\, \,  \sin^2 (q_\nu) \cos^2 (q_\mu/2)  \, \frac{1}{\widehat q^2},
\label{lattnorm}
\ee
where the lattice momentum is given by $ \widehat q_\mu = 2 \sin(q_\mu/2)$, with $q_\mu$ taking values:
\be
q_\mu = \frac{2 \pi m_\mu}{\sqrt{N}}\, ,
\ee
for $m_\mu  = 0, \cdots,  \sqrt{N}-1$. The prime in the sum excludes the zero momentum mode, with $m_\mu  = 0$, $\forall \mu$.
With all this, our final formula for the discretized version of the flow is given by eq.~\eqref{improved-phi}.

    \section{Fitting strategy to determine the flow-based scale}
\label{app:t1_all_flow}

In this appendix we discuss the fitting strategy employed to determine the scales based on the gradient flow and present a detailed account of the extraction of $T_1$. 

 As mentioned in section~\ref{s:tekflow},  our results show that one can use a universal fitting function to describe the flow-time dependence of the energy density, allowing to extract the scale even in the cases where $T_1$ can only be reached by extrapolation. 
 The parameterization of the flow-time dependence we have used, to be described below, relies on the connection between the infinite volume flow and the gradient flow renormalized coupling constant $\lambda_\GF$, c.f. eq.~\eqref{gf-coupling}~\footnote{For the rest of this section and for simplicity we will drop the sub-index $\GF$ and refer to the gradient flow coupling as $\lambda$.}. 

Starting from the renormalization group equation defining the $\beta$ function and integrating it between two different reference scales $t_s$ and $t$ one arrives at:
\be
\int_{\lambda(t_s)}^{\lambda(t)} \frac{dx}{\beta(x)} = \half \log \left ( \frac{t_s}{t} \right).
\label{betaint}
\ee
The left hand side of this equation can be easily integrated using a general parameterization of the $\beta$-function
inspired by the Novikov-Shifman-Vainshtein-Zakharov (NSVZ) $\beta$-function~\cite{Novikov:1985rd}:
\be
\beta(\lambda) = - \frac{b_0 \lambda^2}{1 - \sum_{k=0}^{n_b} a_k \lambda^{k+1}},
\ee
with coefficients chosen to reproduce the universal expansion of the $\beta$-function to second order in $\lambda$, c.f.
$a_0$ is set to $b_1/b_0$, where $b_0$ and $b_1$ stand for the first two universal coefficients of the $\beta$-function of the $\mathcal{N}=1$ SUSY Yang-Mills theory, c.f.  eqs.~\eqref{eqn:b0pietro},~\eqref{eqn:b1pietro} with $N_f=1/2$.

\begin{table}
\centering
    \begin{tabular}{c c c c c c }
    \toprule
    $b$ & $\kappa_a$ & $T_1 (N=289)$ &  $T_1 (N=361)$ &$T_1^0 (N=289)$  & $T_1^0 (N=361)$ \\
     \midrule
       0.340  &    0.185000  & 2.878(2) (46)   & 2.883(2) (58)  &-- & --\\
       0.340  &    0.187500  & 3.209(3) (68)   & 3.188(3) (90)  & 3.2014(12) & 3.1818(10)\\ 
       0.340  &    0.189000  & 3.514(3) (65)   & 3.488(3) (88)  & 3.5108(20) & 3.4785(13)\\
       0.340  &    0.191000  & 4.049(4) (35)   & 4.042(4) (58) & 4.0484(81) & 4.0420(44)\\
       0.340  &    0.192067  & 4.568(6) (58)   &  --&-- &--  \\
       0.340  &    0.193000  & 5.273(7)(160)   & 5.244(8)(175) &-- & -- \\
\midrule
       0.345  &    0.180000  & 3.166(3) (74)   & 3.145(2) (93)  & 3.1737(10) & 3.1639(7)\\
       0.345  &    0.184000  & 3.664(4) (77)   & 3.645(3) (98)  & 3.6698(33) & 3.6552(19) \\
       0.345  &    0.186800  & 4.294(5) (40)   & 4.274(5) (55)  & 4.2804(65)& 4.2709(44) \\
       0.345  &    0.189600  & 5.559(8)(154)   & 5.614(8)(201)& -- & --  \\
\midrule
       0.350  &    0.177500  & 3.730(3) (97)   & 3.737(3)(105)  & 3.7409(36)& 3.7536(20) \\
       0.350  &    0.180000  & 4.109(4) (75)   & 4.003(4)(109) & 4.0898(59)& 4.0116(37) \\
       0.350  &    0.182500  & 4.634(6) (70)   & 4.516(5) (64) & 4.5889(50)& 4.4948() \\
       0.350  &    0.185000  & 5.364(8)(144)   & 5.323(7) (65)   & -- & 5.3155()\\
       0.350  &    0.186378  & 5.909(9)(191)   & -- & --&-- \\
       0.350  &    0.187500  & 6.608(11)(249)  & 6.582(12)(287)& --& -- \\
\midrule
       0.360  &    0.176000  & 5.77(1)(32) & -- &-- &-- \\
       0.360  &    0.178000  & 6.32(1)(31) & -- &-- & --\\
       0.360  &    0.180000  & 7.05(1)(31) & -- &-- & --\\
       0.360  &    0.182000  & 7.85(2)(35) & -- &-- & --\\
       0.360  &    0.183172  & 8.68(2)(44) & -- &-- &-- \\
       0.360  &    0.184000  & 9.16(2)(52) & -- &-- & --\\
    \bottomrule
    \end{tabular}
    \caption{Values of the scale $T_1$ determined from the flow data with $N=289$ and $N=361$. Results labelled as $T_1$ come from a  fit to eq.~\eqref{fit-pert} with 3 free parameters of the $\beta$-function on top of the two universal ones, as described in the text. Values labelled as $T_1^0$ are directly determined by interpolation in the cases in which $T_1$ falls within the scaling window given by eq.~\eqref{fit-window} with $\gamma=0.28$.}
    \label{tab:t1_all_flow}
\end{table}

After integration, one arrives at the following identity:
\be
{\cal H} (t) \equiv \frac{1}{ \lambda(t)}- \frac{1}{\lambda(t_s)  } + a_0 \log \left (\frac{\lambda(t)}{\lambda(t_s)}\right) + \sum_{k=0}^{n_b} \frac{a_k}{k}
\left(\lambda^k (t) - \lambda^k(t_s) \right ) + \frac{b_0}{2} \log \left (\frac{t}{t_s} \right )=0.
\label{fit-pert}
\ee
One can use this identity to fit the numerical results for the flow, leaving as free parameters the non-universal coefficients of the $\beta$-function and the $t_s$ scale. 
In the continuum, at infinite volume, and for one masless  Majorana fermion, this relation is universal and dictated by the non-perturbative $\beta$-function of the $\mathcal{N}=1$ SUSY Yang-Mills theory in the gradient flow scheme. Departures from universality arise due to lattice artefacts, finite volume effects and the fermion mass-dependence of the flow. Our results, presented in section~\ref{s:tekflow}, indicate that violations of universality remain small, within the numerical accuracy we have been able to attain, as long as one stays within the scaling window given by eq.~\eqref{fit-window}.

To implement the fitting procedure and determine the lattice scale $T_1(b,\kappa_a,N)$, we set $\lambda(t_s)=0.05/\mathcal{K}(\infty)$ and minimize the $\chi^2$ function defined as:
\be
\chi^2 = \sum_{T} \left ( \frac{ \mathcal{H} (T)   }{\delta\mathcal{H}(T)} \right) ^2,
\ee  
where:
\be
\delta\mathcal{H}(T) =  b_0 \, \delta \lambda (T)  \, |\beta|^{-1},
\ee  
in a  procedure analogous to the one used to parameterize the step scaling function in 
ref.~\cite{DallaBrida:2019wur}.

Our final determination of $T_1$ is obtained by fitting simultaneously in this way all our simulations at fixed value of $N$, all values of $b$ and all values of the hopping parameter $\kappa_a$. For the final fit, the  $\beta$-function has been parameterized with 3 free coefficients, in addition to the two universal ones. Fits were performed restricting the data for different values of $b$ and $\kappa_a$ to the corresponding scaling window $T\in[1.25, \gamma^2 N/8]$, with $\gamma=0.28$, as shown in figure~\ref{fig:t1_all_flow}. 

Our final results are given in table~\ref{tab:t1_all_flow}.
We give separately results of the scale determined from $N=289$ and $N=361$ simulations, whose compatibility serves as a test of the absence of finite size effects.
 In addition to the values obtained from the universal fitting function, we also provide the values $T_1^0$ obtained by interpolation in those cases in which the scale falls well within the scaling window. We have assigned a systematic error to the final result that covers for the difference between them as well as for other types of fits and fitting ranges. Among those, we have included fits to eq.~\eqref{fit-pert} with only one and two free parameters and also a second-degree polynomial fit performed in a reduced fitting window corresponding to $\gamma=0.22$. In addition, separate fits have also been performed, including joint fits to all datasets at fixed value of the bare coupling $b$. The final systematic error quoted in the table remains in general below a $3\%$ relative error,  going up to about 6$\%$ for the, in physical units,  smaller lattices.
    \section{Meson correlators in the reduced model}
\label{app:meson_correlator}

In this appendix we give technical details about computing 
meson correlators in both the 
fundamental and adjoint representation  and their subsequent use to 
obtain the masses of the lowest state with the corresponding quantum
numbers. The underlying ideas and derivations are given in the text
and references supplied. 
For the case of fundamental representation fermions there have been 
extensive studies done previously~\cite{Gonzalez-Arroyo:2015bya,Perez:2020vbn} using the same techniques
applied in the paper. For that reason we will mostly focus on the adjoint
representation formulas.

The meson correlation functions are expectation values of products of
fermion bilinears separated in time and averaged over space. The correlation function
is computed as a Fourier transform in time 
\begin{equation}\label{eqn:correlatorij}
\mathcal{C}_{AB}^{i j}(n_0)=  \sum_{q_0} \eu^{-\iu q_0  n_0 }
\expval{\Tr\qty[ \mathbf{O}_A^{(i)} D_W^{-1}(q_0)  \mathbf{O}_B^{(j)}
D_W^{-1}(0)]},
\end{equation}
where the trace is over spin, space and colour degrees of freedom. 
In this work we take the time period to be $2\sqrt{N}$ so that the
temporal momentum  takes values $q_0= \pi m/\sqrt{N}$ with integer $m$.
The symbols $A$ and $B$ specify the spin-parity quantum numbers of the 
operator and the indices $i$ and $j$ run over a family of operators
with the same quantum numbers. The symbol $D_W^{-1}(0)$ denotes the
inverse of the Wilson-Dirac operator  of the corresponding fermion (fundamental or adjoint). 
For the reduced model this inverse greatly simplifies. In the case of
the adjoint the operator is simply the one appearing in
eq.~\eqref{eq:AdjWDmatrix}. The inversion  is performed using the BiCGStab algorithm or the Conjugate Gradient algorithm whenever the former does not converge.

 One can define $D_W(p)$ as the corresponding
operator with the substitution
\begin{equation}
U_\mu^{\mathrm{adj}} \longrightarrow U_\mu^{\mathrm{adj}} e^{i p_\mu}.
\end{equation}
These operators and their inverses
can be used to allow  these valence fermions to live in an arbitrary
lattice. One has simply to average over  the
corresponding family of Fourier modes $p$
\begin{equation}
\mathcal{C}_{AB}^{i j}(n_0)=  \sum_{q_0} \eu^{-\iu q_0  n_0 }
\frac{1}{|\Lambda_p|}\sum_{p \in \Lambda_p}
 \expval{\Tr\qty[ \mathbf{O}_A^{(i)}(p) D_W^{-1}(\vec{p},p_0+q_0)
 \mathbf{O}_B^{(j)}(p)
D_W^{-1}(p)]}.
\end{equation}
If $|\Lambda_p|$, the number of elements in $\Lambda_p$, is large,  this
might imply many inversions. In practice,  what we do is to perform this
average stochastically: For each configuration we generate $p$ randomly
and use it to perform the inversion of $D_W(p)$. 

For the case of fundamental fermions things are very similar except for
the expression of the Wilson-Dirac operator. We refer the reader to
the literature for further details~\cite{Gonzalez-Arroyo:2015bya,Perez:2020vbn}.

Now we have to specify the selection of operators $\mathbf{O}_A^{(i)}$
used to obtain the masses.  This is based in applying Wuppertal
smearing~\cite{Gusken:1989ad,Gusken:1989qx,Bali:2016lva} to the bilinear operator combined with 
three-dimensional APE smearing to the link variables~\cite{APE:1987ehd}. The particular
application to the reduced model in the adjoint representation 
is explained below.

\subsection*{Wuppertal fermion smearing in adjoint representation}
\label{subsec:wuppertalsmear}
This amounts  to replacing the fermion  bilinear associated to an
element of the Clifford algebra $\gamma_A$ as follows
\begin{equation}\label{eqn:fsmearing}
\overline{\Psi}\gamma_A\Psi \longrightarrow \overline{\Psi}
\mathbf{O}_A^{(i)} \Psi \equiv 
\overline{\Psi}\gamma_A  (\tilde{M}(\bm{p})) ^{s_i} \Psi,
\end{equation}
where the single step smearing operator is given  by 
\begin{equation}
\tilde{M}(\bm{p})=  \dfrac{1}{1+6 c}
  \qty[\mathbf{I} + c\sum_{k=1}^{3}\qty[\eu^{\iu p_k } \bar{U}^{\mathrm{adj}}_k +
   \eu^{-\iu p_k } (\bar{U}^{\mathrm{adj}}_k )^\dag] ],
     \label{eq:LapH}
\end{equation}
and the $s_i$ are integers. In this work we have used  $c=0.5$ and the following
list of values: 
\begin{table}[H]
   \centering
   \begin{tabular}{ccccccccccc}
\toprule
               $i$ & 0 & 1 & 2 &  3 &  4 &  5 &  6 & 7 & 8 & 9\\\midrule
$s_i$ & 0 & 1 & 4 & 16 & 36 & 64 & 100 & 144 & 196 & 256 \\
\bottomrule
   \end{tabular}
\end{table}
In the case of adjoint fermion we only used up to 6 operators, while for fundamental ones we used up to 9.
The symbol $\bar{U}_k$ appearing in eq.~\eqref{eq:LapH} corresponds to the 
10 times APE-3d smeared link to be explained below.

\subsection*{APE-3d link smearing}
\label{subsec:ape}
This smearing procedure is an iterative one which in our case maps three  spatial SU($N$) matrices onto new ones:
\begin{equation}
 U_i^{(s)} \longrightarrow  U_i^{(s+1)}=
 \mathcal{P}\qty[(1-f)U^{(s)}_i +
 \frac{f}{4}\sum_{j\neq i }
              \qty(z_{ji}^{*} U^{(s)}_j U^{(s)}_i U_j^{{(s)}\dagger} +
	      z_{ji} U_j^{{(s)}\dagger}
	      U^{(s)}_i U^{(s)}_j)],
\end{equation}	      
where $\mathcal{P}$ is an operator projecting on SU($N$). The starting
point of the iteration are the link matrices  $U_i^{(0)}=U_i$. In this
paper, for the case of adjoint fermion correlators, we chose  $f=0.081$ and stopped after 10 iterations
$\bar{U}_i=U^{(10)}_i$, which is then transformed into the adjoint
representation and replaced in eq.~\eqref{eq:LapH}. 
For the adjoint fermion, the projection to SU($N$) matrix is not
required and it is enough to project onto U($N$) as follows
\begin{equation}
\mathcal{P}(W)= W (W^\dag W)^{-1/2}.
\end{equation}
For the case of fundamental meson correlators we applied the same method using $f=0.15$.

\subsection*{GEVP methodology}

We expect the signal to decay in time as an (infinite) sum of exponentials, corresponding to the contribution of the ground state plus other heavier excited states. Given the typical hierarchy in the mass spectrum, we expect the excited states to decay faster than the ground state, whose signal dominates for large-enough time separation. We want to extract the mass of the ground state from an exponential fit in a region where excited states give no systematic contribution. In order to do so, one has to maximize the projection onto the ground state to have a single-exponential decay setting in early on. Fermion smearing in eq.~\eqref{eqn:fsmearing} should provide an improvement in this regard since it increases the overlap onto the ground state wave function. Beyond that, we make use of a variational approach, in the same philosophy of ref.~\cite{Perez:2020vbn}. Using the definition of the correlator matrix given in eq.~\eqref{eqn:correlatorij}, where the index $i$ and $j$ run over the smearing levels, we applied the so-called GEVP method.
Given two timeslices $\tau_0$ and $\tau_1$ (with $\tau_0<\tau_1$) (not to be confused with gradient flow scales) we solve numerically the generalized eigenvalue problem (GEVP)
\begin{equation}
    \mathcal{C}^{ij}(\tau_1) {\bm v}_j^{(n)} = \lambda^{(n)} \mathcal{C}^{ij}(\tau_0) {\bm v}_j^{(n)} \quad\text{at fixed $\tau_0, \tau_1$},
\end{equation}
where ${\bm v}^{(n)}$ and $\lambda^{(n)}$ are the eigenvectors and the eigenvalues, respectively for a given choice of $\tau_0$ and $\tau_1$ (in order to lighten the notation, we also omitted the indices $A$ and $B$ that appeared in eq.~\eqref{eqn:correlatorij}).  In this work we used $\tau_0=a$ and $\tau_1=2a$. Among the basis composed by the eigenvectors {${\bm v}^{(n)}$}, we choose the one whose corresponding eigenvalue is the biggest among the others. Let us denote this maximum eigenvector with ${\bm v}^\text{max}$, which we use to define an \textit{optimal} operator by rotating the original correlator-matrix
\begin{equation}
    \mathcal{C}_\text{opt}(n_0,\tau_1,\tau_0) = {{\bm v}_i^\text{max}}^* \mathcal{C}^{ij}(n_0,\tau_1,\tau_0) {\bm v}_j^\text{max}.
\end{equation}
The ground state mass is extracted from the exponential decay at large time of this correlator.

    \FloatBarrier
    \bibliography{main.bbl}

\end{document}